\documentstyle[amssymb,12pt]{article}

\newlength{\minitwocolumn}
\font\teneufm=eufm10
\font\seveneufm=eufm7
\font\fiveeufm=eufm5
\newfam\eufmfam
\textfont\eufmfam=\teneufm
\scriptfont\eufmfam=\seveneufm
\scriptscriptfont\eufmfam=\fiveeufm

\makeatletter
\@addtoreset{equation}{section}
\makeatother

\newtheorem{thm}{Theorem}[section]
\newtheorem{prop}[thm]{Proposition}

\newtheorem{df}{Definition}[section]

\title{\bf The Intergals of Motion for\\
the Deformed $W$-Algebra $W_{q,t}(\widehat{sl_N})$~II:\\
\begin{large}
\bf Proof of the commutation relations
\end{large}}
\begin{document}
\maketitle
\begin{center}
{\it
Dedicated to 
Professor Tetsuji Miwa
on the occasion on the 60th birthday}
\\~\\
{
T. KOJIMA$~^{\alpha}$~and
~J. SHIRAISHI$~^{\beta}$}
\\~\\
{\it
$~^\alpha$
Department of Mathematics,
College of Science and Technology,
Nihon University,\\
Surugadai, Chiyoda-ku, Tokyo 101-0062, 
JAPAN\\
$~^\beta$
Graduate School of Mathematical Science,
University of Tokyo, \\
Komaba, Meguro-ku, Tokyo, 153-8914,
JAPAN
}
\end{center}
~\\
\begin{abstract}
We explicitly construct
two classes of infinitly many commutative operators
in terms of the deformed $W$-algebra $W_{q,t}(\widehat{sl_N})$,
and give proofs of the commutation relations
of these operators.
We call one of them local integrals of motion
and the other nonlocal one,
since they can be regarded as elliptic deformations of local
and nonlocal integrals of motion for
the $W_N$ algebra \cite{BLZ,BHK}.
\end{abstract}

\newpage

\section{Introduction}

This is a continuation of the papers
\cite{FKSW1, FKSW2}, hereafter referred to
as Part 1
\cite{FKSW1} and 
Part 2 \cite{FKSW2}.
In Part 1 we constructed
two classes of infinitly many commutative operators,
in terms of the deformed Virasoro algebra.
In Part 2 we announced conjecturous formulae of
two classes of infinitly many commutative operators,
in terms of the deformed $W$ algebra
$W_{q,t}(\widehat{sl_N})$,
which is the higher-rank generalization of
Part 1 \cite{FKSW1}.
We call one of them local integrals of motion
and the other nonlocal one,
since they can be regarded as elliptic deformations of local
and nonlocal integrals of motion for
the $W_N$ algebra \cite{BLZ,BHK}.
In this paper
we give
proofs of the commutation relations
for the integrals of motion for
the deformed $W$ algebra
$W_{q,t}(\widehat{sl_N})$.

Let us recall some facts about soliton equation
and its quantization.
B.Feigin and E.Frenkel \cite{FF1}
considered the so-called local integrals of motion $I^{(cl)}$ 
for the Toda field theory 
associated with the root system of finite and affine type
$\{I^{(cl)},H^{(cl)}\}_{P.B.}=0$,
where $H^{(cl)}=\frac{1}{2}\int (e^{\phi(t)}+e^{-\phi(t)})dt$
is the Hamiltonian
of the Toda field theory.
They showed the existence of infinitly many
commutative integrals of motion by a cohomological argumemnt,
and showed that they can be regarded as 
the conservation laws for the generalized KdV equation.
In \cite{FF1} they constructed
the quantum deformtion of the local integrals of motion, too.
In other words they showed the existence of quantum deformation
of the conservation laws of the generalized KdV equation.
After quantization
Gel'fand-Dickij bracket $\{,\}_{P.B.}$ for the second 
Hamiltonian structure of the generalized KdV, 
gives rise to the $W_N$ algebra.
V.Bazhanov et.al \cite{BLZ, BHK}
constructed field theoretical analogue of the commuting
transfer matrix ${\bf T}(z)$, acting on 
the irreducible highest weight 
module of the Virasoro algebra and the $W_3$ algebra.
They constructed this commuting transfer matrix ${\bf T}(z)$
as the trace of the monodromy matrix associated with
the quantum affine symmetry $U_q(\widehat{sl_2})$
and $U_q(\widehat{sl_3})$,
and showed that
the commutatin relation $[{\bf T}(z),{\bf T}(w)]=0$
is a direct consequence of the Yang-Baxter relation.
The coefficients of the asymptotic expansion
of the operator ${\rm log}~{\bf T}(z)$ at $z \to \infty$,
produce the local integrals of motion
for the Virasoro algebra and the $W_3$ algebra, 
which reproduce the conservation laws of the generalized
KdV equation in the classical limit $c_{CFT} \to \infty$.
They call the coefficients of the Taylor expansion
of the operator ${\bf T}(z)$ at $z=0$,
the nonlocal integrals of motion
for the Virasoro algebra and the $W_3$ algebra.
They have explicit integral
representation of the nonlocal
integrals in terms of the screening currents.

The purpose of this paper
is to construct the elliptic version
of the integrals of motion given by 
Bazhanov et.al 
\cite{BLZ, BHK}.
Bazhanov et.al's construction is based on
the free field realization of
the Borel subalgebra ${\cal B}_\pm$ of 
$U_q(\widehat{sl_2})$ and $U_q(\widehat{sl_3})$.
By using this realization
they construct the monodromy matrix 
as the image of the universal
R-matrix $\bar{R} \in {\cal B}_+\otimes {\cal B}_-$,
and make the transfer matrix ${\bf T}(z)$
as the trace of the monodromy matrix.
The universal
R-matrix $\bar{R}$ of the elliptic quantum group
does not exist in ${\cal B}_+\otimes {\cal B}_-$.
Hence it is impossible to construct the elliptic
deformation of the transfer matris ${\bf T}(z)$
as the same manner as \cite{BLZ}.
Our method of construction should be
completely different from 
those of \cite{BLZ, BHK}.
Instead of considering the transfer mtrix ${\bf T}(z)$,
we directly give the integral representations of 
the integrals of motion 
${\cal I}_n, {\cal G}_n$ for the deformed Virasoro algebra.
The commutativity of our integrals of motion
are not understood as a direct consequence
of the Yang-Baxter equation.
They are understood as a consequence of
the commutative subalgebra of the Feigin-Odesskii algebra
\cite{FO}.

The organization of this paper is as follows.
In Section 2, we review
the deformed $W$ algebra,
including free field realization, screening currents
\cite{SKAO, FF2}.
In Section 3, we
give integral representations for the local 
integrals of motion ${\cal I}_n$,
and show the commutation relations :
$$[{\cal I}_m,{\cal I}_n]=
[{\cal I}_m^*,{\cal I}_n^*]=0.$$
Very precisely, in Part 2 \cite{FKSW2},
we only give the Laurent series representation of
the local integrals motion, which is useful
for proofs of the commutation relation and 
Dynkin-automorphism invariance.
In this section we show the integral representations 
and the Laurent series representation
give the same local integrals of motion.
In Section 4, 
we give explicit formulae
for the nonlocal integrals of motion
${\cal G}_n$,
and show  
the commutation relations :
$$[{\cal G}_m,{\cal G}_n]=
[{\cal G}_m^*,{\cal G}_n^*]=
[{\cal G}_m,{\cal G}_n^*]=0,$$
$$[{\cal I}_m,{\cal G}_n]=
[{\cal I}_m^*,{\cal G}_n]=
[{\cal I}_m,{\cal G}_n^*]=
[{\cal I}_m^*,{\cal G}_n^*]=0.$$
We show the commutation relation
$[{\cal I}_m,{\cal G}_n]=0$ using Dynkin-automorphism invariance
$\eta({\cal I}_n)={\cal I}_n$ and $\eta({\cal G}_n)={\cal G}_n$,
which will be shown in the next section.
In Section 5,
we give proofs of
Dynkin-automorphis invariance :
$$\eta({\cal I}_n)={\cal I}_n,~~
\eta({\cal I}_n^*)={\cal I}_n^*,~~
\eta({\cal G}_n)={\cal G}_n,~~
\eta({\cal G}_n^*)={\cal G}_n^*.$$
In Appendix we summarize the normal ordering of 
the basic operators.
%%%%%%%%%%%%%%%%%%%%%%%%%%%%%%%%%%%%
We would like to point out a differnt point between the case of
the deformed Virasoro $Vir_{q,t}
=W_{q,t}(\widehat{sl_2})$ 
and its higher-rank generalization
: the deformed $W_{q,t}(\widehat{sl_N})$, $(N \geqq 3)$.
Basically situations of 
$W_{q,t}(\widehat{sl_N})$,
$(N \geqq 3)$ are
more complicated than those of $Vir_{q,t}
=W_{q,t}(\widehat{sl_2})$.
However, one thing of 
$W_{q,t}(\widehat{sl_N})$,
$(N\geqq 3)$ is simpler than those of 
$Vir_{q,t}=W_{q,t}(\widehat{sl_2})$.
In the case of $Vir_{q,t}=W_{q,t}(\widehat{sl_2})$, 
the integrals of motions ${\cal I}_n$, ${\cal G}_n$
have singularity at $s=N=2$.
Hence we considered the renormalized limits for
the integral of motions ${\cal I}_n$, ${\cal G}_n$
in the lase section of the paper \cite{FKSW1}.
In the case of $W_{q,t}(\widehat{sl_N})$, $(N\geqq 3)$, 
the integrals of motions ${\cal I}_n$, ${\cal G}_n$
do not have singularity at $s=N\geqq 3$.

At the end of Introduction,
we would like to mention about two important degenerating limits
of the deformed $W$ algebra.
One is the CFT-limit\cite{BLZ,BHK} 
and the other is the classical limit\cite{FR}.
In the CFT-limit V.Bazhanov et.al.
\cite{BLZ, BHK} constructed infinitly many integrals of motion
for the Virasoro algebra, as we mentioned above.
In the classical limit,
the deformed Virasoro algebra
degenerates to the Poisson-Virasoro algebra introduced
by E.Frenkel and N.Reshetikhin \cite{FR}.
%In the classical limit,
%E.Frenkel \cite{F} constructed infinitly many integrals of
%motion for the Poisson-Virasoro algebra.

\section{The Deformed $W$-Algebra $W_{q,t}(\widehat{sl_N})$}

In this section we review 
the deformed $W$-algebra and
its screening currents.
We prepare the notations to be used in this paper.
Throughout this paper, we fix generic three parameters 
$0<x<1$, $r \in {\mathbb C}$ and $s \in {\mathbb C}$.
Let us set $z=x^{2u}$.
Let us set $r^*=r-1$.
The symbol $[u]_r$ for ${\rm Re}(r)>0$ stands for 
the Jacobi theta function
\begin{eqnarray}
~[u]_r&=&x^{\frac{u^2}{r}-u}\frac{\Theta_{x^{2r}}(x^{2u})}{(x^{2r};x^{2r})^3},
~~
\Theta_q(z)=
(z;q)_\infty (qz^{-1};q)_\infty (q;q)_\infty,
\end{eqnarray}
where we have used the standard notation
\begin{eqnarray}
(z;q)_\infty=\prod_{j=0}^\infty (1-q^jz).
\end{eqnarray}
We set the parametrizations $\tau, \tau^*$
\begin{eqnarray}
x=e^{-\pi \sqrt{-1}/r\tau}=e^{-\pi \sqrt{-1}/r^* \tau^*}.
\end{eqnarray}
The theta function $[u]_r$ enjoys 
the quasi-periodicity property
\begin{eqnarray}
[u+r]_r=-[u]_r,~~~[u+r\tau]_r=-e^{-\pi \sqrt{-1}\tau-\frac{2\pi \sqrt{-1}}{r}
u}[u]_r.
\end{eqnarray}
%In what follows we use three theta functions
%$[u]_r, [u]_{-r^*}, [u]_s$ which are associated with three periods
%$r, r^*=r-1, s$.
The symbol $[a]$ stands for
\begin{eqnarray} 
[a]=\frac{x^a-x^{-a}}{x-x^{-1}}.
\end{eqnarray}

\subsection{Free Field Realization}

Let $\epsilon_i (1\leqq i \leqq N)$
be an orthonormal basis in ${\mathbb R}^N$
relative to the standard basis in ${\mathbb R}^N$
relative to the standard inner product $(,)$.
Let us set $\bar{\epsilon}_i=\epsilon_i-\epsilon,
\epsilon=\frac{1}{N}\sum_{j=1}^N \epsilon_j$.
We idetify $\epsilon_{N+1}=\epsilon_1$.
Let $P=\sum_{i=1}^N {\mathbb Z}\bar{\epsilon}_i$ the weight lattice.
Let us set $\alpha_i=\bar{\epsilon}_i-\bar{\epsilon}_{i+1}\in P$.

Let $\beta_m^j$ be the oscillators
$(1\leqq j \leqq N, m \in {\mathbb Z}-\{0\})$
with the commutation relations
\begin{eqnarray}
~[\beta_m^i,\beta_n^j]=\left\{
\begin{array}{cc}
m\frac{[(r-1)m]}{[r m]}\frac{[(s-1)m]}{[s m]}\delta_{n+m,0}&
(1\leqq i=j\leqq N)\\
-m\frac{[(r-1)m]}{[r m]}\frac{[m]}{[sm]}x^{sm~{\rm sgn}(i-j)}
\delta_{n+m,0}
&(1\leqq i\neq j \leqq N)
\end{array}
\right.
\end{eqnarray}

We also introduce the zero mode operator $P_\lambda$, $(\lambda \in P)$.
They are ${\mathbb Z}$-linear in $P$
and satisfy
\begin{eqnarray}
[iP_\lambda, Q_\mu]=(\lambda,\mu),~~(\lambda,\mu \in P).
\end{eqnarray}

Let us intrduce the bosonic Fock space ${\cal F}_{l,k} (l,k \in P)$
generated by $\beta_{-m}^j (m>0)$ over the vacuum vector $|l,k \rangle$ :
\begin{eqnarray}
{\cal F}_{l,k}={\mathbb C}[
\{\beta_{-1}^j,\beta_{-2}^j,\cdots\}_{1\leqq j \leqq N}]|l,k\rangle,
\end{eqnarray}
where
\begin{eqnarray}
\beta_m^j|l,k\rangle&=&0,  (m>0),\\
P_\alpha |l,k\rangle&=&
\left(\alpha, \sqrt{\frac{r}{r-1}}l-\sqrt{\frac{r-1}{r}}k\right)|l,k\rangle,\\
|l,k\rangle&=&e^{i\sqrt{\frac{r}{r-1}}Q_l-i\sqrt{\frac{r-1}{r}}Q_k}|0,0\rangle.
\end{eqnarray}

Let us set the Dynkin-diagram automorphism $\eta$ by
\begin{eqnarray}
\eta(\beta_m^1)=x^{-\frac{2s}{N}m}\beta_m^2,\cdots,
\eta(\beta_m^{N-1})=x^{-\frac{2s}{N}m}\beta_m^N,
~\eta(\beta_m^N)=x^{\frac{2s}{N}(N-1)m}\beta_m^1,
\end{eqnarray}
and
$\eta(\epsilon_i)=\epsilon_{i+1},~(1\leqq i \leqq N)$.

\subsection{The Deformed $W$-Algebra}

In this section we give short review of
the deformed $W$-algebra
$W_{q,t}(\widehat{sl_N})$ \cite{AKOS, FF2, Odake}.

\begin{df}~~We set the fundamental operator 
$\Lambda_j(z), (1\leqq j \leqq N)$ by
\begin{eqnarray}
\Lambda_j(z)&=&x^{-2\sqrt{r(r-1)} P_{\bar{\epsilon}_j}}
:\exp
\left(\sum_{m \neq 0} \frac{x^{rm}-x^{-rm}}{m}
\beta_m^j z^{-m}\right):~~(1\leqq j \leqq N).
\end{eqnarray}
\end{df}

\begin{df}
Let us set the operator $T_j(z),~(1\leqq j \leqq N)$ by
\begin{eqnarray}
T_j(z)=\sum_{1\leqq s_1<s_2<\cdots<s_j\leqq N}
:\Lambda_{s_1}(x^{-j+1}z)\Lambda_{s_2}(x^{-j+3}z)\cdots \Lambda_{s_j}(x^{j-1}z):.
\end{eqnarray}
\end{df}

\begin{prop}~~
The actions of $\eta$ on the fundamental
operators $\Lambda_j(z),~(1\leqq j \leqq N)$ are given by
\begin{eqnarray}
\eta(\Lambda_j(z))=\Lambda_{j+1}(x^{\frac{2s}{N}}z),~~(1\leqq j \leqq N-1),~~~
\eta(\Lambda_N(z))=\Lambda_1(x^{\frac{2s}{N}-2s}z).
\end{eqnarray}
\end{prop}

\begin{prop}~~The operators $T_j(z),~(1\leqq j \leqq N)$ satisfy
the following relations.
\begin{eqnarray}
&&f_{i,j}(z_2/z_1)T_i(z_1)T_j(z_2)
-f_{j,i}(z_1/z_2)T_j(z_2)T_i(z_1)\nonumber\\
&=&c
\sum_{k=1}^i
\prod_{l=1}^{k-1}\Delta(x^{2l+1})\times
\left(\delta\left(\frac{x^{j-i+2k}z_2}{z_1}\right)
f_{i-k,j+k}(x^{-j+i})T_{i-k}(x^{-k}z_1)T_{j+k}(x^kz_2)\right.\nonumber\\
&-&\left.
\delta\left(\frac{x^{-j+i-2k}z_2}{z_1}\right)
f_{i-k,j+k}(x^{j-i})T_{i-k}(x^{k}z_1)T_{j+k}(x^{-k}z_2)
\right),~~(1\leqq i \leqq j \leqq N),\label{def:3parameter}
\end{eqnarray}
where we have used the delta-function 
$\delta(z)=\sum_{n \in {\mathbb Z}}z^n$.
Here we set the constnt $c$ and the auxiliary function $\Delta(z)$ by
\begin{eqnarray}
c=-\frac{(1-x^{2r})(1-x^{-2r+2})}{(1-x^2)},~~
\Delta(z)=\frac{(1-x^{2r-1}z)(1-x^{1-2r}z)}{(1-xz)(1-x^{-1}z)}.
\label{def:Delta}
\end{eqnarray}
Here we set the structure functions,
\begin{eqnarray}
f_{i,j}(z)=\exp\left(
\sum_{m=1}^\infty
\frac{1}{m}(1-x^{2rm})(1-x^{-2(r-1)m})
\frac{(1-x^{2m Min(i,j)})(1-x^{2m(s-Max(i,j))})}
{(1-x^{2m})(1-x^{2sm})}x^{|i-j|m}z^m
\right).\nonumber\\
\label{def:structure}
\end{eqnarray}
\end{prop}

Above proposition is one parameter ``$s$'' generalization of
\cite{Odake}.
The proof is given by the same manner.

~\\
{\bf Example}~~For $N=2$ the operators $T_1(z), T_2(z)$ satisfy
\begin{eqnarray}
&&f_{1,1}(z_2/z_1)T_1(z_1)T_1(z_2)-
f_{1,1}(z_1/z_2)T_1(z_2)T_1(z_1)\nonumber\\
&=&c(\delta(x^2z_2/z_1)T_2(x z_2)-
\delta(x^2z_1/z_2)T_2(x^{-1}z_2)),\\
&&f_{1,2}(z_2/z_1)T_1(z_1)T_2(z_2)=
f_{2,1}(z_1/z_2)T_2(z_2)T_1(z_1),\\
&&f_{2,2}(z_2/z_1)T_2(z_1)T_2(z_2)=
f_{2,2}(z_1/z_2)T_2(z_2)T_2(z_1).
\end{eqnarray}
{\bf Example}
~~For $N=3$ the operators $T_1(z), T_2(z), T_3(z)$ satisfy
\begin{eqnarray}
&&f_{1,1}(z_2/z_1)T_1(z_1)T_1(z_2)-
f_{1,1}(z_1/z_2)T_1(z_2)T_1(z_1)\nonumber\\
&=&c(\delta(x^2z_2/z_1)T_2(x z_2)-
\delta(x^2z_1/z_2)T_2(x^{-1}z_2)),\\
&&f_{1,2}(z_2/z_1)T_1(z_1)T_2(z_2)-
f_{2,1}(z_1/z_2)T_2(z_2)T_1(z_1)\nonumber\\
&=&c(\delta(x^3 z_2/z_1)T_3(x z_2)-
\delta(x^3 z_1/z_2)T_3(x^{-1}z_2)),\\
&&f_{2,2}(z_2/z_1)T_2(z_1)T_2(z_2)-
f_{2,2}(z_1/z_2)T_2(z_2)T_2(z_1)\nonumber\\
&=&c f_{1,3}(1)(\delta(x^2 z_2/z_1)T_1(xz_2)T_3(x z_2)-
\delta(x^2 z_1/z_2)T_1(x^{-1}z_2)T_3(x^{-1}z_2)),\\
&&f_{1,3}(z_2/z_1)T_1(z_1)T_3(z_2)=
f_{3,1}(z_1/z_2)T_3(z_2)T_1(z_1),\\
&&f_{2,3}(z_2/z_1)T_2(z_1)T_3(z_2)=
f_{3,2}(z_1/z_2)T_3(z_2)T_2(z_1),\\
&&f_{3,3}(z_2/z_1)T_3(z_1)T_3(z_2)=
f_{3,3}(z_1/z_2)T_3(z_2)T_3(z_1).
\end{eqnarray}
{\bf Example}
~~For $N=4$ the operators $T_1(z), T_2(z), T_3(z), T_4(z)$ satisfy
\begin{eqnarray}
&&f_{1,1}(z_2/z_1)T_1(z_1)T_1(z_2)-
f_{1,1}(z_1/z_2)T_1(z_2)T_1(z_1)\nonumber\\
&=&c(\delta(x^2z_2/z_1)T_2(x z_2)-
\delta(x^2z_1/z_2)T_2(x^{-1}z_2)),\\
&&f_{1,2}(z_2/z_1)T_1(z_1)T_2(z_2)-
f_{2,1}(z_1/z_2)T_2(z_2)T_1(z_1)\nonumber\\
&=&c(\delta(x^3 z_2/z_1)T_3(x z_2)-
\delta(x^3 z_1/z_2)T_3(x^{-1}z_2)),\\
&&f_{1,3}(z_2/z_1)T_1(z_1)T_3(z_2)-
f_{3,1}(z_1/z_2)T_3(z_2)T_1(z_1)\nonumber\\
&=&c
(\delta(x^4 z_2/z_1)T_4(x z_2)-
\delta(x^4 z_1/z_2)T_4(x^{-1}z_2)),\\
&&f_{2,2}(z_2/z_1)T_2(z_1)T_2(z_2)-
f_{2,2}(z_1/z_2)T_2(z_2)T_2(z_1)\nonumber\\
&=&c f_{1,3}(1)(\delta(x^2 z_2/z_1)T_1(x^{-1}z_1)T_3(x z_2)-
\delta(x^2 z_1/z_2)T_1(x z_1)T_3(x^{-1}z_2)),\\
&&f_{2,3}(z_2/z_1)T_2(z_1)T_3(z_2)-
f_{3,2}(z_1/z_2)T_3(z_2)T_2(z_1)\\
&=&c(
\delta(x^3 z_2/z_1)f_{1,4}(x^{-1})T_1(x^{-1}z_1)T_4(x z_2)-
\delta(x^3 z_1/z_2)f_{1,4}(x)T_1(xz_1)T_4(x^{-1}z_2)),\nonumber
\\
%%%%%%%%%%%%%%%%%%%%%%%%%%%%%%%%%%%%%%%%%%%%%%%%
&&f_{1,4}(z_2/z_1)T_1(z_1)T_4(z_2)=
f_{4,1}(z_1/z_2)T_4(z_2)T_1(z_1),\\
&&f_{2,4}(z_2/z_1)T_2(z_1)T_4(z_2)=
f_{4,2}(z_1/z_2)T_4(z_2)T_2(z_1),\\
&&f_{3,4}(z_2/z_1)T_3(z_1)T_4(z_2)=
f_{4,3}(z_1/z_2)T_4(z_2)T_3(z_1),\\
&&f_{4,4}(z_2/z_1)T_4(z_1)T_4(z_2)=
f_{4,4}(z_1/z_2)T_4(z_2)T_4(z_1).
\end{eqnarray}

\begin{df}~~
The deformed $W$-algebra
is defined by the generators $\widehat{T}_m^{(j)},~(1\leqq j \leqq N, m \in {\mathbb Z})$
with the defining relations (\ref{def:3parameter}).
Here we should understand $\widehat{T}_m^{(j)}$ as the Fourier coefficients of the operators
$\widehat{T}_j(z)=\sum_{m\in {\mathbb Z}}\widehat{T}_m^{(j)}z^{-m},~(1\leqq j \leqq N)$.
\end{df}

\subsection{Screening Currents}

In this section we introduce the screening currents 
$E_j(z)$ and $F_j(z)$.

\begin{df}~~~We set the screening currents 
$F_j(z), (1\leqq j \leqq N)$ by
\begin{eqnarray}
F_j(z)&=&e^{i\sqrt{\frac{r-1}{r}}Q_{\alpha_j}}
(x^{(\frac{2s}{N}-1)j}z)^{\sqrt{\frac{r-1}{r}}P_{\alpha_j}+\frac{r-1}{r}}
\nonumber\\
&\times&
:\exp\left(\sum_{m\neq 0}\frac{1}{m}B_m^jz^{-m}\right):,~(1\leqq j\leqq N-1)\\
F_N(z)&=&e^{i\sqrt{\frac{r-1}{r}}Q_{\alpha_N}}
(x^{2s-N}z)^{\sqrt{\frac{r-1}{r}}P_{\bar{\epsilon}_N}+\frac{r-1}{2r}}
z^{-\sqrt{\frac{r-1}{r}}P_{\bar{\epsilon}_1}+\frac{r-1}{2r}}\nonumber\\
&\times&
:\exp\left(
\sum_{m\neq 0}\frac{1}{m}B_m^N
z^{-m}
\right):,
\end{eqnarray}
We set the screening currents $E_j(z), (1\leqq j \leqq N)$ by
\begin{eqnarray}
E_j(z)&=&e^{-i\sqrt{\frac{r}{r-1}}Q_{\alpha_j}}
(x^{(\frac{2s}{N}-1)j}z)
^{-\sqrt{\frac{r}{r-1}}P_{\alpha_j}+\frac{r}{r-1}}
\nonumber\\
&\times&
:\exp\left(-\sum_{m\neq 0}\frac{1}{m}
\frac{[rm]}{[(r-1)m]}
B_m^jz^{-m}\right):,~(1\leqq j\leqq N-1)\\
E_N(z)&=&e^{-i\sqrt{\frac{r}{r-1}}Q_{\alpha_N}}
(x^{2s-N}z)
^{-\sqrt{\frac{r}{r-1}}P_{\bar{\epsilon}_N}+\frac{r}{2(r-1)}}
z^{\sqrt{\frac{r}{r-1}}P_{\bar{\epsilon}_1}+\frac{r}{2(r-1)}}\nonumber\\
&\times&
:\exp\left(-
\sum_{m\neq 0}\frac{1}{m}
\frac{[rm]}{[(r-1)m]}B_m^N
z^{-m}
\right):.
\end{eqnarray}
Here we have set
\begin{eqnarray}
B_m^j&=&(\beta_m^j-\beta_m^{j+1})x^{-\frac{2s}{N}
jm},~(1\leqq j \leqq N-1),\\
B_m^N&=&(x^{-2sm}\beta_m^N-\beta_m^1).
\end{eqnarray}
\end{df}
The screening currents $F_j(z), E_j(z)~
(1\leqq j \leqq N-1)$ have already been studied in
\cite{AJMP, FJMOP}. We introduce new screening
current $F_N(z), E_N(z)$, which can be regarded
as ``affinization'' of screening currents 
$F_j(z), E_j(z)~~(1\leqq j \leqq N-1)$.
The following commutation relations are
convenient for calculations.
\begin{eqnarray}
~[\beta_m^j,B_n^j]&=&m\delta_{m+n,0}\frac{[r^*m]}{[rm]}
x^{(-1+\frac{2s}{N}j)m},~~(1\leqq j \leqq N)\\
~[\beta_m^{j+1},B_n^j]&=&-
m\delta_{m+n,0}\frac{[r^*m]}{[rm]}
x^{(1+\frac{2s}{N}j)m},~~(1\leqq j \leqq N-1)\\
~[\beta_m^1,B_n^N]&=&-m\delta_{m+n,0}\frac{[r^*m]}{[rm]}
x^{m},\\
~[B_m^j,B_n^j]&=&m\delta_{m+n,0}\frac{[r^*m]}{[rm]}
\frac{[2m]}{[m]},~~(1\leqq j \leqq N)\\
~[B_m^{j},B_n^{j+1}]&=&-m\delta_{m+n,0}\frac{[r^*m]}{[rm]}
x^{(-1+\frac{2s}{N})m},~~(1\leqq j \leqq N).
\end{eqnarray}
Here we read $B_m^{N+1}=B_m^1$.
We summarize the commutation relations of 
the screening currents for $N \geqq 3$.
\begin{prop}~~~
The screening currents $F_j(z),~(1\leqq j \leqq N;
N \geqq 3)$ satisfy
the following commutation relations for ${\rm Re}(r)>0$
\begin{eqnarray}
\frac{1}{[u_1-u_2-\frac{s}{N}+1]_r}F_j(z_1)F_{j+1}(z_2)
&=&\frac{1}{[u_2-u_1+\frac{s}{N}]_r}
F_{j+1}(z_2)F_{j}(z_1),~~(1\leqq j \leqq N),\nonumber\\\\
\frac{[u_1-u_2]_r}{[u_1-u_2-1]_r}F_j(z_1)F_{j}(z_2)&=&
\frac{[u_2-u_1]_r}{
[u_2-u_1-1]_r}F_{j}(z_2)F_{j}(z_1),~~(1\leqq j \leqq N),
\end{eqnarray}
\begin{eqnarray}
F_i(z_1)F_j(z_2)&=&F_j(z_2)F_i(z_1),~~(|i-j|\geqq 2).
\end{eqnarray}
We read $F_{N+1}(z)=F_1(z)$. 
%%%%%%%%%%%%%%%%%%%%%%%%%%%%%%%%%%%%%%%%%%%%%%
The screening currents $F_j(z),~(1\leqq j \leqq N;
N \geqq 3)$ satisfy
the following commutation relations
for ${\rm Re}(r)<0$.
\begin{eqnarray}
\frac{1}{[u_1-u_2-\frac{s}{N}]_{-r}}F_j(z_1)F_{j+1}(z_2)
&=&\frac{1}{[u_2-u_1-1+\frac{s}{N}]_{-r}}
F_{j+1}(z_2)F_{j}(z_1),~~(1\leqq j \leqq N),\nonumber\\\\
\frac{[u_1-u_2]_{-r}}{[u_1-u_2+1]_{-r}}
F_j(z_1)F_{j}(z_2)&=&
\frac{[u_2-u_1]_{-r}}{
[u_2-u_1+1]_{-r}}F_{j}(z_2)F_{j}(z_1),~~(1\leqq j \leqq N),
\end{eqnarray}
\begin{eqnarray}
F_i(z_1)F_j(z_2)&=&F_j(z_2)F_i(z_1),~~(|i-j|\geqq 2).
\end{eqnarray}
We read $F_{N+1}(z)=F_1(z)$. \\
%%%%%%%%%%%%%%%%%%%%%%%%%%%%%%%%%%%%%%%%%%%%
The screening currents $E_j(z),~(1\leqq j \leqq N;
N \geqq 3)$ satisfy
the following commutation relations for ${\rm Re}(r^*)>0$
\begin{eqnarray}
\frac{1}{[u_1-u_2-\frac{s}{N}]_{r^*}}
E_j(z_1)E_{j+1}(z_2)
&=&\frac{1}{[u_2-u_1+\frac{s}{N}-1]_{r^*}}
E_{j+1}(z_2)E_{j}(z_1),~~(1\leqq j \leqq N),\nonumber\\\\
\frac{[u_1-u_2]_{r^*}}{[u_1-u_2+1]_{r^*}}
E_j(z_1)E_{j}(z_2)&=&
\frac{[u_2-u_1]_{r^*}}{
[u_2-u_1+1]_{r^*}}
E_{j}(z_2)E_{j}(z_1),~~(1\leqq j \leqq N),
\end{eqnarray}
\begin{eqnarray}
E_i(z_1)E_j(z_2)&=&E_j(z_2)E_i(z_1),~~(|i-j|\geqq 2).
\end{eqnarray}
We read $E_{N+1}(z)=E_1(z)$. \\
The screening currents $E_j(z),~(1\leqq j \leqq N;
N \geqq 3)$ satisfy
the following commutation relations 
for ${\rm Re}(r^*)<0$.
\begin{eqnarray}
\frac{1}{[u_1-u_2-\frac{s}{N}+1]_{-r^*}}E_j(z_1)E_{j+1}(z_2)
&=&\frac{1}{[u_2-u_1+\frac{s}{N}]_{-r^*}}
E_{j+1}(z_2)E_{j}(z_1),~~(1\leqq j \leqq N),\nonumber\\\\
\frac{[u_1-u_2]_{-r^*}}{[u_1-u_2-1]_{-r^*}}
E_j(z_1)E_{j}(z_2)&=&
\frac{[u_2-u_1]_{-r^*}}{
[u_2-u_1-1]_{-r^*}}E_{j}(z_2)E_{j}(z_1),~~(1\leqq j \leqq N),
\nonumber
\\
\\
E_i(z_1)E_j(z_2)&=&E_j(z_2)E_i(z_1),~~(|i-j|\geqq 2).
\end{eqnarray}
\end{prop}

\begin{prop}~~
The screening currents
$E_j(z), F_j(z)$, $(1\leqq j \leqq N; N\geqq 3)$
satisfy the following commutation relation
%for both ${\rm Re}(r^*)>0$ and
${\rm Re}(r)<0$.
\begin{eqnarray}
[E_j(z_1),F_j(z_2)]
&=&\frac{1}{x-x^{-1}}
(\delta(xz_2/z_1){H}_j(x^{r}z_2)
-\delta(xz_1/z_2){H}_j(x^{-r}z_2)),
(1\leqq j \leqq N),\nonumber\\
\\
E_i(z_1)F_j(z_2)&=&F_j(z_2)E_i(z_1),~(1\leqq i \neq j \leqq N).
\end{eqnarray}
Here we have set
\begin{eqnarray}
{H}_j(z)&=&x^{(1-\frac{2s}{N})2j}
e^{-\frac{i}{\sqrt{r r^*}}Q_{\alpha_j}}
(x^{(\frac{2s}{N}-1)j}z)^{-\frac{1}{
\sqrt{r r^*}}P_{\alpha_j}+\frac{1}{r r^*}}\nonumber\\
&\times&:\exp\left(
-\sum_{m\neq 0}\frac{1}{m}\frac{[m]}{[r^*m]}B_m^j z^{-m}\right):
,~~(1\leqq j \leqq N-1),\\
{H}_N(z)&=&x^{2(N-2s)}
e^{-\frac{i}{\sqrt{r r^*}}Q_{\alpha_N}}
(x^{2s-N}z)^{-\frac{1}{\sqrt{r r^*}}P_{{\bar{\epsilon}}_N}
+\frac{1}{2 r r^*}}
z^{-\frac{1}{\sqrt{r r^*}}P_{{\bar{\epsilon}}_1}
+\frac{1}{2 r r^*}}\nonumber\\
&\times&
:\exp\left(
-\sum_{m\neq 0}\frac{1}{m}\frac{[m]}{[r^*m]}B_m^N z^{-m}\right):.
\end{eqnarray}
\end{prop}

\begin{prop}~~~
The actions of $\eta$ on the screenings $F_j(z)$,
$(1\leqq j\leqq N; N \geqq 3)$
are given by
\begin{eqnarray}
&&\eta(F_j(z))=F_{j+1}(z)(x^{\frac{2s}{N}-1})
^{-\sqrt{\frac{r^*}{r}}P_{\alpha_{j+1}}-\frac{r^*}{r}},~~(1\leqq j \leqq N-2),\\
&&\eta(F_{N-1}(z))=F_N(z)
(x^{1-\frac{2s}{N}})^{
\sqrt{\frac{r^*}{r}}P_{\bar{\epsilon}_N}+\frac{r^*}{2r}}
(x^{1-\frac{2s}{N}})^{
-\sqrt{\frac{r^*}{r}}
P_{\bar{\epsilon}_1}+\frac{r^*}{2r}},\\
&&\eta(F_{N}(z))=F_1(z)
(x^{(1-\frac{2s}{N})(N-1)})
^{\sqrt{\frac{r^*}{r}}P_{\bar{\epsilon}_1}+\frac{r^*}{2r}}
(x^{1-\frac{2s}{N}})^{
-\sqrt{\frac{r^*}{r}}
P_{\bar{\epsilon}_2}+\frac{r^*}{2r}}.
\end{eqnarray}
Especially we have
\begin{eqnarray}
\eta(F_1(z_1)F_2(z_2) \cdots F_N(z_N))=F_N(z_1)F_1(z_2)
\cdots F_1(z_N).
\end{eqnarray}
The actions of $\eta$ on the screenings $E_j(z)$,
$(1\leqq j \leqq N; N \geqq 3)$ are given by
\begin{eqnarray}
&&\eta(E_j(z))=E_{j+1}(z)(x^{\frac{2s}{N}-1})
^{\sqrt{\frac{r}{r^*}}P_{\alpha_{j+1}}-\frac{r}{r^*}},~~
(1\leqq j \leqq N-2),\\
&&\eta(E_{N-1}(z))=E_N(z)
(x^{1-\frac{2s}{N}})^{
-\sqrt{\frac{r}{r^*}}P_{\bar{\epsilon}_N}+\frac{r}{2r^*}}
(x^{1-\frac{2s}{N}})^{
\sqrt{\frac{r}{r^*}}
P_{\bar{\epsilon}_1}+\frac{r}{2r^*}},\\
&&\eta(E_{N}(z))=E_1(z)
(x^{(1-\frac{2s}{N})(N-1)})
^{-\sqrt{\frac{r}{r^*}}P_{\bar{\epsilon}_1}+\frac{r}{2r^*}}
(x^{1-\frac{2s}{N}})^{
\sqrt{\frac{r}{r^*}}
P_{\bar{\epsilon}_2}+\frac{r}{2r^*}}.
\end{eqnarray}
Especially we have
\begin{eqnarray}
\eta(E_1(z_1)E_{2}(z_2)\cdots
E_N(z_N))=E_2(z_1)\cdots E_N(z_{N-1})E_1(z_N).
\end{eqnarray}
\end{prop}

\begin{prop}~~The screening currents $F_j(z),~
(1\leqq j\leqq N; N \geqq 3)$
and the fundamental operators 
$\Lambda_j(z),~(1\leqq j \leqq N; N \geqq 3)$ 
commute up to delta-function
$\delta(z)=\sum_{n\in {\mathbb Z}}z^m$. 
\begin{eqnarray}
~[\Lambda_j(z_1),F_j(z_2)]&=&(-x^{r^*}+x^{-r^*})\delta
\left(x^{\frac{2s}{N}j-r}\frac{z_2}{z_1}\right)
{\cal A}_j(x^{\frac{2s}{N}j-r}z_1),~~(1\leqq j \leqq N-1),\nonumber
\\
\\
~[\Lambda_{j+1}(z_1),F_j(z_2)]&=&(x^{r^*}-x^{-r^*})\delta
\left(x^{\frac{2s}{N}j+r}\frac{z_2}{z_1}\right)
{\cal A}_j(x^{\frac{2s}{N}j+r}z_2),
~~(1\leqq j \leqq N-1),\nonumber\\
\\
~[\Lambda_N(z_1),F_N(z_2)]&=&(-x^{r^*}+x^{-r^*})
\delta\left(x^{-r+2s}\frac{z_2}{z_1}\right)
{\cal A}_N(x^{-r}z_2),\\
~[\Lambda_1(z_1),F_N(z_2)]&=&(x^{r^*}-x^{-r^*})
\delta\left(x^{r}\frac{z_2}{z_1}\right)
{\cal A}_N(x^rz_2).
\end{eqnarray}
Here we have set
\begin{eqnarray}
{\cal A}_j(z)&=&
e^{i\sqrt{\frac{r^*}{r}}Q_{\alpha_j}}
x^{-\sqrt{rr^*}(P_{\bar{\epsilon}_j}+P_{\bar{\epsilon}_{j+1}})}
(zx^{-j})^{\sqrt{\frac{r^*}{r}}P_{\alpha_j}+\frac{r^*}{r}}\nonumber\\
&\times&:\exp\left(
\sum_{m\neq 0}
\frac{1}{m}(x^{rm}\beta_m^j-x^{-rm}\beta_m^{j+1})
z^{-m}\right):,~(1\leqq j \leqq N-1),
\end{eqnarray}
\begin{eqnarray}
{\cal A}_N(z)&=&
e^{i\sqrt{\frac{r^*}{r}}Q_{\alpha_N}}
x^{-\sqrt{rr^*}(P_{\bar{\epsilon}_N}+P_{\bar{\epsilon}_{1}})}
(zx^{2s-N})^{\sqrt{\frac{r^*}{r}}P_{\bar{\epsilon}_N}+\frac{r^*}{2r}}
z^{-\sqrt{\frac{r^*}{r}}P_{\bar{\epsilon}_1}+\frac{r^*}{2r}}
\nonumber\\
&\times&:\exp\left(
\sum_{m\neq 0}
\frac{1}{m}(x^{(r-2s)m}\beta_m^N-x^{-rm}\beta_m^{1})
z^{-m}\right):.
\end{eqnarray}
%%%%%%%%%%%%%%%%%%%%%%%%%%%%%%%%%%%%%%%%%%%%%%
%%%%%%%%%%%%%%%%%%%%%%%%%%%%%%%%%%%%%%%%%%%%%%
\begin{eqnarray}
~[\Lambda_j(z_1),E_j(z_2)]&=&(-x^r+x^{-r})\delta
\left(x^{\frac{2s}{N}j+r^*}\frac{z_2}{z_1}\right)
{\cal B}_j(x^{\frac{2s}{N}j+r^*}z_2),~~(1\leqq j \leqq N-1),\nonumber
\\
\\
~[\Lambda_{j+1}(z_1),E_j(z_2)]&=&(x^r-x^{-r})\delta
\left(x^{\frac{2s}{N}j-r^*}\frac{z_2}{z_1}\right)
{\cal B}_j(x^{\frac{2s}{N}j-r^*}z_1),~~(1\leqq j \leqq N-1),\nonumber\\
\\
~[\Lambda_N(z_1),E_N(z_2)]&=&(-x^r+x^{-r})
\delta\left(x^{r^*+2s}\frac{z_2}{z_1}\right)
{\cal B}_N(x^{r^*}z_1),\\
~[\Lambda_1(z_1),E_N(z_2)]&=&(x^r-x^{-r})
\delta\left(x^{-r^*}\frac{z_2}{z_1}\right)
{\cal B}_N(x^{-r^*}z_2).
\end{eqnarray}
Here we have set
\begin{eqnarray}
{\cal B}_j(z)&=&
e^{-i\sqrt{\frac{r}{r^*}}Q_{\alpha_j}}
x^{-\sqrt{rr^*}(P_{\bar{\epsilon}_j}+P_{\bar{\epsilon}_{j+1}})}
(zx^{-j})^{-\sqrt{\frac{r}{r^*}}P_{\alpha_j}+\frac{r}{r^*}}\nonumber\\
&\times&:\exp\left(
-\sum_{m\neq 0}
\frac{1}{m}\frac{[rm]}{[r^*m]}
(x^{-r^*m}\beta_m^j-x^{r^*m}\beta_m^{j+1})
z^{-m}\right):,~(1\leqq j \leqq N-1),\nonumber
\\
\\
{\cal B}_N(z)&=&
e^{-i\sqrt{\frac{r^*}{r}}Q_{\alpha_N}}
x^{-\sqrt{rr^*}(P_{\bar{\epsilon}_N}+P_{\bar{\epsilon}_{1}})}
(zx^{2s-N})^{-\sqrt{\frac{r}{r^*}}P_{\bar{\epsilon}_N}+\frac{r}{2r^*}}
z^{\sqrt{\frac{r}{r^*}}P_{\bar{\epsilon}_1}+\frac{r}{2r^*}}\nonumber\\
&\times&:\exp\left(
\sum_{m\neq 0}
\frac{1}{m}\frac{[rm]}{[r^*m]}
(x^{(-r^*-2s)m}\beta_m^N-x^{r^*m}\beta_m^{1})
z^{-m}\right):.
\end{eqnarray}
\end{prop}

\subsection{Comparsion with another definition}

At first glance, our definition
of the deformed $W$-algebra is different from
those in \cite{AKOS, FF2, Odake}.
In this section we show they are essentially the same thing.
Let us set the element ${\cal C}_m$ by
\begin{eqnarray}
{\cal C}_m=\sum_{j=1}^N x^{(N-2j+1)m}\beta_m^j.
\end{eqnarray}
This element ${\cal C}_m$ is $\eta$-invariant,
$\eta({\cal C}_m)={\cal C}_m$.
Let us divide $\Lambda_j(z)$ into $\Lambda_j^{DWA}(z)$ and ${\cal Z}(z)$.
\begin{eqnarray}
\Lambda_j(z)&=&\Lambda_j^{DWA}(z){\cal Z}(z),
~~(1\leqq j \leqq N),
\end{eqnarray}
where we set
\begin{eqnarray}
\Lambda_j^{DWA}(z)&=&
x^{-2\sqrt{r(r-1)} P_{\bar{\epsilon}_j}}
:\exp
\left(\sum_{m \neq 0} \frac{x^{rm}-x^{-rm}}{m}
\left(\beta_m^j-\frac{[m]_x}{[Nm]_x}{\cal C}_m
 \right)z^{-m}\right):,\\
{\cal Z}(z)&=&:\exp\left(\sum_{m\neq 0}\frac{x^{rm}-x^{-rm}}{m}
\frac{[m]_x}{[Nm]_x}{\cal C}_m z^{-m}\right):.
\end{eqnarray}
Let us set 
\begin{eqnarray}
T_j^{DWA}(z)=\sum_{1\leqq s_1<s_2<\cdots <s_j\leqq N}
:\Lambda_{s_1}^{DWA}(x^{-j+1}z)\Lambda_{s_2}^{DWA}(x^{-j+3}z)
\cdots \Lambda_{s_j}^{DWA}(x^{j-1}z):.
\end{eqnarray}

\begin{prop}~~The bosonic operators $T_j^{DWA}(z),~(1\leqq j \leqq N-1)$ satisfy
the following relations.
\begin{eqnarray}
&&f_{i,j}^{DWA}(z_2/z_1)T_i^{DWA}(z_1)T_j^{DWA}(z_2)
-f_{j,i}^{DWA}(z_1/z_2)T_j^{DWA}(z_2)T_i^{DWA}(z_1)\nonumber\\
&=&c
\sum_{k=1}^i
\prod_{l=1}^{k-1}\Delta(x^{2l+1})\times
\left(\delta\left(\frac{x^{j-i+2k}z_2}{z_1}\right)
f_{i-k,j+k}^{DWA}(x^{-j+i})T_{i-k}^{DWA}(x^{-k}z_1)T_{j+k}^{DWA}(x^kz_2)\right.\nonumber\\
&-&\left.
\delta\left(\frac{x^{-j+i-2k}z_2}{z_1}\right)
f_{i-k,j+k}^{DWA}(x^{j-i})T_{i-k}^{DWA}(x^{k}z_1)T_{j+k}^{DWA}(x^{-k}z_2)
\right),~~(1\leqq i \leqq j \leqq N-1),\nonumber\\
\end{eqnarray}
where $\delta(z)=\sum_{n \in {\mathbb Z}}z^n$.
We should understand $T_N^{DWA}(z)=1, T_{j}^{DWA}(z)=0,~(j>N)$.\\
Here we set the constant 
$c$ and the auxiliary function $\Delta(z)$ in (\ref{def:Delta}).
Here we set the structure functions,
\begin{eqnarray}
&&f_{i,j}^{DWA}(z)=f_{i,j}(z)|_{s=N}\nonumber\\
&=&\exp\left(
\sum_{m=1}^\infty
\frac{1}{m}(1-x^{2rm})(1-x^{-2(r-1)m})
\frac{(1-x^{2m Min(i,j)})(1-x^{2m(N-Max(i,j))})}
{(1-x^{2m})(1-x^{2Nm})}x^{|i-j|m}z^m
\right).\nonumber\\
\end{eqnarray}
\end{prop}

\begin{prop}~~~
The operators $T_j^{DWA}(z)$ and ${\cal Z}(z)$ commutes with each other.
\begin{eqnarray}
T_j^{DWA}(z_1){\cal Z}(z_2)={\cal Z}(z_2)T_j^{DWA}(z_1),~~~
(1\leqq j \leqq N-1).
\end{eqnarray}
\end{prop}

Therefore three parameter deformed $W$-algebra $T_j(z)$ 
is realized as
an extension of 
two parameter deformed $W$-algebra $T_j^{DWA}(z)$
in \cite{AKOS, FF2, Odake}.
Note that upon the specialization $s=N$ we have
\begin{eqnarray}
~[B_n^N,B_m^N]=0,~[B_m^j,B_n^N]=0~{\rm for}~j\neq N.
\end{eqnarray}
Hence we can regard $B_m^N=0$ and $T_j(z)=T_j^{DWA}(z),~T_N^{DWA}(z)=1$.

\section{Local Integrals of Motion}

In this section 
we construct the local integrals of motion ${\cal I}_n$.
We study the generic case :
$0<x<1$, $r \in {\mathbb C}$ and ${\rm Re}(s)>0$.

\subsection{Local Integrals of Motion for $W_{q,t}(\widehat{sl_N})$}

Let us set the function $h(u)$ and $h^*(u)$ by
\begin{eqnarray}
h(u)&=&\frac{[u]_s [u+r]_s}{
[u+1]_s [u+r^*]_s},
~~~
h^*(u)=
\frac{[u]_s [u-r^*]_s}{
[u+1]_s [u-r]_s},
\label{def:h}
\end{eqnarray}
where we have set $z=x^{2u}$. 

\begin{df}~~\\
$\bullet$~~
We define ${\cal I}_n$ for regime ${\rm Re}(s)>2$ and 
${\rm Re}(r^*)<0$ by
\begin{eqnarray}
{\cal I}_n=\int \cdots \int_{C} 
\prod_{j=1}^n \frac{dz_j}{2\pi\sqrt{-1}z_j}
\prod_{1\leqq j<k \leqq n}h(u_k-u_j)
T_1(z_1)\cdots T_1(z_n)~~~(n=1,2,\cdots).
\label{def:LocalIM}
\end{eqnarray}
Here, the contour $C$ encircles $z_j=0$ in such a way that
$z_j=x^{-2+2sl}z_k, x^{-2r^*+2sl}z_k ~(l=0,1,2,\cdots)$ is inside and 
$z_j=x^{2-2sl}z_k, x^{2r^*-2sl}z_k ~(l=0,1,2,\cdots)$ 
is outside for $1\leqq j<k \leqq n$.
We call ${\cal I}_n$ 
the local integrals of motion
for the deformed $W$-algebra.
The definitions of ${\cal I}_n$ for
generic ${\rm Re}(s)>0$ and $r \in {\mathbb C}$ should be
understood as analytic continuation.\\
$\bullet$~~
We define ${\cal I}_n^*$ for regime ${\rm Re}(s)>2$ and 
${\rm Re}(r)>0$ by
\begin{eqnarray}
{\cal I}_n=\int \cdots \int_{C} 
\prod_{j=1}^n \frac{dz_j}{2\pi\sqrt{-1}z_j}
\prod_{1\leqq j<k \leqq n}h^*(u_k-u_j)
T_1(z_1)\cdots T_1(z_n)~~~(n=1,2,\cdots).
\label{def:Local*IM}
\end{eqnarray}
Here, the contour $C$ encircles $z_j=0$ in such a way that
$z_j=x^{-2+2sl}z_k, x^{2r+2sl}z_k ~(l=0,1,2,\cdots)$ is inside and 
$z_j=x^{2-2sl}z_k, x^{-2r-2sl}z_k ~(l=0,1,2,\cdots)$ 
is outside for $1\leqq j<k \leqq n$.
We call ${\cal I}_n^*$ 
the local integrals of motion
for the deformed $W$-algebra.
The definitions of ${\cal I}_n^*$ for
generic ${\rm Re}(s)>0$ and $r \in {\mathbb C}$ should be
understood as analytic continuation.
\end{df}

The following is one of
{\bf Main Results} of this paper.

\begin{thm}~~~\label{thm:Local-Com}
The local integrals of motion 
${\cal I}_n$
commute with each other
\begin{eqnarray}
~[{\cal I}_n,{\cal I}_m]=0~~~(m,n=1,2,\cdots ).
\end{eqnarray}
The local integrals of motion 
${\cal I}_n^*$
commute with each other
\begin{eqnarray}
~[{\cal I}_n^*,{\cal I}_m^*]=0~~~(m,n=1,2,\cdots ).
\end{eqnarray}
\end{thm}

\subsection{Laurent-Series Formulae}

In this subsection we prepare another formulae of 
the local integrals of motion ${\cal I}_n$.
Because the integral contour of the definition of
the local integrals of motion ${\cal I}_n$
is not annulus. {\it i.e.} $|x^{-p}z_k|<|z_j|<|x^pz_k|$,
the defining relations of the deformed $W$-algebra
(\ref{def:3parameter})
should be used carefully. 
Hence, in order to show
the commutation relations 
$[{\cal I}_m,{\cal I}_n]=0$,
it is better for us to deform the integral representations
of the local integrals of motion ${\cal I}_n$
to another formulae, in which 
the defining relations of the deformed $W$-algebra
(\ref{def:3parameter})
can be used safely.

Let us set the auxiliary function $s(z), s^*(z)$ by
$h(u)=s(z)f_{11}(z)$, $h^*(u)=s^*(z)f_{11}(z)$,
$(z=x^{2u})$
where $h(u), h^*(u)$ 
and $f_{11}(z)$ are given in the previous section.
We have explicitly
\begin{eqnarray}
s(z)&=&x^{-2r^*}\frac{(z;x^{2s})_\infty 
(x^{2s-2r}z;x^{2s})_\infty }
{(x^{2s-2}z;x^{2s})_\infty 
(x^{-2r^*}z;x^{2s})_\infty }\times
\frac{(1/z;x^{2s})_\infty 
(x^{2s-2r}/z;x^{2s})_\infty }
{(x^{2s-2}/z;x^{2s})_\infty 
(x^{-2r^*}/z;x^{2s})_\infty },\\
s^*(z)&=&x^{-2r^*}\frac{(z;x^{2s})_\infty 
(x^{2s+2r^*}z;x^{2s})_\infty }
{(x^{2s-2}z;x^{2s})_\infty 
(x^{2r}z;x^{2s})_\infty }\times
\frac{(1/z;x^{2s})_\infty 
(x^{2s+2r^*}/z;x^{2s})_\infty }
{(x^{2s-2}/z;x^{2s})_\infty 
(x^{2r}/z;x^{2s})_\infty }.
\end{eqnarray}
Let us set
the auxiliary functions $g_{i,j}(z)$ 
by fusion procedure
\begin{eqnarray}
g_{i,1}(z)=g_{1,1}(x^{-i+1}z)g_{1,1}(x^{-i+3}z)\cdots g_{1,1}(x^{i-1}z),
\nonumber\\
g_{i,j}(z)=g_{i,1}(x^{-j+1}z)g_{i,1}(x^{-j+3}z)\cdots g_{i,1}(x^{j-1}z).
\end{eqnarray}
where $g_{11}(z)=f_{11}(z)$ 
is the structure function
of the deformed $W$-algebra defined in 
(\ref{def:3parameter}).
\begin{eqnarray}
f_{1,1}(z)=\frac{1}{1-z}
\frac{(x^{2s-2}z;x^{2s})_\infty 
(x^{2r}z;x^{2s})_\infty
(x^{-2r^*}z;x^{2s})_\infty
}{(x^2z;x^{2s})_\infty 
(x^{2r^*+2s}z;x^{2s})_\infty
(x^{2s-2r}z;x^{2s})_\infty
}.
\end{eqnarray}
The structure functions $f_{1,j}(z)$ and $g_{1,j}(z)$
have the following relations
\begin{eqnarray}
g_{1,j}(z)=\Delta(x^{-j+2}z)\Delta(x^{-j+4}z)\cdots \Delta(x^{j-2}z)f_{1,j}(z).
\end{eqnarray}
Here $\Delta(z)$ is given by $\Delta(z)=\frac{(1-x^{r+r^*}z)
(1-x^{-r-r^*}z)
}{(1-xz)(1-x^{-1}z)}$.

Let us set the formal power series 
${\cal A}(z_1,z_2,\cdots,z_n)$ by
\begin{eqnarray}
{\cal A}(z_1,z_2,\cdots,z_n)=\sum_{k_1,\cdots,k_n \in {\mathbb Z}}
a_{k_1,\cdots,k_n}z_1^{k_1}z_2^{k_2}\cdots z_n^{k_n}.
\end{eqnarray}
We define the symbol $[\cdots]_{1,z_1\cdots z_n}$ by
\begin{eqnarray}
\left[
{\cal A}(z_1,z_2,\cdots,z_n)
\right]_{1,z_1\cdots z_n}=a_{0,0,\cdots ,0}.
\end{eqnarray}
Let us set
$D=\{(z_1,\cdots,z_n)\in {\mathbb C}^n|
\sum_{k_1,\cdots,k_n \in {\mathbb Z}}
|a_{k_1,\cdots,k_n}z_1^{k_1}z_2^{k_2}\cdots z_n^{k_n}|<+\infty
\}$.
When we assume closed curve $J$ is contained in $D$,
we have
\begin{eqnarray}
[{\cal A}(z_1,z_2,\cdots,z_n)]_{1,z_1 \cdots z_n}=\int \cdots \int_J 
\prod_{j=1}^n \frac{dz_j}{2\pi\sqrt{-1}z_j}
{\cal A}(z_1,z_2,\cdots,z_n).
\end{eqnarray}

Let us set the auxiliary functions, $s_{11}(z)=s(z),~
h_{11}(z)=h(u)$,$(z=x^{2u})$ and
\begin{eqnarray}
s_{i,1}(z)=s_{1,1}(x^{-i+1}z)s_{1,1}(x^{-i+3}z)\cdots s_{1,1}(x^{i-1}z),
\nonumber\\
s_{i,j}(z)=s_{i,1}(x^{-j+1}z)s_{i,1}(x^{-j+3}z)\cdots s_{i,1}(x^{j-1}z),\\
h_{i,1}(z)=h_{1,1}(x^{-i+1}z)h_{1,1}(x^{-i+3}z)\cdots h_{1,1}(x^{i-1}z),
\nonumber\\
h_{i,j}(z)=h_{i,1}(x^{-j+1}z)h_{i,1}(x^{-j+3}z)\cdots h_{i,1}(x^{j-1}z),
\end{eqnarray}
and
\begin{eqnarray}
s_{i,1}^*(z)=s_{1,1}^*(x^{-i+1}z)s_{1,1}^*(x^{-i+3}z)\cdots 
s_{1,1}^*(x^{i-1}z),
\nonumber\\
s_{i,j}^*(z)=s_{i,1}^*(x^{-j+1}z)s_{i,1}^*(x^{-j+3}z)
\cdots s_{i,1}^*(x^{j-1}z),\\
h_{i,1}^*(z)=h_{1,1}^*(x^{-i+1}z)h_{1,1}^*(x^{-i+3}z)
\cdots h_{1,1}^*(x^{i-1}z),
\nonumber\\
h_{i,j}^*(z)=h_{i,1}^*(x^{-j+1}z)h_{i,1}^*(x^{-j+3}z)
\cdots h_{i,1}^*(x^{j-1}z).
\end{eqnarray}

In what follows we use the notation of the ordered product
\begin{eqnarray}
\prod_{\longrightarrow \atop{l \in L}}T_1(z_l)=T_1(z_{l_1})T_1(z_{l_2})
\cdots T_1(z_{l_n}),~~~(L=\{l_1,\cdots,l_n|l_1<l_2<\cdots<l_n\}).
\end{eqnarray}

\begin{thm}~~~\label{thm:Local-Laurent}
For ${\rm Re}(s)>N$ and ${\rm Re}(r^*)<0$,
the local integrals of motion ${\cal I}_n$ are written as
\begin{eqnarray}
{\cal I}_n&=&\left[\prod_{1\leqq j<k \leqq n}s(z_k/z_j)
{\cal O}_n(z_1, z_2, \cdots,z_n)\right]_{1,z_1\cdots z_n}.
\end{eqnarray}
For ${\rm Re}(s)>N$ and ${\rm Re}(r)>0$,
the local integrals of motion ${\cal I}_n^*$ are written as
\begin{eqnarray}
{\cal I}_n^*&=&\left[\prod_{1\leqq j<k \leqq n}s^*(z_k/z_j)
{\cal O}_n(z_1, z_2, \cdots,z_n)\right]_{1,z_1\cdots z_n}.
\end{eqnarray}
Here we set the operator ${\cal O}_n(z_1,z_2,\cdots,z_n)$ by
\begin{eqnarray}
&&{\cal O}_n(z_1,z_2,\cdots,z_n)=
\sum_{\alpha_1,\alpha_2,\alpha_3,\cdots,\alpha_N\geqq 0
\atop{\alpha_1+2\alpha_2+3\alpha_3+\cdots+N\alpha_N=n}}
\sum_{
\left\{A_j^{(s)}\right\}_{s=1,\cdots,N
\atop{j=1,\cdots,\alpha_s}}
\atop{
A_{j}^{(s)}\subset
\{1,2,\cdots,n\},~~
|A_j^{(s)}|=s,~~
\cup_{s=1}^N \cup_{j=1}^{\alpha_s}
A_j^{(s)}=\{1,2,\cdots,n\}
\atop{Min(A_1^{(s)})<Min(A_2^{(s)})<\cdots<Min(A_{\alpha_s}^{(s)})}
}}\nonumber\\
&\times&
\prod_{\longrightarrow
\atop{j \in A_{Min}^{(1)}}}T_1(z_j)
\prod_{\longrightarrow
\atop{j \in A_{Min}^{(2)}}}T_2(x^{-1}z_j)
\cdots
\prod_{\longrightarrow
\atop{j \in A_{Min}^{(t)}}}T_t(x^{-1+t-2[\frac{t}{2}]}z_j)
\cdots
\prod_{\longrightarrow
\atop{j \in A_{Min}^{(N)}}}T_N(x^{-1+N-2[\frac{N}{2}]}z_j)\nonumber\\
&\times&
\prod_{t=1}^N
\left((-c)^{t-1}
\prod_{u=1}^{t-1}
\Delta(x^{2u+1})^{t-u-1}\right)^{\alpha_t}
\prod_{t=1}^N
\prod_{j=1
\atop{j_1=A_{j,1}^{(t)}
\atop{\cdots
\atop{j_t=A_{j,t}^{(t)}}
}}}^{\alpha_t}
\sum_{\sigma \in S_t
\atop{\sigma(1)=1}}
\prod_{u=1
\atop{u \neq [\frac{t}{2}]+1}}^t
\delta\left(\frac{x^2z_{j_{\sigma(u+1)}}}{z_{j_{\sigma(u)}}}\right)
\nonumber\\
&\times&
\prod_{t=1}^N \prod_{j<k
\atop{j,k \in A_{Min}^{(t)}}}g_{t,t}\left(\frac{z_k}{z_j}\right)
\prod_{1\leqq t<u \leqq N}\prod_{j\in A_{Min}^{(t)}
\atop{k\in A_{Min}^{(u)}}}g_{t,u}\left(
x^{u-t-2[\frac{u}{2}]+2[\frac{t}{2}]}\frac{z_k}{z_j}
\right).
\end{eqnarray}
Here we have set the constant $c$ and the function
$\Delta(z)$ in (\ref{def:Delta}).
When the index set $A_j^{(t)}=
\{j_1,j_2,\cdots, j_t|j_1<j_2<\cdots<j_t\}$,
$(1\leqq t \leqq N, 1\leqq j \leqq \alpha_t)$,
we set $A_{j,k}^{(t)}=j_k$, and
$A_{Min}^{(t)}=\{A_{1,1}^{(t)},A_{2,1}^{(t)},
\cdots, A_{\alpha_t,1}^{(t)}\}$.
Here we should understand $z_{j_{\sigma(t+1)}}=
z_{j_{\sigma(1)}}$ in the delta-function
$
\delta\left(\frac{x^2z_{j_{\sigma(t+1)}}}
{z_{j_{\sigma(t)}}}\right)$.
\end{thm}

~\\
{\bf Example}~~~
We summarize the operators ${\cal O}_n$ very explicitly. 
\begin{eqnarray}
{\cal O}_1(z)&=&T_1(z),\\
{\cal O}_2(z_1,z_2)&=&g_{1,1}(z_2/z_1)T_1(z_1)T_1(z_2)-
c\delta(x^2z_2/z_1)T_2(x^{-1}z_1),\\
{\cal O}_3(z_1,z_2,z_3)&=&
g_{11}(z_2/z_1)g_{1,1}(z_3/z_1)g_{1,1}(z_3/z_2)T_1(z_1)T_1(z_2)T_1(z_3)\nonumber\\
&-&cg_{1,2}(x^{-1}z_2/z_1)T_1(z_1)
\delta(x^2z_3/z_2)T_2(x^{-1}z_2)\nonumber\\
&-&cg_{1,2}(x^{-1}z_1/z_2)
T_1(z_2)\delta(x^2z_3/z_1)T_2(x^{-1}z_1)\nonumber\\
&-&c
g_{1,2}(x^{-1}z_1/z_3)T_1(z_3)
\delta(x^2z_2/z_1)T_2(x^{-1}z_1)\nonumber\\
&+&c^2\Delta(x^3)
(\delta(x^2z_2/z_1)\delta(x^2z_1/z_3)
+\delta(x^2z_1/z_2)\delta(x^2z_3/z_1))T_3(z_1).
\end{eqnarray}
\begin{eqnarray}
{\cal O}_4(z_1,z_2,z_3,z_4)&=&
\prod_{1\leqq j<k \leqq 4}g_{11}(z_k/z_j)
T_1(z_1)T_1(z_2)T_1(z_3)T_1(z_4)\nonumber\\
&-&cg_{11}(z_2/z_1)g_{12}(x^{-1}z_3/z_1)g_{12}(x^{-1}z_3/z_2)
T_1(z_1)T_1(z_2)T_2(x^{-1}z_3)\delta(x^2z_4/z_3)\nonumber\\
&-&cg_{11}(z_3/z_1)g_{12}(x^{-1}z_2/z_1)g_{12}(x^{-1}z_2/z_3)
T_1(z_1)T_1(z_3)T_2(x^{-1}z_2)\delta(x^2z_4/z_2)\nonumber\\
&-&cg_{11}(z_4/z_1)g_{12}(x^{-1}z_2/z_1)g_{12}(x^{-1}z_2/z_4)
T_1(z_1)T_1(z_4)T_2(x^{-1}z_2)\delta(x^2z_3/z_2)\nonumber\\
&-&cg_{11}(z_3/z_2)g_{12}(x^{-1}z_1/z_2)g_{12}(x^{-1}z_1/z_3)
T_1(z_2)T_1(z_3)T_2(x^{-1}z_1)\delta(x^2z_4/z_1)\nonumber\\
&-&cg_{11}(z_4/z_2)g_{12}(x^{-1}z_1/z_2)g_{12}(x^{-1}z_1/z_4)
T_1(z_2)T_1(z_4)T_2(x^{-1}z_1)\delta(x^2z_3/z_1)\nonumber\\
&-&cg_{11}(z_4/z_3)g_{12}(x^{-1}z_1/z_3)g_{12}(x^{-1}z_1/z_4)
T_1(z_3)T_1(z_4)T_2(x^{-1}z_1)\delta(x^2z_2/z_1)\nonumber\\
&+&c^2 g_{22}(z_3/z_1)\delta(x^2z_2/z_1)\delta(x^2z_4/z_3)
T_2(x^{-1}z_1)T_2(x^{-1}z_3)\nonumber\\
&+&c^2 g_{22}(z_2/z_1)\delta(x^2z_3/z_1)\delta(x^2z_4/z_2)
T_2(x^{-1}z_1)T_2(x^{-1}z_2)\nonumber\\
&+&c^2 g_{22}(z_2/z_1)\delta(x^2z_4/z_1)\delta(x^2z_3/z_2)
T_2(x^{-1}z_1)T_2(x^{-1}z_2)\nonumber\\
&+&c^2\Delta(x^3) 
g_{13}(z_2/z_1)
T_1(z_1)T_3(z_2)
(\delta(x^2z_2/z_3)\delta(x^2z_4/z_2)
+\delta(x^2z_2/z_4)\delta(x^2z_3/z_2))\nonumber
\\&+&c^2\Delta(x^3) 
g_{13}(z_1/z_2)
T_1(z_2)T_3(z_1)
(\delta(x^2z_1/z_3)\delta(x^2z_4/z_1)
+\delta(x^2z_1/z_4)\delta(x^2z_3/z_1))\nonumber
\\&+&c^2\Delta(x^3) 
g_{13}(z_1/z_3)
T_1(z_3)T_3(z_1)
(\delta(x^2z_1/z_2)\delta(x^2z_4/z_1)
+\delta(x^2z_1/z_4)\delta(x^2z_2/z_1))\nonumber
\\
&+&c^2\Delta(x^3) 
g_{13}(z_1/z_4)
T_1(z_4)T_3(z_1)
(\delta(x^2z_1/z_2)\delta(x^2z_3/z_1)
+\delta(x^2z_1/z_3)\delta(x^2z_2/z_1))\nonumber\\
&-&c^3\Delta(x^3)\Delta(x^5)^2
T_4(x^{-1}z_1)\nonumber\\
&\times&(
\delta(x^2z_1/z_2)\delta(x^2z_3/z_1)\delta(x^2z_4/z_3)
+
\delta(x^2z_1/z_2)\delta(x^2z_4/z_1)\delta(x^2z_3/z_4)\nonumber\\
&+&
\delta(x^2z_1/z_3)\delta(x^2z_2/z_1)\delta(x^2z_4/z_2)
+
\delta(x^2z_1/z_3)\delta(x^2z_4/z_1)\delta(x^2z_2/z_4)\nonumber\\
&+&
\delta(x^2z_1/z_4)\delta(x^2z_2/z_1)\delta(x^2z_3/z_2)
+
\delta(x^2z_1/z_4)\delta(x^2z_2/z_1)\delta(x^2z_2/z_3))\nonumber
\end{eqnarray}
We should understand above as $T_j(z)=0$ for $j>N$.

%%%%%%%%%%%%%%%%%%%%%%%%%%%%%%%%%%%%%%%%%%%%%%%%%%%
%%%%%%%%%%%%%%%%%%%%%%%%%%%%%%%%%%%%%%%%%%%%%%%%%%%
\subsection{Weakly sense equality}

In order to show thorem,
we introduce a ``weak sense'' equality

\begin{df}~~
We say the operators ${\cal P}(z_1,z_2,\cdots, z_n)$
and ${\cal Q}(z_1,z_2,\cdots,z_n)$ are equal in 
the "weak sense" if
\begin{eqnarray}
\prod_{1\leqq i<j \leqq n}s(z_j/z_i){\cal P}(z_1,z_2,\cdots,z_n)=
\prod_{1\leqq i<j \leqq n}s(z_j/z_i){\cal Q}(z_1,z_2,\cdots,z_n).
\end{eqnarray}
We write
${\cal P}(z_1,z_2,\cdots,z_n)\sim {\cal Q}(z_1,z_2,\cdots,z_n)$, 
showing the weak equality.
\end{df}
For example 
$\delta(z_1/z_2)\sim 0$
and
$\frac{1}{z_1-z_2}\delta(z_1/z_2)\sim 0$.

\begin{prop}~~~~\label{prop:dW-Delta1}
The following relations hold in the weak sense for 
$1\leqq j \leqq N$
\begin{eqnarray}
&&\left(g_{1,j}\left(
\frac{x^{-1+j-2[\frac{j}{2}]}
w_1}{z_1}\right)
T_1(z_1)T_j(x^{-1+j-2[\frac{j}{2}]}w_1)-
g_{j,1}\left(
\frac{x^{1-j+2[\frac{j}{2}]}z_1}{w_1}\right)
T_j(x^{-1+j-2[\frac{j}{2}]}w_1)T_1(z_1)\right)\nonumber\\
&\times&\sum_{\sigma \in S_j\atop{\sigma(1)=1}}
\prod_{t=1
\atop{t\neq [\frac{j}{2}]+1}}^j
\delta\left(\frac{x^2w_{\sigma(t+1)}}{w_{\sigma(t)}}\right)
\sim
c\prod_{t=1}^{j-1} \Delta(x^{2t+1})
\sum_{\sigma \in S_j\atop{\sigma(1)=1}}
\prod_{t=1
\atop{t\neq [\frac{j}{2}]+1}}^j
\delta\left(\frac{x^2w_{\sigma(t+1)}}{w_{\sigma(t)}}\right)
\nonumber\\
&\times&
\left(\delta\left(\frac{x^{2j-2[\frac{j}{2}]}w_1}{z_1}\right)
T_{j+1}(x^{j-2[\frac{j}{2}]}w_1)-
\delta\left(\frac{x^{-2-2[\frac{j}{2}]}w_1}{z_1}\right)
T_{j+1}(x^{j-2-2[\frac{j}{2}]}w_1)\right).
\end{eqnarray} 
We should understand $T_{N+1}(z)=0$ and
$w_{\sigma(j+1)}=w_{\sigma(1)}$ in
the delta-function
$\delta\left(\frac{x^2w_{\sigma(j+1)}}
{w_{\sigma(j)}}\right)$.
\end{prop}
{\it Proof}~~
We explain the mechanism by the simplest case for $N \geqq 3$.
\begin{eqnarray}
&&\left(g_{1,2}(x^{-1}z_2/z_1)T_1(z_1)T_2(x^{-1}z_2)-
g_{2,1}(xz_1/z_2)T_2(x^{-1}z_2)T_1(z_1)\right)\delta(x^2z_3/z_2)
\nonumber\\
&=&
g_{1,2}(x^{-1}z_2/z_1)c(\delta(z_2/z_1)-\delta(x^{-2}z_2/z_1))
\delta(x^2z_3/z_2)T_1(z_1)T_2(x^{-1}z_2)\\
&+&\Delta(xz_1/z_2)\delta(x^2z_3/z_2)
(f_{1,2}(x^{-1}z_2/z_1)T_1(z_1)T_2(x^{-1}z_2)-
f_{2,1}(xz_1/z_2)T_2(x^{-1}z_2)T_1(z_1)).\nonumber
\end{eqnarray}
Here we have used $g_{1,2}(z)=\Delta(z)f_{1,2}(z)$ and
\begin{eqnarray}
\Delta(z)-\Delta(z^{-1})=c(\delta(xz)-\delta(x^{-1}z)),~~~
\Delta(z)=\frac{(1-x^{2r-1}z)(1-x^{-2r+1}z)}{(1-xz)(1-x^{-1}z)}.
\end{eqnarray}
Using $\delta(z_1/z_2)\sim 0$ and $\delta(x^2z_1/z_2)\delta(x^2z_3/z_2)
\sim 0$ , $\Delta(x^3)=\Delta(x^{-3})$,
and the defining relation of the deformed $W$-algebra 
(\ref{def:3parameter}),
we get this proposition.
~~~
Q.E.D.

~\\
As the same manner as above, we have
the following proposition.

\begin{prop}~~~~\label{prop:dW-Delta2}
The following relations hold in the weak sense for $i,j \geqq 2$
\begin{eqnarray}
&&g_{i,j}
\left(\frac{x^{j-i-2[\frac{j-i}{2}]}w_1}{z_1}\right)
T_i(x^{-1+i-2[\frac{i}{2}]}z_1)
T_j(x^{-1+j-2[\frac{j}{2}]}w_1)\nonumber\\
&\times&
\sum_{\sigma \in S_j\atop{\sigma(1)=1}}
\prod_{t=1
\atop{t\neq [\frac{i}{2}]+1}}^i
\delta\left(\frac{x^2z_{\sigma(t+1)}}{z_{\sigma(t)}}\right)
\sum_{\sigma \in S_j\atop{\sigma(1)=1}}
\prod_{t=1
\atop{t\neq [\frac{j}{2}]+1}}^j
\delta\left(\frac{x^2w_{\sigma(t+1)}}{w_{\sigma(t)}}\right)\nonumber\\
&\sim&
g_{j,i}\left(\frac{x^{i-j-2[\frac{i-j}{2}]}z_1}{w_1}\right)
T_j(x^{-1+j-2[\frac{j}{2}]}w_1)
T_i(x^{-1+i-2[\frac{i}{2}]}z_1)\nonumber\\
&\times&
\sum_{\sigma \in S_j\atop{\sigma(1)=1}}
\prod_{t=1
\atop{t\neq [\frac{i}{2}]+1}}^i
\delta\left(\frac{x^2z_{\sigma(t+1)}}{z_{\sigma(t)}}\right)
\sum_{\sigma \in S_j\atop{\sigma(1)=1}}
\prod_{t=1
\atop{t\neq [\frac{j}{2}]+1}}^j
\delta\left(\frac{x^2w_{\sigma(t+1)}}{w_{\sigma(t)}}\right).
\end{eqnarray} 
\end{prop}
~\\
{\bf Example}~~
The operators $T_1(z),T_2(z),T_3(z)$
satisfy
\begin{eqnarray}
&&g_{1,1}
\left(\frac{z_2}{z_1}\right)
T_1(z_1)T_1(z_2)
-
g_{1,1}\left(\frac{z_1}{z_2}\right)
T_1(z_2)T_1(z_1)\nonumber\\
&\sim&
c
\left(T_2(x^{-1}z_1)\delta\left(\frac{x^2z_2}{z_1}\right)-
T_2(xz_1)\delta\left(\frac{x^2z_1}{z_2}\right)\right),\\
&&
\left(g_{1,2}\left(\frac{x^{-1}z_2}{z_1}\right)
T_1(z_1)T_2(x^{-1}z_2)
-g_{2,1}
\left(\frac{xz_1}{z_2}\right)
T_2(x^{-1}z_2)T_1(z_1)\right)
\delta\left(\frac{x^2z_3}{z_2}\right)\nonumber\\
&\sim&
c\Delta(x^3)
\left(
T_3(z_2)\delta\left(\frac{x^2z_2}{z_1}\right)-
T_3(z_3)\delta\left(\frac{x^2z_1}{z_3}\right)
\right)
\delta
\left(\frac{x^2z_3}{z_2}\right),\\
&&
\left(g_{2,2}\left(\frac{w_1}{z_1}\right)
T_2(z_1)T_2(w_1)-g_{2,2}\left(\frac{z_1}{w_1}\right)
T_2(w_1)T_2(z_1)\right)\nonumber\\
&\times&
\delta\left(\frac{x^2z_1}{z_2}\right)
\delta\left(\frac{x^2w_1}{w_2}\right)\sim 0,\\
&&
\left(g_{1,3}\left(\frac{w_1}{z_1}\right)
T_1(z_1)T_3(w_1)
-
g_{3,1}\left(\frac{z_1}{w_1}\right)T_3(w_1)T_1(z_1)\right)
\nonumber\\
&\times&
\left(\delta\left(\frac{x^2w_1}{w_2}\right)
\delta\left(\frac{x^2w_3}{w_1}\right)+
\delta\left(\frac{x^2w_1}{w_3}\right)
\delta\left(\frac{x^2w_2}{w_1}\right)\right)\sim 0,\\
&&\left(
g_{2,3}\left(\frac{xw_1}{z_1}\right)
T_2(x^{-1}z_1)T_3(w_1)-
g_{3,2}\left(\frac{x^{-1}z_1}{z_2}\right)
T_3(z_2)T_2(x^{-1}z_1)\right)\nonumber\\
&\times&
\delta\left(\frac{x^2z_1}{z_2}\right)
\left(\delta\left(\frac{x^2w_1}{w_2}\right)
\delta\left(\frac{x^2w_3}{w_1}\right)+
\delta\left(\frac{x^2w_1}{w_3}\right)
\delta\left(\frac{x^2w_2}{w_1}\right)\right)
\sim 0,\\
&&\left(g_{3,3}\left(\frac{w_1}{z_1}\right)
T_3(z_1)T_3(w_1)-
g_{3,3}\left(\frac{w_1}{z_1}\right)
T_3(w_1)T_3(z_1)\right)\nonumber\\
&\times&
\left(\delta\left(\frac{x^2w_1}{w_2}\right)
\delta\left(\frac{x^2w_3}{w_1}\right)+
\delta\left(\frac{x^2w_1}{w_3}\right)
\delta\left(\frac{x^2w_2}{w_1}\right)\right)
\nonumber\\
&\times&
\left(\delta\left(\frac{x^2w_1}{w_2}\right)
\delta\left(\frac{x^2w_3}{w_1}\right)+
\delta\left(\frac{x^2w_1}{w_3}\right)
\delta\left(\frac{x^2w_2}{w_1}\right)\right)\sim 0.
\end{eqnarray}
We should understand $T_j(z)=0$ for $j>N$.

Let us introduce $S_n$-invariance in the ``weak sense''.
\begin{df}~
We call 
the operator ${\cal P}(z_1,z_2,\cdots,z_n)$ is $S_n$-invariant 
in the "weak sense"
if 
\begin{eqnarray}
{\cal P}(z_1,z_2,\cdots,z_n)\sim{\cal P}(z_{\sigma(1)},z_{\sigma(2)},
\cdots,z_{\sigma(n)}),~~(\sigma \in S_n).
\end{eqnarray} 
\end{df}
{\bf Example}~~~The operator ${\cal O}_2(z_1,z_2)=
g_{11}(z_2/z_1)T_1(z_1)T_1(z_2)-
c\delta(x^2z_2/z_1)T_2(x^{-1}z_1)$
is $S_2$-invariant.

\begin{thm}~~~\label{thm:Sn-InvLocal}
The operator ${\cal O}_n$ defined in
Theorem 
\ref{thm:Local-Laurent} is $S_n$-invariant
in the weak sense.
\begin{eqnarray}
{\cal O}_n(z_1,z_2,\cdots,z_n)
\sim{\cal O}_n(z_{\sigma(1)},z_{\sigma(2)},
\cdots,z_{\sigma(n)})~~~
(\sigma \in S_n).
\end{eqnarray}
\end{thm}

This theorem plays an important role in proof of
the main theorem \ref{thm:Local-Com}.
We will show above theorem in the next section.

\subsection{Proof of $S_n$-Invariance 
for ${\cal O}_n(z_1,\cdots,z_n)$}

In this section we give proof of theorem \ref{thm:Sn-InvLocal}.
Proof for special case $\widehat{sl_2}$
is summarized in \cite{FKSW1}.
By straightforward calculations we have
the following proposition.

\begin{prop}~~The following relation holds in weakly sense.
\begin{eqnarray}
&&\prod_{1\leqq j<k \leqq M}g_{1,1}(z_k/z_j)
\prod_{\longrightarrow
\atop{1\leqq j \leqq M}}T_1(z_j)-(z_1\leftrightarrow z_2)\nonumber\\
&\sim&
\sum_{t=0}^M
\sum_{3\leqq j_3<j_4<\cdots<j_{t+2}\leqq M}
(-1)^t c^{t+1}
\prod_{u=1}^t \Delta(x^{2u+1})^{t+1-u}\nonumber\\
&\times&
\prod_{3\leqq j<k \leqq M
\atop{j,k\neq j_3,\cdots,j_{t+2}}}
g_{1,1}\left(\frac{z_k}{z_j}\right)
\prod_{3 \leqq j \leqq M
\atop{j\neq j_3,\cdots,j_{t+2}}}
g_{1,t+2}\left(x^{-1+t-2[\frac{t}{2}]}\frac{z_1}{z_j}\right)
\prod_{\longrightarrow
\atop{3\leqq j \leqq M
\atop{j\neq j_3,\cdots,j_{t+2}}}}
T_1(z_j)\nonumber\\
&\times&
T_{t+2}(x^{-1+t-2[\frac{t}{2}]}z_1)
\sum_{\sigma \in S_{t+2}
\atop{\sigma(1)=1}}
\prod_{u=1
\atop{u \neq [\frac{t}{2}]+2
\atop{j_1=1,j_2=2}}}^{t+2}
\delta\left(\frac{x^2z_{j_{\sigma(u+1)}}}{
z_{j_{\sigma(u)}}}\right)
-(z_1\leftrightarrow z_2).
\end{eqnarray}
We should understand $T_j(z)=0~(j>N)$.
\end{prop}

~\\
{\it Proof of Theorem \ref{thm:Sn-InvLocal}}~~
At first 
we consider $\widehat{sl_3}$ case for reader's convenience.
The operator ${\cal O}_n$ for $\widehat{sl_3}$
is written very explicitly as following.
\begin{eqnarray}
&&{\cal O}_n(z_1,z_2,z_3,\cdots,z_n)\nonumber\\
&=&\sum_{\alpha_1,\alpha_2,\alpha_3 \geqq 0
\atop{\alpha_1+2\alpha_2+3\alpha_3=n}}
(-c)^{\alpha_2+2 \alpha_3}\Delta(x^3)^{\alpha_3}
\sum_{
\left\{A_j^{(s)}\right\}_{s=1,\cdots,3
\atop{j=1,\cdots,\alpha_s}}
\atop{
A_j^{(s)} \subset \{1,2,\cdots,n\},~
|A_j^{(s)}|=s,~
\cup_{s=1}^3 \cup_{j=1}^{\alpha_s}
A_j^{(s)}=\{1,2,\cdots,n\}
\atop{Min(A_1^{(s)})<Min(A_2^{(s)})<
\cdots<Min(A_{\alpha_s}^{(s)})}}}\nonumber\\
&\times&
\prod_{\longrightarrow
\atop{j \in A_{Min}^{(1)}}}T_1(z_j)
\prod_{\longrightarrow
\atop{j \in A_{Min}^{(2)}}}T_2(x^{-1}z_j)
\prod_{\longrightarrow
\atop{j \in A_{Min}^{(3)}}}T_3(z_j)\nonumber\\
&\times&
\prod_{j=1
\atop{j_1=A_{j,1}^{(2)}
\atop{j_2=A_{j,2}^{(2)}}}}^{\alpha_2}
\delta\left(\frac{x^2z_{j_2}}{z_{j_1}}\right)
\prod_{j=1
\atop{j_1=A_{j,1}^{(3)}
\atop{j_2=A_{j,2}^{(3)}
\atop{j_3=A_{j,3}^{(3)}
}}}
}^{\alpha_3}\left(
\delta\left(\frac{x^2z_{j_2}}{z_{j_1}}\right)
\delta\left(\frac{x^2z_{j_1}}{z_{j_3}}\right)+
\delta\left(\frac{x^2z_{j_3}}{z_{j_1}}\right)
\delta\left(\frac{x^2z_{j_1}}{z_{j_2}}\right)\right)\nonumber\\
&\times&
\prod_{t=1}^3 
\prod_{1\leqq j<k \leqq n
\atop{j,k \in A_{Min}^{(t)}}
}g_{t,t}\left(\frac{z_k}{z_j}\right)
\prod_{j \in A_{Min}^{(1)}
\atop{k \in A_{Min}^{(2)}}}
g_{1,2}\left(\frac{x^{-1}z_k}{z_j}\right)
\prod_{j \in A_{Min}^{(1)}
\atop{k \in A_{Min}^{(3)}}}
g_{1,3}\left(\frac{z_k}{z_j}\right)
\prod_{j \in A_{Min}^{(2)}
\atop{k \in A_{Min}^{(3)}}}
g_{2,3}\left(\frac{xz_k}{z_j}\right).\nonumber\\
\end{eqnarray}
%%%%%%%%%%%%%%%%%%%%%%%%%%%%%
%\begin{prop}~~
%\begin{eqnarray}
%&&\prod_{1\leqq j<k \leqq M}g_{1,1}(z_k/z_j)
%\prod_{\longrightarrow
%\atop{1\leqq j \leqq M}}T_1(z_j)-(z_1\leftrightarrow z_2)]
%\nonumber\\
%&\sim&
%\prod_{3\leqq j<k \leqq M}g_{1,1}(z_k/z_j)
%\prod_{3 \leqq j \leqq M}g_{1,2}(x^{-1}z_1/z_j)
%\prod_{\longrightarrow
%\atop{3\leqq j \leqq M}}T_1(z_j)\cdot
%T_2(x^{-1}z_1)\delta\left(\frac{x^2z_2}{z_1}\right)
%\nonumber\\
%&-&
%c\Delta(x^3)
%\sum_{i=3}^M \prod_{3\leqq j<k \leqq M
%\atop{j,k\neq i}}
%g_{1,1}(z_k/z_j)
%\prod_{k=3
%\atop{k \neq i}}^M
%g_{1,3}(z_1/z_k)
%\prod_{\longrightarrow
%\atop{3\leqq j \leqq M
%\atop{j \neq i}}
%}T_1(z_j)\cdot
%T_3(z_1)\nonumber\\
%&\times&
%\left(\delta\left(\frac{x^2z_1}{z_i}\right)
%\delta\left(
%\frac{x^2z_2}{z_1}
%\right)+
%\delta\left(\frac{x^2z_1}{z_2}\right)
%\delta\left(
%\frac{x^2z_i}{z_1}
%\right)
%\right)-(z_1\leftrightarrow z_2).
%\end{eqnarray}
%\end{prop}
In order to show $S_n$-invariance, it is enough to show
the case of the permutations $\sigma=(i,i+1)$ 
for $1\leqq i \leqq n-1$.
Let us study the permutation $\sigma=(i,i+1)$.
Because of the cancellations,
the differnce ${\cal O}_n(\cdots,z_i,z_{i+1},\cdots)
-{\cal O}_n(\cdots,z_{i+1},z_i,\cdots)$ has simplification.
We don't have to consider every summation
$\sum_{
\left\{A_j^{(s)}\right\}_{s=1,\cdots,3
\atop{j=1,\cdots,\alpha_s}}}$
in the definition of ${\cal O}_n$.
We only have to consider the summation of 
the following three cases for any $\sigma=(i,i+1)$ 
\begin{eqnarray}
&&(1)~\{i,i+1\}\subset \cup_{j=1}^{\alpha_1}A_j^{(1)},\nonumber\\
&&(2)~A_J^{(2)}=\{i,i+1\} ~~{\rm for~some}~J,\nonumber\\
&&(3)~A_K^{(3)}=\{i,i+1,j|j>i+1\}~~{\rm for~some}~K.\nonumber
\end{eqnarray}
We have
\begin{eqnarray}
&&{\cal O}_n(z_1,\cdots,z_i,z_{i+1},\cdots,z_n)
-{\cal O}_n(z_1,\cdots,z_{i+1},z_i,\cdots,z_n)\nonumber\\
&=&\widetilde{{\cal O}}_n(z_1,\cdots,z_i,z_{i+1},\cdots,z_n)
-\widetilde{{\cal O}}_n(z_1,\cdots,z_{i+1},z_i,\cdots,z_n).
\end{eqnarray}
Here we have set
\begin{eqnarray}
&&\widetilde{{\cal O}}_n(z_1,z_2,z_3,\cdots,z_{n})=
\sum_{\alpha_1,\alpha_2,\alpha_3 \geqq 0
\atop{\alpha_1+2\alpha_2+3\alpha_3=n}}
(-c)^{\alpha_2+2\alpha_3}\Delta(x^3)^{\alpha_3}
\nonumber\\
&\times&
\left(
\sum_{
\left\{A_j^{(s)}\right\}_{s=1,2,3
\atop{j=1,\cdots,\alpha_s}}
\atop{\{i,i+1\}\subset \cup_{j=1}^{\alpha_s}A_j^{(1)}}}
+
\sum_{
\left\{A_j^{(s)}\right\}_{s=1,2,3
\atop{j=1,\cdots,\alpha_s}}
\atop{A_J^{(2)}=\{i,i+1\}~{\rm for~some}~J}}
+
\sum_{
\left\{A_j^{(s)}\right\}_{s=1,2,3
\atop{j=1,\cdots,\alpha_s}}
\atop{A_K^{(3)}=\{i,i+1,j|i+1<j\}~{\rm for~some~}K}}
\right)\nonumber\\
&\times&
\prod_{\longrightarrow
\atop{j \in A_{Min}^{(1)}}}T_1(z_j)
\prod_{\longrightarrow
\atop{j \in A_{Min}^{(2)}}}T_2(x^{-1}z_j)
\prod_{\longrightarrow
\atop{j \in A_{Min}^{(3)}}}T_3(z_j)\nonumber\\
&\times&
\prod_{j=1
\atop{j_1=A_{j,1}^{(2)}
\atop{j_2=A_{j,2}^{(2)}}}}^{\alpha_2}
\delta\left(\frac{x^2z_{j_2}}{z_{j_1}}\right)
\prod_{j=1
\atop{j_1=A_{j,1}^{(3)}
\atop{j_2=A_{j,2}^{(3)}
\atop{j_3=A_{j,3}^{(3)}
}}}
}^{\alpha_3}\left(
\delta\left(\frac{x^2z_{j_2}}{z_{j_1}}\right)
\delta\left(\frac{x^2z_{j_1}}{z_{j_3}}\right)+
\delta\left(\frac{x^2z_{j_3}}{z_{j_1}}\right)
\delta\left(\frac{x^2z_{j_1}}{z_{j_2}}\right)\right)\nonumber\\
&\times&
\prod_{t=1}^3 
\prod_{1\leqq j<k \leqq n
\atop{j,k \in A_{Min}^{(t)}}
}g_{t,t}\left(\frac{z_k}{z_j}\right)
\prod_{j \in A_{Min}^{(1)}
\atop{k \in A_{Min}^{(2)}}}
g_{1,2}\left(\frac{x^{-1}z_k}{z_j}\right)
\prod_{j \in A_{Min}^{(1)}
\atop{k \in A_{Min}^{(3)}}}
g_{1,3}\left(\frac{z_k}{z_j}\right)
\prod_{j \in A_{Min}^{(2)}
\atop{k \in A_{Min}^{(3)}}}
g_{2,3}\left(\frac{xz_k}{z_j}\right).\nonumber\\
\end{eqnarray}
Let us consider the formulae relating to the first summation
in $\widetilde{{\cal O}}_n(z_1,\cdots,z_i,z_{i+1},\cdots,z_n)-
\widetilde{{\cal O}}_n(z_1,\cdots,z_{i+1},z_{i},\cdots,z_n)$.
Let us start from
\begin{eqnarray}
&&\sum_{\alpha_1,\alpha_2,\alpha_3 \geqq 0
\atop{\alpha_1+2\alpha_2+3\alpha_3=n}}
(-c)^{\alpha_2+2\alpha_3}\Delta(x^3)^{\alpha_3}
\sum_{
\left\{A_j^{(s)}\right\}_{s=1,2,3
\atop{j=1,\cdots,\alpha_s}}
\atop{\{i,i+1\}\subset \cup_{j=1}^{\alpha_s}A_j^{(1)}}}
\nonumber\\
&\times&
\prod_{\longrightarrow
\atop{j \in A_{Min}^{(1)}
\atop{j<i}}}
T_1(z_j)
\left(g_{11}(z_{i+1}/z_i)T_1(z_i)T_1(z_{i+1})
-g_{11}(z_{i+1}/z_i)T_1(z_{i+1})T_1(z_i)\right)
\prod_{\longrightarrow
\atop{j \in A_{Min}^{(1)}
\atop{i+1<j}
}}T_1(z_j)
\nonumber\\
&\times&
\prod_{\longrightarrow
\atop{j \in A_{Min}^{(2)}}}T_2(x^{-1}z_j)
\prod_{\longrightarrow
\atop{j \in A_{Min}^{(3)}}}T_3(z_j)\nonumber\\
&\times&
\prod_{j=1
\atop{j_1=A_{j,1}^{(2)}
\atop{j_2=A_{j,2}^{(2)}}}}^{\alpha_2}
\delta\left(\frac{x^2z_{j_2}}{z_{j_1}}\right)
\prod_{j=1
\atop{j_1=A_{j,1}^{(3)}
\atop{j_2=A_{j,2}^{(3)}
\atop{j_3=A_{j,3}^{(3)}
}}}
}^{\alpha_3}\left(
\delta\left(\frac{x^2z_{j_2}}{z_{j_1}}\right)
\delta\left(\frac{x^2z_{j_1}}{z_{j_3}}\right)+
\delta\left(\frac{x^2z_{j_3}}{z_{j_1}}\right)
\delta\left(\frac{x^2z_{j_1}}{z_{j_2}}\right)\right)\nonumber\\
&\times&
\prod_{t=1}^3 
\prod_{1\leqq j<k \leqq n
\atop{j,k \in A_{Min}^{(t)}}
}g_{t,t}\left(\frac{z_k}{z_j}\right)
\prod_{j \in A_{Min}^{(1)}
\atop{k \in A_{Min}^{(2)}}}
g_{1,2}\left(\frac{x^{-1}z_k}{z_j}\right)
\prod_{j \in A_{Min}^{(1)}
\atop{k \in A_{Min}^{(3)}}}
g_{1,3}\left(\frac{z_k}{z_j}\right)
\prod_{j \in A_{Min}^{(2)}
\atop{k \in A_{Min}^{(3)}}}
g_{2,3}\left(\frac{xz_k}{z_j}\right).
\end{eqnarray}
By applying the weakly sense
relation in Proposition \ref{prop:dW-Delta1},
let us change the ordering of
$g_{11}(z_{i+1}/z_i)T_1(z_i)T_1(z_{i+1})
-g_{11}(z_{i+1}/z_i)T_1(z_{i+1})T_1(z_i)$
and
$T_1(z_j)$ for $j>i+1~(j \in A_{Min}^{(1)})$.
We have
\begin{eqnarray}
&&\sum_{\alpha_1,\alpha_2,\alpha_3 \geqq 0
\atop{\alpha_1+2\alpha_2+3\alpha_3=n}}
(-c)^{\alpha_2+2\alpha_3}\Delta(x^3)^{\alpha_3}
\sum_{
\left\{A_j^{(s)}\right\}_{s=1,2,3
\atop{j=1,\cdots,\alpha_s}}
\atop{\{i,i+1\}\subset \cup_{j=1}^{\alpha_s}A_j^{(1)}}}
\nonumber\\
&\times&
\prod_{\longrightarrow
\atop{j \in A_{Min}^{(1)}-\{i,i+1\}}}
T_1(z_j)
T_2(x^{-1}z_i)
\prod_{\longrightarrow
\atop{j \in A_{Min}^{(2)}}}T_2(x^{-1}z_j)
\prod_{\longrightarrow
\atop{j \in A_{Min}^{(3)}}}T_3(z_j)\nonumber\\
&\times&
\delta\left(\frac{x^2z_{i+1}}{z_{i}}\right)
\prod_{j=1
\atop{j_1=A_{j,1}^{(2)}
\atop{j_2=A_{j,2}^{(2)}}}}^{\alpha_2}
\delta\left(\frac{x^2z_{j_2}}{z_{j_1}}\right)
\prod_{j=1
\atop{j_1=A_{j,1}^{(3)}
\atop{j_2=A_{j,2}^{(3)}
\atop{j_3=A_{j,3}^{(3)}
}}}
}^{\alpha_3}\left(
\delta\left(\frac{x^2z_{j_2}}{z_{j_1}}\right)
\delta\left(\frac{x^2z_{j_1}}{z_{j_3}}\right)+
\delta\left(\frac{x^2z_{j_3}}{z_{j_1}}\right)
\delta\left(\frac{x^2z_{j_1}}{z_{j_2}}\right)\right)\nonumber\\
&\times& 
\prod_{1\leqq j<k \leqq n
\atop{j,k \in A_{Min}^{(1)}-\{i,i+1\}}
}g_{1,1}\left(\frac{z_k}{z_j}\right)
\prod_{1\leqq j<k \leqq n
\atop{j,k \in A_{Min}^{(2)}}
}g_{2,2}\left(\frac{z_k}{z_j}\right)
\prod_{j \in A_{Min}^{(2)}}g_{2,2}\left(\frac{z_j}{z_i}\right)
\prod_{1\leqq j<k \leqq n
\atop{j,k \in A_{Min}^{(3)}}
}g_{3,3}\left(\frac{z_k}{z_j}\right)
\nonumber\\
&\times&
\prod_{j \in A_{Min}^{(1)}-\{i,i+1\}
\atop{k \in A_{Min}^{(2)}\cup\{i\}}}
g_{1,2}\left(\frac{x^{-1}z_k}{z_j}\right)
\prod_{j \in A_{Min}^{(1)}-\{i,i+1\}
\atop{k \in A_{Min}^{(3)}}}
g_{1,3}\left(\frac{z_k}{z_j}\right)
\prod_{j \in A_{Min}^{(2)}\cup\{i\}
\atop{k \in A_{Min}^{(3)}}}
g_{2,3}\left(\frac{xz_k}{z_j}\right)\nonumber\\
%%%%%%%%%%%%%%%%%%%%%%%%%%%%%%%%%%%%%%%%%%%%%%%%%%%%%%%%%%%
%%%%%%%%%%%%%%%%%%%%%%%%%%%%%%%%%%%%%%%%%%%%%%%%%%%%%%%%%%%
&-&
\sum_{\alpha_1,\alpha_2,\alpha_3 \geqq 0
\atop{\alpha_1+2\alpha_2+3\alpha_3=n}}
\sum_{
\left\{A_j^{(s)}\right\}_{s=1,2,3
\atop{j=1,\cdots,\alpha_s}}
\atop{\{i,i+1\}\subset \cup_{j=1}^{\alpha_s}A_j^{(1)}}}
\sum_{j\in A_{Min}^{(1)}
\atop{i+1<j}}
(-c)^{\alpha_2+2(\alpha_3+1)}\Delta(x^3)^{\alpha_3+1}
\nonumber\\
&\times&
\prod_{\longrightarrow
\atop{j \in A_{Min}^{(1)}-\{i,i+1\}}}
T_1(z_j)
T_3(z_i)\left(
\delta\left(\frac{x^2z_{i+1}}{z_i}\right)
\delta\left(\frac{x^2z_{z_i}}{z_j}\right)+
\delta\left(\frac{x^2z_{j_j}}{z_{i}}\right)
\delta\left(\frac{x^2z_{i}}{z_{i+1}}\right)\right)
\nonumber\\
&\times&\prod_{\longrightarrow
\atop{j \in A_{Min}^{(2)}}}T_2(x^{-1}z_j)
\prod_{\longrightarrow
\atop{j \in A_{Min}^{(3)}}}T_3(z_j)\nonumber\\
&\times&
\prod_{j=1
\atop{j_1=A_{j,1}^{(2)}
\atop{j_2=A_{j,2}^{(2)}}}}^{\alpha_2}
\delta\left(\frac{x^2z_{j_2}}{z_{j_1}}\right)
\prod_{j=1
\atop{j_1=A_{j,1}^{(3)}
\atop{j_2=A_{j,2}^{(3)}
\atop{j_3=A_{j,3}^{(3)}
}}}
}^{\alpha_3}
\left(
\delta\left(\frac{x^2z_{j_2}}{z_{j_1}}\right)
\delta\left(\frac{x^2z_{j_1}}{z_{j_3}}\right)+
\delta\left(\frac{x^2z_{j_3}}{z_{j_1}}\right)
\delta\left(\frac{x^2z_{j_1}}{z_{j_2}}\right)\right)\nonumber\\
&\times& 
\prod_{1\leqq j<k \leqq n
\atop{j,k \in A_{Min}^{(1)}-\{i,i+1\}}
}g_{1,1}\left(\frac{z_k}{z_j}\right)
\prod_{1\leqq j<k \leqq n
\atop{j,k \in A_{Min}^{(2)}}
}g_{2,2}\left(\frac{z_k}{z_j}\right)
\prod_{1\leqq j<k \leqq n
\atop{j,k \in A_{Min}^{(3)}}
}g_{3,3}\left(\frac{z_k}{z_j}\right)
\prod_{j \in A_{Min}^{(3)}}g_{3,3}\left(\frac{z_j}{z_i}\right)
\nonumber\\
&\times&
\prod_{j \in A_{Min}^{(1)}-\{i,i+1\}
\atop{k \in A_{Min}^{(2)}}}
g_{1,2}\left(\frac{x^{-1}z_k}{z_j}\right)
\prod_{j \in A_{Min}^{(1)}-\{i,i+1\}
\atop{k \in A_{Min}^{(3)}\cup\{i\}}}
g_{1,3}\left(\frac{z_k}{z_j}\right)
\prod_{j \in A_{Min}^{(2)}
\atop{k \in A_{Min}^{(3)}\cup\{i\}}}
g_{2,3}\left(\frac{xz_k}{z_j}\right)
\prod_{j\in A_{Min}^{(2)}}g_{3,2}\left(\frac{z_j}{z_i}\right)\nonumber\\
&-&(z_i\leftrightarrow z_{i+1}).
\end{eqnarray}
%%%%%%%%%%%%%%%%%%%%%%%%%%%%%%%%%%%%%%%%%%%%%%%%%%%%%%%%%
%%%%%%%%%%%%%%%%%%%%%%%%%%%%%%%%%%%%%%%%%%%%%%%%%%%%%%%%%
By using the weakly sense relations in 
Proposition \ref{prop:dW-Delta2},
we move the operators $T_2(z), T_3(z)$ to the right. 
By changing variables $\{A_j^{(s)}\}$ 
to $\{B_j^{(s)}\}$, we have
\begin{eqnarray}
&-&\sum_{\beta_1,\beta_2,\beta_3 \geqq 0
\atop{\beta_1+2\beta_2+3\beta_3=n}}
(-c)^{\beta_2+2\beta_3}\Delta(x^3)^{\beta_3}
\left(
\sum_{
\left\{B_j^{(s)}\right\}_{s=1,2,3
\atop{j=1,\cdots,\beta_s}}
\atop{B_J^{(2)}}=\{i,i+1\}~{\rm for~some~}J}
+
\sum_{
\left\{B_j^{(s)}\right\}_{s=1,2,3
\atop{j=1,\cdots,\alpha_s}}
\atop{B_K^{(3)}=\{i,i+1,j|i+1<j\}~{\rm for~some~}K}}\right)
\nonumber\\
&\times&
\prod_{\longrightarrow
\atop{j \in B_{Min}^{(1)}}}
T_1(z_j)
\prod_{\longrightarrow
\atop{j \in B_{Min}^{(2)}}}T_2(x^{-1}z_j)
\prod_{\longrightarrow
\atop{j \in B_{Min}^{(3)}}}T_3(z_j)\nonumber\\
&\times&
\prod_{j=1
\atop{j_1=B_{j,1}^{(2)}
\atop{j_2=B_{j,2}^{(2)}}}}^{\beta_2}
\delta\left(\frac{x^2z_{j_2}}{z_{j_1}}\right)
\prod_{j=1
\atop{j_1=B_{j,1}^{(3)}
\atop{j_2=B_{j,2}^{(3)}
\atop{j_3=B_{j,3}^{(3)}
}}}}^{\beta_3}\left(
\delta\left(\frac{x^2z_{j_2}}{z_{j_1}}\right)
\delta\left(\frac{x^2z_{j_1}}{z_{j_3}}\right)+
\delta\left(\frac{x^2z_{j_3}}{z_{j_1}}\right)
\delta\left(\frac{x^2z_{j_1}}{z_{j_2}}\right)\right)\nonumber\\
&\times& 
\prod_{s=1}^3
\prod_{1\leqq j<k \leqq n
\atop{j,k \in B_{Min}^{(s)}}
}g_{s,s}\left(\frac{z_k}{z_j}\right)
\prod_{j \in B_{Min}^{(1)}
\atop{k \in B_{Min}^{(2)}}}
g_{1,2}\left(\frac{x^{-1}z_k}{z_j}\right)
\prod_{j \in B_{Min}^{(1)}
\atop{k \in B_{Min}^{(3)}}}
g_{1,3}\left(\frac{z_k}{z_j}\right)
\prod_{j \in B_{Min}^{(2)}
\atop{k \in B_{Min}^{(3)}}}
g_{2,3}\left(\frac{xz_k}{z_j}\right)\nonumber\\
&-&(z_i\leftrightarrow z_{i+1}).
\end{eqnarray}
This is exactly the same as the second and the third summation
up to signature.
Now we have shown $S_n$-invriance of
${\cal O}_n$ in ``the weak sense''.

For the second we summareize the proof for $\widehat{sl_N}$.
Formulae are more complicated , however the idea of the proof
is the same.
In order to show $S_n$-invariance, it is enough to show
the case of the permutations $\sigma=(i,i+1)$ 
for $1\leqq i \leqq n-1$.
Because of the cancellations,
the differnce ${\cal O}_n(\cdots,z_i,z_{i+1},\cdots)
-{\cal O}_n(\cdots,z_{i+1},z_i,\cdots)$ has simplification.
We don't have to consider every summation
$\sum_{
\left\{A_j^{(s)}\right\}_{s=1,\cdots,N
\atop{j=1,\cdots,\alpha_s}}}$
in the definition of ${\cal O}_n$.
We only have to consider the summation of 
the following $N$-cases for $\sigma=(i,i+1)$ 
\begin{eqnarray}
&&(1)~\{i,i+1\}\subset \cup_{j=1}^{\alpha_1}A_j^{(1)},\nonumber\\
&&(2)~A_J^{(2)}=\{i,i+1\} ~~{\rm for~some}~J,\nonumber\\
&&(3)~A_J^{(3)}=\{i,i+1,j_3|i+1<j_3\} ~~{\rm for~some}~J,\nonumber\\
&& ~~~~~~~~~~~~~~~~~~~~~~~~~~~~~~~~~
\cdots \cdots \cdots\nonumber\\
&&(s)~A_J^{(s)}=\{i,i+1,j_3,\cdots,j_s|i+1<j_3<\cdots
<j_s\}~~{\rm for~some}~J,\nonumber\\
&&~~~~~~~~~~~~~~~~~~~~~~~~~~~~~~~~~~~~~~~~
 \cdots \cdots \cdots\nonumber\\
&&(N)~A_J^{(N)}=\{i,i+1,j_3,j_4,\cdots,j_N|i+1<j_3<j_4<\cdots
<j_N\}~~{\rm for~some}~J.\nonumber
\end{eqnarray}
We have
\begin{eqnarray}
&&{\cal O}_n(z_1,\cdots,z_i,z_{i+1},\cdots,z_n)
-{\cal O}_n(z_1,\cdots,z_{i+1},z_i,\cdots,z_n)\nonumber\\
&=&\widetilde{{\cal O}}_n(z_1,\cdots,z_i,z_{i+1},\cdots,z_n)
-\widetilde{{\cal O}}_n(z_1,\cdots,z_{i+1},z_i,\cdots,z_n).
\end{eqnarray}
Here we have set
\begin{eqnarray}
&&\widetilde{{\cal O}}_n(z_1,\cdots,z_i,z_{i+1},\cdots,z_{n})=
\sum_{\alpha_1,\alpha_2,\cdots,\alpha_N \geqq 0
\atop{\alpha_1+2\alpha_2+\cdots+N\alpha_N=n}}
\prod_{t=1}^N
\left(
(-c)^{t-1}\prod_{u=1}^{t-2}
\Delta(x^{2u+1})^{t-u-1}
\right)^{\alpha_t}
\nonumber\\
&\times&
\left(
\sum_{
\left\{A_j^{(s)}\right\}_{s=1,\cdots,N
\atop{j=1,\cdots,\alpha_s}}
\atop{\{i,i+1\}\subset \cup_{j=1}^{\alpha_s}A_j^{(1)}}}
+
\sum_{t=2}^N
\sum_{
\left\{A_j^{(s)}\right\}_{s=1,\cdots,N
\atop{j=1,\cdots,\alpha_s}}
\atop{A_J^{(t)}=\{i,i+1,j_3,\cdots,j_t|
i+1<j_3<\cdots<j_t\}~{\rm for~some}~J}}
\right)
\prod_{\longrightarrow
\atop{1\leqq s \leqq N}}
\prod_{\longrightarrow
\atop{j \in A_{Min}^{(s)}}}
T_s(x^{-1+s+[\frac{s}{2}]}z_j)\nonumber\\
&\times&
\prod_{t=1}^N
\prod_{j=1
\atop{j_1=A_{j,1}^{(t)}
\atop{\cdots
\atop{j_t=A_{j,t}^{(t)}}}}}^{\alpha_t}
\sum_{\sigma \in S_t
\atop{\sigma(1)=1}}
\prod_{u=1
\atop{u \neq [\frac{t}{2}]+1}}^t
\delta
\left(\frac{x^2z_{j_{\sigma(u+1)}}}{z_{j_{\sigma(u)}}}\right)
\nonumber\\
&\times&
\prod_{t=1}^N 
\prod_{1\leqq j<k \leqq n
\atop{j,k \in A_{Min}^{(t)}}}
g_{t,t}\left(\frac{z_k}{z_j}\right)
\prod_{1\leqq t<u \leqq N}
\prod_{j \in A_{Min}^{(t)}
\atop{k \in A_{Min}^{(u)}}}
g_{t,u}
\left(x^{u-t-2[\frac{u-t}{2}]}\frac{z_k}{z_j}\right).
\end{eqnarray}
Let us consider the formulae relating to the first term
in $\widetilde{{\cal O}}_n(z_1,\cdots,z_i,z_{i+1},\cdots,z_n)-
\widetilde{{\cal O}}_n(z_1,\cdots,z_{i+1},z_{i},\cdots,z_n)$.
Let us start from
%%%%%%%%%%%%%%%%%%%%%%%%%%%%%%%%%%%%%%%%%%%%%
\begin{eqnarray}
&&
\sum_{\alpha_1\geqq 2~{\rm and}~\alpha_2,\cdots,\alpha_N \geqq 0
\atop{\alpha_1+2\alpha_2+\cdots+N\alpha_N=n}}
\prod_{t=1}^N
\left(
(-c)^{t-1}\prod_{u=1}^{t-2}
\Delta(x^{2u+1})^{t-u-1}
\right)^{\alpha_t}
\sum_{
\left\{A_j^{(s)}\right\}_{s=1,\cdots,N
\atop{j=1,\cdots,\alpha_s}}
\atop{\{i,i+1\}\subset \cup_{j=1}^{\alpha_s}A_j^{(1)}}}
\nonumber\\
&\times&
\prod_{\longrightarrow
\atop{j \in A_{Min}^{(1)}
\atop{j<i}}}
T_1(z_j)
\cdot
T_1(z_i)
T_1(z_{i+1})\cdot
\prod_{\longrightarrow
\atop{j \in A_{Min}^{(1)}
\atop{i+1<j}}}
T_1(z_j)
\prod_{\longrightarrow
\atop{2\leqq s \leqq N}}
\prod_{\longrightarrow
\atop{j \in A_{Min}^{(s)}}}
T_s(x^{-1+s+[\frac{s}{2}]}z_j)\nonumber\\
&\times&
\prod_{t=2}^N
\prod_{j=1
\atop{j_1=A_{j,1}^{(t)}
\atop{\cdots
\atop{j_t=A_{j,t}^{(t)}}}}}^{\alpha_t}
\sum_{\sigma \in S_t
\atop{\sigma(1)=1}}
\prod_{u=1
\atop{u \neq [\frac{t}{2}]+1}}^t
\delta
\left(\frac{x^2z_{j_{\sigma(u+1)}}}{z_{j_{\sigma(u)}}}\right)
\nonumber\\
&\times&
\prod_{t=1}^N 
\prod_{1\leqq j<k \leqq n
\atop{j,k \in A_{Min}^{(t)}}}
g_{t,t}\left(\frac{z_k}{z_j}\right)
\prod_{1\leqq t<u \leqq N}
\prod_{j \in A_{Min}^{(t)}
\atop{k \in A_{Min}^{(u)}}}
g_{t,u}
\left(x^{u-t-2[\frac{u-t}{2}]}\frac{z_k}{z_j}\right)
-(z_i \leftrightarrow z_{i+1}).
\end{eqnarray}
%%%%%%%%%%%%%%%%%%%%%%%%%%%%%%%%%%%%%%%%%
By using the weakly sense relations in 
Proposition \ref{prop:dW-Delta11}
we change the ordering of
$T_1(z_i)T_1(z_{i+1})$
and
$\prod_{\longrightarrow
\atop{j \in A_{Min}^{(1)}
\atop{i+1<j}}}T_1(z_j)$.
We have
\begin{eqnarray}
&&-
\sum_{\alpha_1+2\alpha_2+\cdots+N\alpha_N=n
\atop{\alpha_1 \geqq 2~{\rm and}~
\alpha_2,\cdots,\alpha_N \geqq 0}}
\sum_{t=2}^N
\sum_{\{A_j^{(s)}\}
\atop{\{i,i+1\}\subset
A_{Min}^{(1)}}}
\sum_{i+1<j_3<\cdots<j_t
\atop{j_3,\cdots, j_t \in A_{Min}^{(1)}}}
\nonumber\\
&\times&
\prod_{s=1\atop{s\neq t}}^N
\left((-c)^{s-1}
\prod_{u=1}^{s-2}
\Delta(x^{2u+1})^{s-u-1}\right)^{\alpha_s}
\left((-c)^{t-1}
\prod_{u=1}^{t-2}
\Delta(x^{2u+1})^{t-u-1}\right)^{\alpha_t+1}
\nonumber\\
&\times&
\prod_{\longrightarrow
\atop{j \in A_{Min}^{(1)}-\{i,i+1\}}}T_1(z_j)
T_t(x^{-1+t-2[\frac{t}{2}]}z_i)
\prod_{\longrightarrow
\atop{2\leqq s \leqq N}}
\prod_{\longrightarrow
\atop{j \in A_{Min}^{(s)}}}
T_s(x^{-1+s-2[\frac{s}{2}]}z_j)\nonumber
\\
&\times&
\sum_{\sigma \in S_t
\atop{\sigma(1)=1}}
\prod_{u=1
\atop{
u\neq [\frac{t}{2}]+1
\atop{j_1=1,j_2=2}}}^t
\delta\left(\frac{x^2z_{j_{\sigma(u+1)}}}{z_{j_{\sigma(u)}}}
\right)
\prod_{s=2}^N
\prod_{j=1
\atop{j_1=A_{j,1}^{(s)},
\cdots,j_s=A_{j,s}^{(s)}}
}^{\alpha_s}
\sum_{\sigma \in S_s
\atop{\sigma(1)=1}}
\prod_{u=1
\atop{
u\neq [\frac{s}{2}]+1}}^s
\delta\left(\frac{x^2z_{j_{\sigma(u+1)}}}{z_{j_{\sigma(u)}}}
\right)\nonumber
\\
&\times&
\prod_{j<k
\atop{j,k \in A_{Min}^{(1)}-\{i,i+1\}}}
g_{1,1}(z_k/z_j)
\prod_{s=2}^N
\prod_{j<k
\atop{j,k\in A_{Min}^{(s)}}}
g_{s,s}(z_k/z_j)
\prod_{2\leqq t<u \leqq N}
\prod_{j\in A_{Min}^{(t)}
\atop{k \in A_{Min}^{(u)}}}
g_{t,u}\left(x^{u-t-2[\frac{u-t}{2}]}z_k/z_j\right)\nonumber\\
&\times&
\prod_{s=2}^N
\prod_{j\in A_{Min}^{(1)}-\{i,i+1\}
\atop{
k \in A_{Min}^{(s)}}}
g_{1,s}\left(
x^{-1+s-2[\frac{s}{2}]}\frac{z_k}{z_j}
\right)
\prod_{j\in A_{Min}^{(1)}-\{i,i+1\}}
g_{1,t}\left(x^{-1+t-2[\frac{t}{2}]}\frac{z_i}{z_j}\right)
\nonumber\\
&\times&\prod_{s=2}^N
\prod_{j\in A_{Min}^{(s)}}
g_{t,s}\left(
x^{s-t-2[\frac{s-t}{2}]}
\frac{z_j}{z_i}
\right)-(z_i\leftrightarrow z_{i+1}).
\end{eqnarray}
%%%%%%%%%%%%%%%%%%%%%%%%%%%%%%%%%%%%%%%%
We change the summation variables $\{A_j^{(s)}\}$ in
\begin{eqnarray}
\sum_{0\leqq t \leqq N-2}
\sum_{\alpha_1+2\alpha_2+\cdots+N\alpha_N=n
\atop{
\alpha_1\geqq 2~{\rm and}~
\alpha_2,\cdots,\alpha_N \geqq 0}}
\sum_{
\{A_j^{(s)}\}_{s=1,\cdots,N
\atop{j=1,\cdots,\alpha_s}}
\atop{\{i,i+1\}\subset A_{Min}^{(1)}}}
\sum_{i+1<j_3<\cdots<j_{t+2}
\atop{j_3,\cdots,j_{t+2}\in A_{Min}^{(1)}}}
\end{eqnarray}
to the following $\{B_j^{(s)}\}$,
\begin{eqnarray}
\sum_{0\leqq t \leqq N-2}
\sum_{\beta_1+2\beta_2+\cdots+N\beta_N=n
\atop{
\beta_1,
\beta_2,\cdots,\beta_N \geqq 0}}
\sum_{
\{B_j^{(s)}\}_{s=1,\cdots,N
\atop{j=1,\cdots,\beta_s}}
\atop{B_J^{(t)}=\{i,i+1,j_3,\cdots,j_{t+2}|
i+1<j_3<\cdots<j_{t+2}\}
~{\rm for~some~}J
}}.
\end{eqnarray}
Similtaneously let us use the weakly sense relations
in Proposition \ref{prop:dW-Delta2} on
the commutation relations between
$T_i(z)$ and $T_j(w)$ for $i,j \geqq 2$
and make the ordering
\begin{eqnarray}
\prod_{\longrightarrow
\atop{1\leqq s \leqq N}}\prod_{\longrightarrow
\atop{j\in B_{Min}^{(s)}}}T_j(x^{-1+s-2[\frac{s}{2}]}z_j),
\nonumber
\end{eqnarray}
where
$B_{Min}^{(s)}=
\{Min(B_1^{(s)}),\cdots,Min(B_{\alpha_s}^{(s)})\}$.
We have
exactly the same summation of the second to $N$-th terms
of $\widetilde{\cal O}_n(\cdots,z_i,z_{i+1},\cdots)-
\widetilde{\cal O}_n(\cdots,z_{i+1},z_i,\cdots)$ up to signature.
Now we have shown theorem for $\widehat{sl_N}$ case.~~~
Q.E.D.

%%%%%%%%%%%%%%%%%%%%%%%%%%%%%%%%%%%%%%%%%%%%%%
%%%%%%%%%%%%%%%%%%%%%%%%%%%%%%%%%%%%%%%%%%%%%%
%%%%%%%%%%%%%%%%%%%%%%%%%%%%%%%%%%%%%%%%%%%%%%

\subsection{Derivation of Laurent-Series Formulae}

In this section we give proof of Theorem 
\ref{thm:Local-Laurent}.

~
\\
{\it Proof of Theorem \ref{thm:Local-Laurent}}~~~
At first we give proof for $\widehat{sl_3}$ case.
We start from the integral representation
${\cal I}_n$ in (\ref{def:LocalIM}).
Let us pay attention to the poles $z_{J_1}=x^{-2}z_1,
(2\leqq J_1 \leqq n)$.
We have
\begin{eqnarray}
{\cal I}_n&=&\int\cdots \int_{C(1)}
\prod_{j=1}^n \frac{dz_j}{2\pi\sqrt{-1}z_j}
\prod_{1\leqq j<k\leqq n}h(u_k-u_j)\prod_{\longrightarrow
\atop{1\leqq j \leqq n}}T_1(z_j)\\
&-&\sum_{J_1=2}^n
\int \cdots \int_{\widehat{C}(J_1)}
\prod_{j=1
\atop{j\neq J_1}}^n 
\frac{dz_j}{2\pi\sqrt{-1}z_j}
\int_{C_{x^{-2}z_1}}
\frac{dz_{J_1}}{2\pi\sqrt{-1}z_{J_1}}
\prod_{1\leqq j<k\leqq n}h(u_k-u_j)\prod_{\longrightarrow
\atop{1\leqq j \leqq n}}T_1(z_j).\nonumber
\end{eqnarray}
Here we have set
\begin{eqnarray}
C(1)&:&|x^{-2}z_k|<|z_1|<|x^{2-2s}z_k|,~(2\leqq k \leqq n)\nonumber
\\
&&|x^{-2}z_k|<|z_j|<|x^2z_k|,~(2\leqq j <k\leqq n),\\
\widehat{C}(J_1)&:&
|x^{-2}z_k|<|z_1|<|x^{2-2s}z_k|,~(2\leqq k 
\leqq J_1-1)\nonumber
\\
&&|x^{-2}z_k|<|z_1|<|x^2z_k|,~(J_1+1\leqq k\leqq n),\nonumber
\\
&&|x^{-2}z_k|<|z_j|<|x^2z_k|,~(2\leqq j<k\leqq n;j,k\neq J_1).
\end{eqnarray}
Here $C_{x^{-2}z_1}$ is a small circle
which encircle $x^{-2}z_1$ anticlockwise.
The region $\{(z_1,z_k)\in {\mathbb C}^2|
|x^{-2}z_k|<|z_1|<|x^{2-2s}z_k|\}$ for $2\leqq k \leqq J_1$,
are annulus.
Hence the defining relations of the deformed $W$-algebra
can be used. 
Let us change the ordering of $T_1(z_1)$ and $T_1(z_k)$
for $2\leqq k \leqq J_1-1$, and take the residue of
$T_1(z_1)T_1(z_{J_1})$
at $z_{J_1}=x^{-2}z_1$.
We have
\begin{eqnarray}
&&\int \cdots \int_{\widehat{C}(J_1)}
\prod_{j=1
\atop{j\neq J_1}}^n 
\frac{dz_j}{2\pi\sqrt{-1}z_j}
\int_{C_{x^{-2}z_1}}
\frac{dz_{J_1}}{2\pi\sqrt{-1}z_{J_1}}
\prod_{1\leqq j<k\leqq n}h(u_k-u_j)\prod_{\longrightarrow
\atop{1\leqq j \leqq n}}T_1(z_j)\nonumber\\
&=&c
\int \cdots \int_{{C}(J_1)}
\prod_{j=1
\atop{j\neq J_1}}^n 
\frac{dz_j}{2\pi\sqrt{-1}z_j}
\prod_{\longrightarrow
\atop{2\leqq j\leqq J_1-1}}T_1(z_j)
\cdot
T_2(x^{-1}z_1)
\cdot
\prod_{\longrightarrow
\atop{J_1+1\leqq j\leqq n}}T_1(z_j)\nonumber\\
&\times&
\prod_{2\leqq j<k \leqq n\atop{j,k\neq J_1}}
h_{11}(u_k-u_j)
\prod_{j=2}^{J_1-1}h_{12}\left(u_1-u_j-\frac{1}{2}\right)
\prod_{j=J_1+1}^{n}h_{21}\left(u_j-u_1+\frac{1}{2}\right).
\label{eq1}
\end{eqnarray}
Here we have set
\begin{eqnarray}
C(J_1)&:& |x^{-2}z_j|<|z_1|<|x^4z_j|,~
(2\leqq j \leqq n; j\neq J_1),\nonumber\\
&& |x^{-2}z_k|<|z_j|<|x^2z_k|,~
(2\leqq j<k  \leqq n; j,k\neq J_1).
\end{eqnarray}
Let us pay attention to the poles at $z_{J_2}=x^2z_1$,
$(2\leqq J_2 \leqq n; J_2\neq J_1)$.
We deform the RHS of (\ref{eq1}) to the following.
\begin{eqnarray}
&&c\int \cdots \int_{{C}(J_1)(J_1)}
\prod_{j=1
\atop{j\neq J_1}}^n 
\frac{dz_j}{2\pi\sqrt{-1}z_j}
\prod_{\longrightarrow
\atop{2\leqq j\leqq J_1-1}}T_1(z_j)
\cdot
T_2(x^{-1}z_1)
\cdot
\prod_{\longrightarrow
\atop{J_1+1\leqq j\leqq n}}T_1(z_j)\nonumber\\
&\times&
\prod_{2\leqq j<k \leqq n\atop{j,k\neq J_1}}
h_{11}(u_k-u_j)
\prod_{j=2}^{J_1-1}h_{12}\left(u_1-u_j-\frac{1}{2}\right)
\prod_{j=J_1+1}^{n}h_{21}\left(u_j-u_1+\frac{1}{2}\right)
\nonumber\\
&-&c\sum_{J_2=2
\atop{J_2 \neq J_1}}^n
\int \cdots \int_{\widehat{C}(J_1)(J_2)}
\prod_{j=1
\atop{j\neq J_1,J_2}}^n 
\frac{dz_j}{2\pi\sqrt{-1}z_j}
\int_{C_{x^{2}z_1}}\frac{dz_{J_2}}{2\pi\sqrt{-1}z_{J_2}}
\nonumber\\
&\times&\prod_{\longrightarrow
\atop{2\leqq j\leqq J_1-1}}T_1(z_j)
\cdot
T_2(x^{-1}z_1)
\cdot
\prod_{\longrightarrow
\atop{J_1+1\leqq j\leqq n}}T_1(z_j)\nonumber\\
&\times&
\prod_{2\leqq j<k \leqq n\atop{j,k\neq J_1}}
h_{11}(u_k-u_j)
\prod_{j=2}^{J_1-1}h_{12}\left(u_1-u_j-\frac{1}{2}\right)
\prod_{j=J_1+1}^{n}h_{21}\left(u_j-u_1+\frac{1}{2}\right).
\end{eqnarray}
Here we have set
\begin{eqnarray}
C(J_1)(J_1)&:&|x^{-2+2s}z_j|<|z_1|<|x^4z_j|,
~(2\leqq j \leqq n; j \neq J_1),
\nonumber\\
&&|x^{-2}z_k|<|z_j|<|x^2z_k|,~
(2\leqq j < k \leqq n; j,k \neq J_1).
\end{eqnarray}
For $2\leqq J_2<J_1 \leqq n$ we set
\begin{eqnarray}
\widehat{C}(J_1)(J_2)&:&|x^{-2+2s}z_j|<|z_1|<|x^4z_j|,~
(J_2 \leqq j \leqq J_1-1),\nonumber\\
&&|x^{-2}z_j|<|z_1|<|x^4z_j|,~
(2 \leqq j \leqq J_2-1~{\rm or}~
J_1+1 \leqq j \leqq n),\nonumber\\
&&|x^{-2}z_k|<|z_j|<|x^2z_k|,~
(2\leqq j < k \leqq n; j,k \neq J_1).
\end{eqnarray}
For $2\leqq J_1<J_2\leqq n$ we set
\begin{eqnarray}
\widehat{C}(J_1)(J_2)&:&|x^{-2+2s}z_j|<|z_1|<|x^4z_j|,~
(2 \leqq j \leqq J_2; j \neq J_1),\nonumber\\
&&|x^{-2}z_j|<|z_1|<|x^4z_j|,~
(J_2+1 \leqq j \leqq n),\nonumber\\
&&|x^{-2}z_k|<|z_j|<|x^2z_k|,~
(2\leqq j < k \leqq n; j,k \neq J_1).
\end{eqnarray}
The above formulae 
for this 
integrand $C(J_1)(J_2)$ holds for ${\rm Re}(s)>N \geqq 3$.
For $N=2$ another treatment should be done.
Let us study the first term $c\int \cdots \int_{C(J_1)(J_1)}
\prod_{j\neq J_1}\frac{dz_j}{2\pi \sqrt{-1}}$
See the integral contour $C(J_1)(J_1)$.
The region $\{(z_1,z_j)\in {\mathbb C}^2|
|x^{-2+2s}z_j|<|z_1|<|x^{4}z_j|\}$ for $j \neq J_1$
are annulus.
Hence the defining relations of the deformed $W$-algebra
can be used. 
By using the weakly sense relation in
Proposition \ref{prop:dW-Delta1},
we deform the first term to the following.
\begin{eqnarray}
&&c\int \cdots \int_{{C}(J_1)(J_1)}
\prod_{j=1
\atop{j\neq J_1}}^n 
\frac{dz_j}{2\pi\sqrt{-1}z_j}
\prod_{\longrightarrow
\atop{2\leqq j\leqq n
\atop{j\neq J_1}
}}T_1(z_j)\cdot
T_2(x^{-1}z_1)\nonumber\\
&&\times
\prod_{2\leqq j<k \leqq n\atop{j,k\neq J_1}}
h_{11}(u_k-u_j)
\prod_{j=2\atop{j \neq J_1}}^{n}h_{12}\left(u_1-u_j-\frac{1}{2}\right).
\end{eqnarray}
Let us study the second term
$-c\sum_{J_2 \neq 2,J_1}\int \cdots \int_{\widehat{C}(J_1)(J_2)}
\prod_{
j\neq J_1,J_2}
\frac{dz_j}{2\pi\sqrt{-1}z_j}
\int_{C_{x^{2}z_1}}\frac{dz_{J_2}}{2\pi\sqrt{-1}z_{J_2}}$
See the integral contour $\widehat{C}(J_1)(J_2)$.
The region $\{(z_1,z_j)\in {\mathbb C}^2|
|x^{-2+2s}z_j|<|z_1|<|x^{4}z_j|\}$ for $2\leqq j \leqq J_1$,
are annulus for ${\rm Re}(s)>N=3$.
Hence the defining relations of the deformed $W$-algebra
can be used. 
Let us change the ordering of $T_1(z_{J_2})$ and $T_1(z_k)$
and make the product of the operators $T_1(z_{J_2})T_2(x^{-1}z_1)$
or $T_2(x^{-1}z_1)T_1(z_{J_2})$.
Let us
take the residue of
$T_1(z_{J_2})T_2(x^{-1}z_{1})$ and
$T_2(x^{-1}z_1)T_1(z_{J_2})$
at $z_{J_2}=x^{2}z_1$ by regarding the weakly sense
equation in Proposition \ref{prop:dW-Delta1}.
We have
\begin{eqnarray}
&&c^2 \Delta(x^3)
\sum_{J_2=2\atop{J_2 \neq J_1}}^n
\int \cdots \int_{C(J_1)(J_2)}\prod_{j=1 \atop{j \neq J_1,J_2}}^n
\frac{dz_j}{2 \pi \sqrt{-1} z_j}
\prod_{
\longrightarrow
\atop{2 \leqq j \leqq J_1-1
\atop{j \neq J_2}}}
T_1(z_j)
\cdot
T_3(z_1)
\cdot
\prod_{\longrightarrow
\atop{J_1+1 \leqq j \leqq n
\atop{j \neq J_2}}}
T_1(z_j)\nonumber\\
&\times&
\prod_{
2\leqq j<k \leqq n
\atop{j,k
\neq J_1,J_2}}
h_{11}(u_k-u_j)
\prod_{j=2
\atop{j \neq J_2}}^{J_1-1}h_{13}\left(u_1-u_j\right)
\prod_{j=J_1+1
\atop{j \neq J_2}}^{n}h_{31}(u_j-u_1).
\end{eqnarray}
Here we have set 
\begin{eqnarray}
C(J_1)(J_2)&:&|x^{-4+2s}z_j|<|z_1|<|x^{4-2s}z_j|,~(2\leqq j \leqq n;
j\neq J_1,J_2),\nonumber
\\
&&|x^{-2}z_k|<|z_j|<|x^2z_k|,~(2\leqq j<k \leqq n; j\neq J_1,J_2).
\label{eq2}
\end{eqnarray}
This integral contur $C(J_1)(J_2)$ holds only for $N=3$ case.
For $N \geq 4$ case another treatment should be done.
The region $\{(z_1,z_j)\in {\mathbb C}^2|
|x^{-4+2s}z_j|<|z_1|<|x^4z_j|\}$
are annulus. We move
$T_3(z_j)$ to the right, and get 
\begin{eqnarray}
&&c^2 \Delta(x^3)
\sum_{J_2=2\atop{J_2 \neq J_1}}^n
\int \cdots \int_{C(J_1)(J_2)}\prod_{j=1 \atop{j \neq J_1,J_2}}^n
\frac{dz_j}{2 \pi \sqrt{-1} z_j}
\prod_{
\longrightarrow
\atop{2 \leqq j \leqq n
\atop{j \neq J_1,J_2}}}
T_1(z_j)
\cdot
T_3(z_1)
\nonumber\\
&\times&
\prod_{
2\leqq j<k \leqq n
\atop{j,k
\neq J_1,J_2}}
h_{11}(u_k-u_j)
\prod_{j=2
\atop{j \neq J_1,J_2}}^{n}
h_{13}\left(u_1-u_j\right).
\end{eqnarray}
Summing up every terms, we have
\begin{eqnarray}
{\cal I}_n&=&\left(
\sum_{A^{(1)}=\{1\}
\atop{A^{(2)}=A^{(3)}=\phi
}}
-c
\sum_{A^{(2)}=\{1,j\}\atop
{A^{(1)}=A^{(3)}=\phi}}
+2! c^2\Delta(x^3)
\sum_{A^{(3)}=\{1,j,k\}\atop
{A^{(1)}=A^{(2)}=\phi}}
\right)\nonumber\\
&\times&
\int \cdots \int_{C\{A^{(1)},A^{(2)},A^{(3)},A_c\}}
\prod_{j\in A_c\cup A_{Min}^{(1)}
\cup A_{Min}^{(2)} \cup A_{Min}^{(3)}}
\frac{dz_j}{2\pi \sqrt{-1}z_j}
\nonumber\\
&\times&
\prod_{\longrightarrow
\atop{j\in A_c \cup A_{Min}^{(1)}}}T_1(z_j)
\prod_{\longrightarrow
\atop{j\in A_{Min}^{(2)}}}T_2(x^{-1}z_j)
\prod_{\longrightarrow
\atop{j\in A_{MIn}^{(3)}}}T_3(z_j)
\prod_{j<k
\atop{j,k \in A_c \cup A_{Min}^{(1)}}}
h_{11}(u_k-u_j)\nonumber\\
&\times&\prod_{k \in A_{Min}^{(2)}}
\prod_{j \in A_{Min}^{(1)}\cup A_c}
h_{12}\left(u_k-u_j-\frac{1}{2}\right)
\prod_{k \in A_{Min}^{(3)}}
\prod_{j \in A_{Min}^{(1)}\cup A_c}
h_{13}\left(u_k-u_j\right).
\end{eqnarray}
Here we set $A_c=\{1,2,\cdots,n\}-A^{(1)}\cup
A^{(2)}\cup A^{(3)}$.
We have set $A_{Min}^{(t)}=\{j_1\}
$ for $A^{(t)}=\{j_1<j_2<\cdots<j_t\}$.
Here we have set
$C\{A^{(1)},A^{(2)},A^{(3)},A_c\}$ by
\begin{eqnarray}
&&|x^{-2}z_k|<|z_1|<|x^{2-2s}z_k|,~(k\in A_c
~{\rm for}~
A^{(1)}\neq \phi),
\nonumber\\
&&
|x^{-2+2s}z_k|<|z_1|<|x^{4}z_k|,~(k\in A_c
~{\rm for}~
A^{(2)}\neq \phi),
\nonumber
\\
&&
|x^{-4+2s}z_k|<|z_1|<|x^{4-2s}z_k|,~(k\in A_c
~{\rm for}~ 
A^{(3)}\neq \phi),
\nonumber\\
&&|x^{-2}z_k|<|z_j|<|x^2z_k|,~(j<k; j,k \in A_c).
\end{eqnarray}
Next we deform the part $
\prod_{\longrightarrow
\atop{j \in A_c}}T_1(z_j)$.
Let us take the residue at $z_J=x^{-2}z_2$,
and continue similar calculations as above.
%We have
%\begin{eqnarray}
%{\cal I}_n&=&
%\sum_{\alpha_1,\alpha_2,\alpha_3 \geqq 0
%\atop{\alpha_1+2\alpha_2+3\alpha_3=n}}
%(-c)^{\alpha_2+2 \alpha_3}\Delta(x^3)^{\alpha_3}
%(2!)^{\alpha_3}
%\sum_{
%\left\{A_j^{(s)}\right\}_{s=1,\cdots,3
%\atop{j=1,\cdots,\alpha_s}}
%\atop{
%A_j^{(s)} \subset \{1,2,\cdots,n\},~
%|A_j^{(s)}|=s,~
%\cup_{s=1}^3 \cup_{j=1}^{\alpha_s}
%A_j^{(s)}=\{1,2,\cdots,n\}
%\atop{Min(A_1^{(s)})<Min(A_2^{(s)})<
%\cdots<Min(A_{\alpha_s}^{(s)})}}}\nonumber\\
%&\times&
%\int \cdots \int_{C\{A_{Min}^{(1)},A_{Min}^{(2)},
%A_{Min}^{(3)}\}}
%\prod_{j\in A_{Min}^{(1)}\cup A_{Min}^{(2)}
%\cup A_{Min}^{(3)}}
%\frac{dz_j}{2\pi\sqrt{-1}z_j}
%\prod_{t=1}^3 
%\prod_{1\leqq j<k \leqq n
%\atop{j,k \in A_{Min}^{(t)}}}
%h_{t,t}\left(u_k-u_j\right)
%\nonumber\\
%&\times&
%\prod_{j \in A_{Min}^{(1)}
%\atop{k \in A_{Min}^{(2)}}}
%h_{1,2}\left(u_k-u_j-\frac{1}{2}\right)
%\prod_{j \in A_{Min}^{(1)}
%\atop{k \in A_{Min}^{(3)}}}
%h_{1,3}\left(u_k-u_j\right)
%\prod_{j \in A_{Min}^{(2)}
%\atop{k \in A_{Min}^{(3)}}}
%h_{2,3}\left(u_k-u_j+\frac{1}{2}\right)
%\nonumber\\
%&\times&
%\prod_{\longrightarrow
%\atop{j \in A_{Min}^{(1)}}}T_1(z_j)
%\prod_{\longrightarrow
%\atop{j \in A_{Min}^{(2)}}}T_2(x^{-1}z_j)
%\prod_{\longrightarrow
%\atop{j \in A_{Min}^{(3)}}}T_3(z_j).
%\end{eqnarray}
%Here we have set
%$A_{Min}^{(t)}=\{Min(A_1^{(t)}),
%Min(A_2^{(t)}),\cdots,Min(A_{\alpha_t}^{(t)})\}$.
%The integral contour
%$C\{A_{Min}^{(1)},A_{Min}^{(2)},A_{Min}^{(3)}\}$ is summarized in
%(\ref{}) in proof for $\widehat{sl_N}$ case.
We use
the weakly sense equations in Proposition \ref{prop:dW-Delta2}
and change the ordering of $T_2(x^{-1}w)$ and $T_3(w)$,
without taking residues.
Now we have shown 
theorem for $\widehat{sl_3}$.

Now we begin Proof for $\widehat{sl_N}$ case.
Let us start from
the integral representation ${\cal I}_n$.
Proof for generl $\widehat{sl_N}$ case is similar as
those for $\widehat{sl_3}$ case.
However it is not exactlly the same.
For example the integral contour 
$C(J_1)(J_2)$ of the equation (\ref{eq2})
should be changed for $N \geq 4$ to the following.
\begin{eqnarray}
&&|x^{-2}z_k|<|z_1|<|x^{2-2s}z_k|,~(k\in A_c
~{\rm for}~
A^{(1)}\neq \phi),
\nonumber\\
&&
|x^{-2+2s}z_k|<|z_1|<|x^{4}z_k|,~(k\in A_c
~{\rm for}~
A^{(2)}\neq \phi),
\nonumber
\\
&&
|x^{-4}z_k|<|z_1|<|x^{4-2s}z_k|,~(k\in A_c
~{\rm for}~ 
A^{(3)}\neq \phi),
\nonumber\\
&&|x^{-2}z_k|<|z_j|<|x^2z_k|,~(j<k; j,k \in A_c).
\end{eqnarray}
Here we have to take the residue at $z_k=x^{-4}z_1$
for $k \in A_c$.
For proof for $\widehat{sl_N}$ case,
we have to take the residue deeper.
Taking the residue relating to variable $z_1$ deeper,
we have
\begin{eqnarray}
{\cal I}_n&=&\int\cdots \int_{
C\{A^{(1)}_{Min}=\{1\},
A_{Min}^{(2)}=\phi,\cdots,A_{Min}^{(N)}=\phi,A_c=\{2,\cdots,n\}\}}
\prod_{1\leqq j<k\leqq n}h(u_k-u_j) 
\prod_{\longrightarrow
\atop{1\leqq j\leqq n}}T_1(z_j)\nonumber\\
&+&\sum_{k=1}^{Min(N,n)}(-c)^{k-1}
(k-1)!
\prod_{u=1}^{k-1}
\Delta(x^{2u+1})^{k-u-1}
\sum_{A^{(k)}=\{j_1=1,j_2,\cdots,j_k\}
\atop{A^{(s)}=\phi,~(s\neq k)}}\nonumber\\
&\times&
\int \cdots \int_{C\{A_{Min}^{(1)},\cdots,A_{Min}^{(N)},A_c\}}
\prod_{j \in A_{Min}^{(1)} \cup \cdots A_{Min}^{(N)}\cup A_c}
\frac{dz_j}{2\pi \sqrt{-1}z_j}
\nonumber\\
&\times&
\prod_{\longrightarrow
\atop{j \in A_{Min}^{(1)}\cup A_c}}T_1(z_j)
\prod_{\longrightarrow
\atop{j \in A_{Min}^{(2)}}}T_2(x^{-1}z_j)
\cdots
\prod_{\longrightarrow
\atop{j \in A_{Min}^{(N)}}}T_N\left(x^{-1+N-2[\frac{N}{2}]}z_j\right)
\nonumber\\
&\times&
\prod_{j<k\atop{j,k \in A_{Min}^{(1)}}}h_{11}(u_k-u_j)
\prod_{t=2}^N
\prod_{j \in A_{Min}^{(t)}}
\prod_{k \in A_{Min}^{(1)}\cup A_c}h_{1,t}
\left(u_j-u_k+\frac{t-1}{2}-[\frac{t}{2}]\right).
\end{eqnarray}
Here we have set $C\{A_{Min}^{(1)},A_{Min}^{(2)},\cdots,A_{Min}^{(N)},A_c\}$
by
\begin{eqnarray}
&&|x^{-2J-2}z_k|<|z_1|<|x^{2J+2-2s}z_k|,~(k\in A_c; A^{(2J+1)}\neq \phi; 
J<\frac{N}{2}-1),\nonumber\\
&&|x^{-2J-2+2s}z_k|<|z_1|<|x^{2J+2}z_k|,~(k\in A_c; A^{(2J)}\neq \phi; 
J<\frac{N}{2}-3),\nonumber\\
&&|x^{-N+2s}z_k|<|z_1|<|x^{N-2s}z_k|,~(k\in A_c; A^{(N-1)}\neq \phi; 
N~{\rm even}),\nonumber\\
&&|x^{-N-1+2s}z_k|<|z_1|<|x^{N+1-2s}z_k|,~(k\in A_c; A^{(N-1)}\neq \phi; 
N~{\rm odd}),\nonumber\\
&&|x^{-2}z_k|<|z_j|<|x^2z_k|,~(j<k; j,k \in A_c).
\end{eqnarray}
Next we deform the part 
$\prod_{\longrightarrow
\atop{j \in A_c}}T_1(z_j)$.
Let us take the residue at $z_J=x^{-2}z_2$,
and continue similar calculations as above.
We use
the weakly sense equations in
Propositions \ref{prop:dW-Delta1} and 
\ref{prop:dW-Delta2}
and change the ordering of $T_i(z)$ and $T_j(w)$
for $i,j\geqq 2$,
without taking residues.
We get Theorem for $\widehat{sl_N}$ case.
Calculations for ${\cal I}_n^*$ are given by similar way.
~~~
Q.E.D.

\subsection{Proof of $[{\cal I}_m,{\cal I}_n]=0$}

In this section we show the commutation relation
$[{\cal I}_m,{\cal I}_n]=0$.

\begin{prop}~~
The folloing theta identity holds.
\begin{eqnarray}
\sum_{\sigma \in S_{m+n}}
\prod_{j=1}^n \prod_{k=n+1}^{n+m}
\frac{1}{h(u_{\sigma(k)}-u_{\sigma(j)})}
=
\sum_{\sigma \in S_{m+n}}
\prod_{j=1}^m \prod_{k=m+1}^{n+m}
\frac{1}{h(u_{\sigma(k)}-u_{\sigma(j)})}.
\end{eqnarray}
\begin{eqnarray}
\sum_{\sigma \in S_{m+n}}
\prod_{j=1}^n \prod_{k=n+1}^{n+m}
\frac{1}{h^*(u_{\sigma(k)}-u_{\sigma(j)})}
=
\sum_{\sigma \in S_{m+n}}
\prod_{j=1}^m \prod_{k=m+1}^{n+m}
\frac{1}{h^*(u_{\sigma(k)}-u_{\sigma(j)})}.
\end{eqnarray}
Here $h(u)$ and $h^*(u)$ are given in (\ref{def:h}).
\label{prop:theta-Local}
\end{prop}
This theta identity was written in \cite{FO}
without proof.
We have already summarized a proof of the theta identity
in \cite{FKSW1}.
In order to make this paper self-contained,
we re-summarize the proof, here.

~\\
{\it Proof}~~~Let us set
\begin{eqnarray}
{\rm LHS}(n,m)&=&\sum_{J \subset \{1,2,\cdots,n+m\}
\atop{|J|=n}}
\prod_{j \in J}
\prod_{k \notin J}
\frac{[u_k-u_j+1]_s[u_k-u_j+r^*]_s}{[u_{k}-u_{j}]_s[u_k-u_j+r]_s},
\\
{\rm RHS}(n,m)&=&\sum_{J^c \subset \{1,2,\cdots,n+m\}
\atop{|J^c|=m}}
\prod_{j \in J^c}
\prod_{k \notin J^c}
\frac{[u_k-u_j+1]_s[u_k-u_j+r^*]_s}{[u_{k}-u_{j}]_s[u_k-u_j+r]_s}.
\end{eqnarray}
We will show ${\rm LHS}(n,m)={\rm RHS}(n,m)$.
LHS$(n,m)$ and RHS$(n,m)$ are an elliptic functions.
Therefore, from Liouville thorem, it is enough 
to check whether all the residues of LHS$(n,m)$ and RHS$(n,m)$ coincide or not.
Candidates of poles are
$u_\alpha=u_\beta~(\alpha \neq \beta)$ 
and $u_\alpha=u_\beta-r~(\alpha \neq \beta)$.
Let us consider $u_\alpha=u_\beta~(\alpha \neq \beta)$ 
\begin{eqnarray}
{\rm LHS}(n,m)&=&\left(
\frac{[u_\alpha-u_\beta+1]_s
[u_\alpha-u_\beta+r^*]_s}{[u_\alpha-u_\beta]_s
[u_\alpha-u_\beta-r]_s}+
\frac{[u_\beta-u_\alpha+1]_s[u_\beta-u_\alpha+r^*]_s}{
[u_\beta-u_\alpha]_s[u_\beta-u_\alpha-r]_s}\right)
\nonumber\\
&\times&\sum_{J\cup J^c\{1,\cdots,n+m\}-\{\alpha,\beta\}
\atop{|J|=n-1}}
\prod_{j \in J}
\prod_{k \notin J}
\frac{[u_k-u_j+1]_s[u_k-u_j+r^*]_s}{[u_{k}-u_{j}]_s[u_k-u_j+r]_s}
\end{eqnarray}
Hence we have ${\rm Res}_{u_\alpha=u_\beta}{\rm LHS}(n,m)=0$.
As the same manner
we have ${\rm Res}_{u_\alpha=u_\beta}{\rm RHS}(n,m)=0$.
Therefore $u_\alpha=u_{\beta}$ is not pole.
We only have to consider poles 
$u_\alpha=u_\beta~(\alpha \neq \beta)$.
We show the ${\rm LHS}(n,m)={\rm RHS}(n,m)$ by the induction
of the number $n+m$.
We assume $n>m\geqq 1$ without loosing generality.
At first we show the starting point $n>m=1$.
\begin{eqnarray}
\sum_{k=1}^{n+1}\prod_{j=1
\atop{j\neq k}}^{n+1}\frac{
[u_j-u_k+1]_s[u_j-u_k+r^*]_s}{[u_j-u_k]_s[u_j-u_k+r]_s}
=
\sum_{k=1}^{n+1}\prod_{j=1
\atop{j\neq k}}^{n+1}\frac{
[u_k-u_j+1]_s[u_k-u_j+r^*]_s}{[u_k-u_j]_s[u_k-u_j+r]_s}
\end{eqnarray}
Both LHS$(n,1)$ and RHS$(n,1)$ have simple poles
at $u_\alpha=u_\beta-r~(\alpha \neq \beta)$
modulo ${\mathbb Z}+{\mathbb Z}\tau$.
Beacuse both LHS$(n,1)$ and RHS$(n,1)$ are symmetric with respect with
$u_1,u_2,\cdots,u_{n+1}$,
it is enough to check the pole at $u_2=u_1-r$. We have
\begin{eqnarray}
&&{\rm Res}_{~u_2=u_1-r}
{\rm LHS}(n,1)={\rm Res}_{~u_2=u_1-r}
{\rm RHS}(n,1)\nonumber\\
&&=
{\rm Res}_{u=0}\frac{[-r^*]_s[-1]_s}{[-r]_s[u]_s}
\prod_{j=3}^{n+1}\frac{[u_j-u_1+1]_s[u_j-u_1+r^*]_s}
{[u_j-u_1]_s[u_j-u_1+r]_s}.
\end{eqnarray}
We have shown $n>m=1$ case.
We show general $n>m\geqq 1$ case.
We assume the equation ${\rm LHS}(n,1)={\rm RHS}(n,1)$
for some $(m,n)$.
Beacuse both LHS$(n+1,m+1)$ and RHS$(n+1,m+1)$ are symmetric with respect with
$u_1,u_2,\cdots,u_{n+m+2}$,
it is enough to check the pole at $u_2=u_1-r$.
Let us take the residue at $u_2=u_1-r$ for $(m+1,n+1)$.
\begin{eqnarray}
{\rm Res}_{~u_2=u_1-r}
&&\left(
\sum_{J \subset \{1,2,\cdots,n+m+2\}
\atop{|J|=n+1}}
\prod_{j \in J}
\prod_{k \notin J}
\frac{[u_k-u_j+1]_s[u_k-u_j+r^*]_s}{[u_{k}-u_{j}]_s[u_k-u_j+r]_s}
\right.
\nonumber\\
&&\left.
-\sum_{J^c \subset \{1,2,\cdots,n+m+2\}
\atop{|J^c|=m+1}}
\prod_{j \in J^c}
\prod_{k \notin J^c}
\frac{[u_k-u_j+1]_s[u_k-u_j+r^*]_s}{[u_{k}-u_{j}]_s[u_k-u_j+r]_s}.
\right)\nonumber\\
&=&
\prod_{j=3}^{n+m+2}
\frac{[u_j-u_1+1]_s[u_j-u_1+r^*]_s}{
[u_j-u_1]_s[u_j-u_1+r]_s}\\
&\times&\left(
\sum_{J \subset \{3,4,\cdots,n+m+2\}
\atop{|J|=n}}
\prod_{j \in J}
\prod_{k \notin J}
\frac{[u_k-u_j+1]_s[u_k-u_j+r^*]_s}{[u_{k}-u_{j}]_s[u_k-u_j+r]_s}
\right.
\nonumber\\
&&\left.
-\sum_{J^c \subset \{3,4,\cdots,n+m+2\}
\atop{|J^c|=m}}
\prod_{j \in J^c}
\prod_{k \notin J^c}
\frac{[u_k-u_j+1]_s[u_k-u_j+r^*]_s}{[u_{k}-u_{j}]_s[u_k-u_j+r]_s}.
\right)=0.\nonumber
\end{eqnarray}
We have used the assumption of induction 
${\rm LHS}(n,m)={\rm RHS}(n,m)$.
~~~Q.E.D.

\begin{prop}~~The following weakly sense equation holds.
\begin{eqnarray}
{\cal O}_n(z_1,\cdots,z_n)
{\cal O}_m(z_{n+1},\cdots,z_{n+m})
\sim
\prod_{1\leqq j \leqq n
\atop{n+1\leqq k \leqq n+m}}
\frac{1}{g_{11}(z_k/z_j)}
{\cal O}_{m+n}(z_1,\cdots,z_{n+m}).
\end{eqnarray}
\end{prop}
{\it Proof}~~~This is direct consequence of
the following explicit folrmulae
\begin{eqnarray}
&&{\cal O}_{n+m}(z_1,\cdots,z_{n+m})\sim
\prod_{1\leqq j \leqq n
\atop{n+1\leqq k \leqq n+m}}g_{11}(z_k/z_j)
{\cal O}_n(z_1,\cdots,z_n)
{\cal O}_m(z_{n+1},\cdots,z_{n+m})\nonumber\\
&+&
\sum_{
\alpha,\alpha_2,\cdots,\alpha_N\geqq 0
\atop{
\alpha_1+2\alpha_2+\cdots+N\alpha_N=n
\atop{
2\leqq 2\alpha_2+\cdots+N\alpha_N \leqq n}
}}\prod_{t=1}^N
\left((-c)^{t-1}
\prod_{u=1}^{t-1}
\Delta(x^{2u+1})^{t-u-1}\right)^{\alpha_t}
\nonumber\\
&\times&\sum_{
\{L_j^{(s)}\}_{s=1,\cdots,N\atop{j=1\cdots,\alpha_s}},
\{R_j^{(s)}\}_{s=1,\cdots,N\atop{j=1\cdots,\alpha_s}}
\atop{\sum_{s=2}^N |L_1^{(s)}||R_1^{(s)}|\geqq 1}}
\prod_{t=1}^N
\prod_{j=1
\atop{j_1=A_{j,1}^{(t)}
\atop{\cdots
\atop{j_t=A_{j,t}^{(t)}}
}}}^{\alpha_t}
\sum_{\sigma \in S_t
\atop{\sigma(1)=1}}
\prod_{u=1
\atop{u \neq [\frac{t}{2}]+1}}^t
\delta\left(\frac{x^2z_{j_{\sigma(u+1)}}}{z_{j_{\sigma(u)}}}\right)
\nonumber\\
&\times&
\prod_{t=1}^N \prod_{j<k
\atop{j,k \in A_{Min}^{(t)}}}g_{t,t}\left(\frac{z_k}{z_j}\right)
\prod_{1\leqq t<u \leqq N}\prod_{j\in A_{Min}^{(t)}
\atop{k\in A_{Min}^{(u)}}}g_{t,u}\left(
x^{u-t-2[\frac{u}{2}]+2[\frac{t}{2}]}\frac{z_k}{z_j}
\right).
\end{eqnarray}
Here the summation
$\sum_{
\{L_j^{(s)}\}_{s=1,\cdots,N\atop{j=1\cdots,\alpha_s}},
\{R_j^{(s)}\}_{s=1,\cdots,N\atop{j=1\cdots,\alpha_s}}
\atop{\sum_{s=2}^N |L_1^{(s)}||R_1^{(s)}|\geqq 1}}$
is taken over the conditions that
\begin{eqnarray}
&&\cup_{s=1}^N \cup_{j=1}^{\alpha_s}L_j^{(s)}=\{1,2,\cdots,m\},
L_i^{(s)}\cap L_j^{(s)}=\phi,(i\neq j),\nonumber\\
&&Min(L_1^{(s)})<Min(L_2^{(s)})<\cdots<Min (L_{\alpha_s}^{(s)}),
\nonumber\\
&&\cup_{s=1}^N \cup_{j=1}^{\alpha_s}R_j^{(s)}=\{m+1,m+2,\cdots,m+n\},
R_i^{(s)}\cap R_j^{(s)}=\phi,(i\neq j),\nonumber\\
&&Min(R_1^{(s)})<Min(R_2^{(s)})<\cdots<Min (R_{\alpha_s}^{(s)}).
\end{eqnarray}
Here we have set
$A_j^{(s)}=L_j^{(s)}\cup R_j^{(s)}$.
We have set $A_{j,k}^{(s)}=j_k$ for $A_j^{(s)}=\{j_1<j_2<\cdots<j_s\}$,
and $A_{Min}^{(s)}=\{A_{1,1}^{(s)},A_{2,1}^{(s)},\cdot, 
A_{\alpha_s,1}^{(s)}\}$.
We want to point out that
every term of the summation $
\sum_{
\{L_j^{(s)}\}_{s=1,\cdots,N\atop{j=1\cdots,\alpha_s}},
\{R_j^{(s)}\}_{s=1,\cdots,N\atop{j=1\cdots,\alpha_s}}}$
has the delta-function 
$\delta(x^2z_k/z_j),~(1\leqq j \leqq m, m+1\leqq k \leqq m+n)$.
Dividing $\prod_{1\leqq j \leqq n
\atop{n+1\leqq k \leqq n+m}}g_{11}(z_k/z_j)$ to both sides
and using $1/g_{11}(x^{-2})=0$, we have this proposition .
~~~
Q.E.D.

~\\
{\it Proof of Theorem \ref{thm:Local-Com}}~~
At first we restrict ourself to the regime,
${\rm Re}(s)>N$ and ${\rm Re }(r)<0$, in order to use
the power series formulae of the local integrals of motion,
${\cal I}_n$.
In Proposition \ref{thm:Sn-InvLocal}
we have shown for $\sigma \in S_n$
\begin{eqnarray}
\prod_{1\leqq j<k \leqq n}s(z_k/z_j)
{\cal O}_{n}(z_1,\cdots, z_{n})
=
\prod_{1\leqq j<k \leqq n}
s(z_{\sigma(k)}/z_{\sigma(j)})
{\cal O}_{n}
(z_{\sigma(1)},\cdots, z_{\sigma(n)}).
\end{eqnarray}
Hence we have
\begin{eqnarray}
&&{\cal I}_n \cdot {\cal I}_m\nonumber\\
&=&
\left[\prod_{1\leqq j<k \leqq n}s(z_k/z_j){\cal O}_n(z_1,\cdots,z_n)
\prod_{n+1\leqq j<k \leqq n+m}s(z_k/z_j)
{\cal O}_m(z_{n+1},\cdots,z_{n+m})\right]_{1,z_1\cdots z_{n+m}}\nonumber
\\&=&
\left[\frac{1}{(n+m)!}
\sum_{\sigma \in S_{n+m}}
\prod_{j=1}^n
\prod_{k=n+1}^{n+m}
\frac{1}{h(u_{\sigma(k)}-u_{\sigma(j)})}
\prod_{1\leqq j<k \leqq n+m}s(z_k/z_j)
{\cal O}_{n+m}(z_1,\cdots, z_{n+m})
\right]_{1,z_1 \cdots z_{n+m}}.\nonumber\\
\end{eqnarray}
Hence the commutation relation
${\cal I}_n \cdot {\cal I}_m={\cal I}_m\cdot
{\cal I}_n$ is reduced to the theta identity in Proposition
\ref{prop:theta-Local}.
\begin{eqnarray}
\sum_{\sigma \in S_{m+n}}
\prod_{j=1}^n \prod_{k=n+1}^{n+m}
\frac{1}{h(u_{\sigma(k)}-u_{\sigma(j)})}
=
\sum_{\sigma \in S_{m+n}}
\prod_{j=1}^m \prod_{k=m+1}^{n+m}
\frac{1}{h(u_{\sigma(k)}-u_{\sigma(j)})}.
\end{eqnarray}
Proof of the commutation relation $[{\cal I}_m^*,
{\cal I}_n^*]=0$ is given as slmilar way.
Here we omit details for ${\cal I}_n^*$.
~~~
Q.E.D.

%%%%%%%%%%%%%%%%%%%%%%%%%%%%%%%%%%%%%
%%%%%%%%%%%%%%%%%%%%%%%%%%%%%%%%%%%%%
%%%%%%%%%%%%%%%%%%%%%%%%%%%%%%%%%%%%%

\section{Nonlocal Integrals of Motion}

In this section we give explicit formulae of
the nonlocal integrals of motion.
We study generic case :
$0<x<1, {\rm Re}(r)\neq 0$ and $s \in {\mathbb C}$
(resp. $0<x<1$, ${\rm Re}(r^*)\neq 0$ and $s \in {\mathbb C}$).

\subsection{Nonlocal Integrals of Motion}

We explicitly construct
the nonlocal integrals of motion
and state the main results for $N \geqq 3$.
The results for $N=2$ is summarized in \cite{FKSW1}.

\begin{df}~~~\\
$\bullet$~For the regime
${\rm Re}(r)>0$ and $0<{\rm Re}(s)<N$,
we define a family of operators ${\cal G}_m,~(m=1,2,\cdots)$ by
\begin{eqnarray}
{\cal G}_m&=&\prod_{t=1}^N \prod_{j=1}^m 
\oint_C \frac{dz_j^{(t)}}{2\pi \sqrt{-1}z_j^{(t)}}
F_1(z_1^{(1)})\cdots F_1(z_m^{(1)})
F_2(z_1^{(2)})\cdots F_2(z_m^{(2)})\cdots
F_N(z_1^{(N)})\cdots F_N(z_m^{(N)})\nonumber\\
&\times&
\frac{\displaystyle
\prod_{t=1}^N \prod_{1\leqq i<j \leqq m}
\left[u_i^{(t)}-u_j^{(t)}\right]_r
\left[u_j^{(t)}-u_i^{(t)}-1\right]_r
}{
\displaystyle
\prod_{t=1}^{N-1}\prod_{i,j=1}^m 
\left[u_i^{(t)}-u_j^{(t+1)}+1-\frac{s}{N}\right]_r
\prod_{i,j=1}^m 
\left[u_i^{(1)}-u_j^{(N)}+\frac{s}{N}\right]_r}\nonumber\\
&\times&
\vartheta\left(\sum_{j=1}^m u_j^{(1)}\right|
\left.\sum_{j=1}^m u_j^{(2)}\right|
\cdots
\left|\sum_{j=1}^m u_j^{(N)}\right).
\end{eqnarray}
Here we have set
the theta function
$\vartheta(u^{(1)}|u^{(2)}|\cdots|u^{(N)})$ by 
\begin{eqnarray}
&&
\vartheta(u^{(1)}|\cdots|u^{(t)}+r|\cdots|u^{(N)})
=\vartheta(u^{(1)}|\cdots|u^{(t)}|\cdots|u^{(N)}),~~(1\leqq t \leqq N)\\
&&
\vartheta(u^{(1)}|\cdots|u^{(t)}+r\tau|\cdots|u^{(N)})\nonumber\\
&=&
e^{-2\pi i \tau+\frac{2\pi i}{r}(u_{t-1}-2u_t+u_{t+1}+\sqrt{r(r-1)}P_{\alpha_t})}
\vartheta(u^{(1)}|\cdots|u^{(t)}|\cdots|u^{(N)}),~~(1\leqq t \leqq N),\\
&&\vartheta(u^{(1)}+k|\cdots|u^{(N)}+k)=
\vartheta(u^{(1)}|\cdots|u^{(N)}),~~(k\in {\mathbb C}),\\
&&\eta(\vartheta(u^{(1)}|\cdots|u^{(N)}))=
\vartheta(u^{(N)}|u^{(1)}|\cdots|u^{(N-1)}).
\end{eqnarray}
Here the integral contour $C$ is given by
\begin{eqnarray}
&&|x^{\frac{2s}{N}}z_j^{(t+1)}|
<|z_i^{(t)}|<|x^{-2+\frac{2s}{N}}z_j^{(t+1)}|,~~(1\leqq t \leqq N-1, 1\leqq i,j \leqq m),\\
&&
|x^{2-\frac{2s}{N}}z_j^{(1)}|<|z_i^{(N)}|<|x^{-\frac{2s}{N}}z_j^{(1)}|,
~~(1\leqq i,j \leqq m).
\end{eqnarray}
For generic $s \in {\mathbb C}$, the definition of ${\cal G}_n$
should be understood as analytic continuation.
We call the operator ${\cal G}_n$ 
the nonlocal integrals of motion
for the deformed $W$-algebra $W_{q,t}(\widehat{sl_N})$.
\\
%%%%%%%%%%%%%%%%%%%%%%%%%%%%%%%%%%%%%
$\bullet$~
~For the regime ${\rm Re}(r)<0$ and $0<{\rm Re}(s)<N$,
we define a family of operators ${\cal G}_m,~(m=1,2,\cdots)$ by
\begin{eqnarray}
{\cal G}_m&=&\prod_{t=1}^N \prod_{j=1}^m 
\oint_C \frac{dz_j^{(t)}}{2\pi \sqrt{-1}z_j^{(t)}}
F_1(z_1^{(1)})\cdots F_1(z_m^{(1)})
F_2(z_1^{(2)})\cdots F_2(z_m^{(2)})\cdots
F_N(z_1^{(N)})\cdots F_N(z_m^{(N)})\nonumber\\
&\times&
\frac{\displaystyle
\prod_{t=1}^N \prod_{1\leqq i<j \leqq m}
\left[u_i^{(t)}-u_j^{(t)}\right]_{-r}
\left[u_j^{(t)}-u_i^{(t)}+1\right]_{-r}
}{
\displaystyle
\prod_{t=1}^{N-1}\prod_{i,j=1}^m 
\left[u_i^{(t)}-u_j^{(t+1)}-\frac{s}{N}\right]_{-r}
\prod_{i,j=1}^m 
\left[u_i^{(1)}-u_j^{(N)}-1+\frac{s}{N}\right]_{-r}}\nonumber\\
&\times&
\vartheta\left(\sum_{j=1}^m u_j^{(1)}\right|
\left.\sum_{j=1}^m u_j^{(2)}\right|
\cdots
\left|\sum_{j=1}^m u_j^{(N)}\right).
\end{eqnarray}
Here we have set
the theta function
$\vartheta(u^{(1)}|u^{(2)}|\cdots|u^{(N)})$ by 
\begin{eqnarray}
&&
\vartheta(u^{(1)}|\cdots|u^{(t)}+r|\cdots|u^{(N)})
=\vartheta(u^{(1)}|\cdots|u^{(t)}|\cdots|u^{(N)}),~~(1\leqq t \leqq N)\\
&&
\vartheta(u^{(1)}|\cdots|u^{(t)}-r\tau|\cdots|u^{(N)})\nonumber\\
&=&
e^{-2\pi i \tau-\frac{2\pi i}{r}(u_{t-1}-2u_t+u_{t+1}+\sqrt{r(r-1)}P_{\alpha_t})}
\vartheta(u^{(1)}|\cdots|u^{(t)}|\cdots|u^{(N)}),~~(1\leqq t \leqq N),\\
&&\vartheta(u^{(1)}+k|\cdots|u^{(N)}+k)=
\vartheta(u^{(1)}|\cdots|u^{(N)}),~~(k\in {\mathbb C}),\\
&&\eta(\vartheta(u^{(1)}|\cdots|u^{(N)}))=
\vartheta(u^{(N)}|u^{(1)}|\cdots|u^{(N-1)}).
\end{eqnarray}
Here the integral contour $C$ is given by
\begin{eqnarray}
&&|x^{-2+\frac{2s}{N}}z_j^{(t+1)}|
<|z_i^{(t)}|<|x^{\frac{2s}{N}}z_j^{(t+1)}|,~~(1\leqq t \leqq N-1, 1\leqq i,j \leqq m),\\
&&
|x^{-\frac{2s}{N}}z_j^{(1)}|<|z_i^{(N)}|<
|x^{2-\frac{2s}{N}}z_j^{(1)}|,
~~(1\leqq i,j \leqq m).
\end{eqnarray}
For generic $s \in {\mathbb C}$, the definition of ${\cal G}_n$
should be understood as analytic continuation.
We call the operator ${\cal G}_n$ 
the nonlocal integrals of motion
for the deformed $W$-algebra $W_{q,t}(\widehat{sl_N})$.
\\
%%%%%%%%%%%%%%%%%%%%%%%%%%%%%%%%%%%%%%%%%%%%%%%%
$\bullet$~
~For ${\rm Re}(r^*)>0$ and $0<{\rm Re}(s)<N$,
we define a family of operators ${\cal G}^*_m,~(m=1,2,\cdots)$ by
\begin{eqnarray}
{\cal G}^*_m&=&\prod_{t=1}^N \prod_{j=1}^m 
\oint_{C^*} \frac{dz_j^{(t)}}{2\pi \sqrt{-1}z_j^{(t)}}
E_1(z_1^{(1)})\cdots E_1(z_m^{(1)})
E_2(z_1^{(2)})\cdots E_2(z_m^{(2)})\cdots
E_N(z_1^{(N)})\cdots E_N(z_m^{(N)})\nonumber\\
&\times&
\frac{\displaystyle
\prod_{t=1}^N \prod_{1\leqq i<j \leqq m}
\left[u_i^{(t)}-u_j^{(t)}\right]_{r^*}
\left[u_j^{(t)}-u_i^{(t)}+1\right]_{r^*}
}{
\displaystyle
\prod_{t=1}^{N-1}\prod_{i,j=1}^m 
\left[u_i^{(t)}-u_j^{(t+1)}-\frac{s}{N}\right]_{r^*}
\prod_{i,j=1}^m 
\left[u_i^{(1)}-u_j^{(N)}-1+\frac{s}{N}\right]_{r^*}}\nonumber\\
&\times&
\vartheta^*\left(\sum_{j=1}^m u_j^{(1)}\right|
\left.\sum_{j=1}^m u_j^{(2)}\right|
\cdots
\left|\sum_{j=1}^m u_j^{(N)}\right).
\end{eqnarray}
Here we have set
the theta function
$\vartheta^*(u^{(1)}|u^{(2)}|\cdots|u^{(N)})$ by 
\begin{eqnarray}
&&
\vartheta^*(u^{(1)}|\cdots|u^{(t)}+r|\cdots|u^{(N)})
=\vartheta^*(u^{(1)}|\cdots|u^{(t)}|\cdots|u^{(N)}),~~(1\leqq t \leqq N)\\
&&
\vartheta^*(u^{(1)}|\cdots|u^{(t)}+r^*
\tau|\cdots|u^{(N)})\nonumber\\
&=&
e^{-2\pi i \tau+\frac{2\pi i}{r^*}
(u_{t-1}-2u_t+u_{t+1}+\sqrt{r(r-1)}P_{\alpha_t})}
\vartheta^*(u^{(1)}|\cdots|u^{(t)}|\cdots|u^{(N)}),~~(1\leqq t \leqq N),\\
&&\vartheta^*(u^{(1)}+k|\cdots|u^{(N)}+k)=
\vartheta^*(u^{(1)}|\cdots|u^{(N)}),~~(k\in {\mathbb C}),\\
&&\eta(\vartheta^*(u^{(1)}|\cdots|u^{(N)}))=
\vartheta^*(u^{(N)}|u^{(1)}|\cdots|u^{(N-1)}).
\end{eqnarray}
Here the integral contour $C^*$ is given by
\begin{eqnarray}
&&|x^{-2+\frac{2s}{N}}z_j^{(t+1)}|
<|z_i^{(t)}|<|x^{\frac{2s}{N}}z_j^{(t+1)}|,~~(1\leqq t \leqq N-1, 1\leqq i,j \leqq m),\\
&&
|x^{-\frac{2s}{N}}z_j^{(1)}|<|z_i^{(N)}|<|x^{2
-\frac{2s}{N}}z_j^{(1)}|,
~~(1\leqq i,j \leqq m).
\end{eqnarray}
For generic $s \in {\mathbb C}$, the definition of ${\cal G}_n$
should be understood as analytic continuation.
We call the operator ${\cal G}_n$ 
the nonlocal integrals of motion
for the deformed $W$-algebra $W_{q,t}(\widehat{sl_N})$.
\\
%%%%%%%%%%%%%%%%%%%%%%%%%%%%%%%%%%%%%%%
$\bullet$~
~For ${\rm Re}(r^*)<0$ 
and $0<{\rm Re}(s)<N$,
we define a family of operators 
${\cal G}^*_m,~(m=1,2,\cdots)$ by
\begin{eqnarray}
{\cal G}^*_m&=&\prod_{t=1}^N \prod_{j=1}^m 
\oint_{C^*} \frac{dz_j^{(t)}}{2\pi \sqrt{-1}z_j^{(t)}}
E_1(z_1^{(1)})\cdots E_1(z_m^{(1)})
E_2(z_1^{(2)})\cdots E_2(z_m^{(2)})\cdots
E_N(z_1^{(N)})\cdots E_N(z_m^{(N)})\nonumber\\
&\times&
\frac{\displaystyle
\prod_{t=1}^N \prod_{1\leqq i<j \leqq m}
\left[u_i^{(t)}-u_j^{(t)}\right]_{-r^*}
\left[u_j^{(t)}-u_i^{(t)}-1\right]_{-r^*}
}{
\displaystyle
\prod_{t=1}^{N-1}\prod_{i,j=1}^m 
\left[u_i^{(t)}-u_j^{(t+1)}+1-\frac{s}{N}\right]_{-r^*}
\prod_{i,j=1}^m 
\left[u_i^{(1)}-u_j^{(N)}+\frac{s}{N}\right]_{-r^*}}\nonumber\\
&\times&
\vartheta^*
\left(\sum_{j=1}^m u_j^{(1)}\right|
\left.\sum_{j=1}^m u_j^{(2)}\right|
\cdots
\left|\sum_{j=1}^m u_j^{(N)}\right).
\end{eqnarray}
Here we have set
the theta function
$\vartheta^*(u^{(1)}|u^{(2)}|\cdots|u^{(N)})$ by 
\begin{eqnarray}
&&
\vartheta^*(u^{(1)}|\cdots|u^{(t)}+r|\cdots|u^{(N)})
=\vartheta^*(u^{(1)}|\cdots|u^{(t)}|\cdots|u^{(N)}),~~(1\leqq t \leqq N)\\
&&
\vartheta^*(u^{(1)}|\cdots|u^{(t)}r^*\tau|
\cdots|u^{(N)})\nonumber\\
&=&
e^{-2\pi i \tau-\frac{2\pi i}{r^*}
(u_{t-1}-2u_t+u_{t+1}+\sqrt{r(r-1)}P_{\alpha_t})}
\vartheta^*(u^{(1)}|\cdots|u^{(t)}|\cdots|u^{(N)}),~~(1\leqq t \leqq N),\\
&&\vartheta^*(u^{(1)}+k|\cdots|u^{(N)}+k)=
\vartheta^*(u^{(1)}|\cdots|u^{(N)}),~~(k\in {\mathbb C}),\\
&&\eta(\vartheta^*(u^{(1)}|\cdots|u^{(N)}))=
\vartheta^*(u^{(N)}|u^{(1)}|\cdots|u^{(N-1)}).
\end{eqnarray}
Here the integral contour $C^*$ is given by
\begin{eqnarray}
&&|x^{\frac{2s}{N}}z_j^{(t+1)}|
<|z_i^{(t)}|<|x^{-2+\frac{2s}{N}}z_j^{(t+1)}|,~~(1\leqq t \leqq N-1, 1\leqq i,j \leqq m),\\
&&
|x^{2-\frac{2s}{N}}z_j^{(1)}|<|z_i^{(N)}|<|x^{-\frac{2s}{N}}z_j^{(1)}|,
~~(1\leqq i,j \leqq m).
\end{eqnarray}
For generic $s \in {\mathbb C}$, the definition of ${\cal G}_n$
should be understood as analytic continuation.
We call the operator ${\cal G}_n$ 
the nonlocal integrals of motion
for the deformed $W$-algebra $W_{q,t}(\widehat{sl_N})$.
\end{df}

We summarize explicit formulae for the integrand function
$\vartheta(u^{(1)}|u^{(2)}|\cdots|u^{(N)})$.

\begin{prop}~~
For $\alpha_1,\alpha_2,\cdots,\alpha_N
\in {\mathbb C}$ and ${\rm Re}(r)>0$, 
we set the theta function $\widetilde
{\vartheta}_\alpha
(u^{(1)}|u^{(2)}|\cdots|u^{(N)})$ by
\begin{eqnarray}
\widetilde
{\vartheta}_{\alpha}
(u^{(1)}|u^{(2)}|\cdots|u^{(N)})
&=&[u^{(1)}-u^{(2)}-\sqrt{rr^*}P_{\bar{\epsilon}_2}
+\alpha_1P_{\bar{\epsilon}_1}
+\alpha_2P_{\bar{\epsilon}_2}+\cdots
+\alpha_NP_{\bar{\epsilon}_N}]_r\nonumber\\
&\times&
[u^{(2)}-u^{(3)}-\sqrt{rr^*}P_{\bar{\epsilon}_3}
+\alpha_1P_{\bar{\epsilon}_1}
+\alpha_2P_{\bar{\epsilon}_2}+\cdots
+\alpha_NP_{\bar{\epsilon}_N}]_r\nonumber\\
&\times&\cdots\nonumber\\
&\times&
[u^{(N)}-u^{(1)}-\sqrt{rr^*}
P_{\bar{\epsilon}_1}
+\alpha_1P_{\bar{\epsilon}_1}
+\alpha_2P_{\bar{\epsilon}_2}+\cdots
+\alpha_NP_{\bar{\epsilon}_N}]_r.\nonumber\\
\end{eqnarray}
This theta function $\widetilde
{\vartheta}_\alpha
(u^{(1)}|\cdots|u^{(N)})$ 
satisfies
the conditions
\begin{eqnarray}
&&
\widetilde{\vartheta}_\alpha
(u^{(1)}|\cdots|u^{(t)}+r|\cdots|u^{(N)})
=\widetilde{\vartheta}_\alpha
(u^{(1)}|\cdots|u^{(t)}|\cdots|u^{(N)}),~~(1\leqq t \leqq N)\\
&&
\widetilde{\vartheta}_\alpha
(u^{(1)}|\cdots|u^{(t)}+r\tau|\cdots|u^{(N)})\nonumber\\
&=&
e^{-2\pi i \tau+\frac{2\pi i}{r}(u_{t-1}-2u_t+u_{t+1}+\sqrt{r(r-1)}P_{\alpha_t})}
\widetilde{\vartheta}_\alpha(u^{(1)}|\cdots|u^{(t)}|\cdots|u^{(N)}),~~(1\leqq t \leqq N),\\
&&
\widetilde{\vartheta}_\alpha(u^{(1)}+k|\cdots|u^{(N)}+k)=
\widetilde{\vartheta}_\alpha(u^{(1)}|\cdots|u^{(N)}),~~(k\in {\mathbb C}),\\
&&\eta(\widetilde{\vartheta}_\alpha(u^{(1)}|\cdots|u^{(N)}))=
\widetilde{\vartheta}_\alpha(u^{(N)}|u^{(1)}|\cdots|u^{(N-1)}).
\end{eqnarray}
\end{prop}
{\it Proof}~~~
Let us set
$
\widetilde
{\vartheta}
(u^{(1)}|\cdots|u^{(N)})=
[u^{(1)}-u^{(2)}+\hat{\pi}_{1,2}]_r
[u^{(2)}-u^{(3)}+\hat{\pi}_{2,3}]_r
\cdots
[u^{(N)}-u^{(1)}+\hat{\pi}_{N,1}]_r$.
The second quasi-periodic condition
is equivalent with
\begin{eqnarray}
\left(\begin{array}{cccccc}
1&0&0&0&\cdots&-1\\
-1&1&0&0&\cdots&0\\
0&-1&1&0&\cdots&0\\
\cdots&\cdots&\cdots&\cdots&\cdots&\\
0&0&0&0&-1&1\\
\end{array}\right)
\left(\begin{array}{c}
\hat{\pi}_{1,2}\\
\hat{\pi}_{2,3}\\\hat{\pi}_{1,2}
\\
\cdots\\
\hat{\pi}_{N,1}\\
\end{array}\right)=
\sqrt{rr^*}\left(\begin{array}{c}
{P}_{\alpha_1}\\
{P}_{\alpha_2}\\
{P}_{\alpha_3}\\
\cdots\\
{P}_{\alpha_N}\\
\end{array}\right).
\end{eqnarray}
Hence we have the general solution for $\hat{\pi}_{i,i+1}$ by
\begin{eqnarray}
\left(\begin{array}{c}
\hat{\pi}_{1,2}\\
\hat{\pi}_{2,3}\\
\cdots\\
\hat{\pi}_{N,1}\\
\end{array}\right)=
\left(\begin{array}{c}
-\sqrt{rr^*}P_{\bar{\epsilon}_2}
+\alpha_1P_{\bar{\epsilon}_1}
+\alpha_2P_{\bar{\epsilon}_2}+\cdots
+\alpha_NP_{\bar{\epsilon}_N}
\\
-\sqrt{rr^*}P_{\bar{\epsilon}_3}
+\alpha_1P_{\bar{\epsilon}_1}
+\alpha_2P_{\bar{\epsilon}_2}+\cdots
+\alpha_NP_{\bar{\epsilon}_N}
\\
\cdots\\
-\sqrt{rr^*}P_{\bar{\epsilon}_1}
+\alpha_1P_{\bar{\epsilon}_1}
+\alpha_2P_{\bar{\epsilon}_2}+\cdots
+\alpha_NP_{\bar{\epsilon}_N}
\end{array}\right),
\end{eqnarray}
where $\alpha_1, \alpha_2, \cdots,\alpha_N \in {\mathbb C}$.
Other conditions are trivial.
~~~
Q.E.D.

~\\
{\bf Example}~~For $N=2$, $m=1$ and ${\rm Re}(r)>0$ case, 
we have
\begin{eqnarray}
{\cal G}_1=
\int \int_C \frac{dz_1}{2\pi \sqrt{-1}z_1}
\frac{dz_2}{2\pi \sqrt{-1}z_2}
F_1(z_1)F_2(z_2)\frac{\vartheta(u_1|u_2)}{[u_1-u_2+\frac{s}{2}]_r
[u_1-u_2-\frac{s}{2}+1]_r}.
\end{eqnarray}
Here $C$ is given by
$$|x^{s}z_2|<|z_1|<|x^{-2+s}z_2|.$$

~\\
{\bf Example}~~For $N=3$, $m=1$ and ${\rm Re}(r)>0$ case, 
we have
\begin{eqnarray}
{\cal G}_1&=&
\int \int \int_C \frac{dz_1}{2\pi \sqrt{-1}z_1}
\frac{dz_2}{2\pi \sqrt{-1}z_2}\frac{dz_3}{2\pi \sqrt{-1}z_3}
F_1(z_1)F_2(z_2)F_3(z_3)\nonumber\\
&\times&
\frac{\vartheta(u_1|u_2|u_3)}{
[u_1-u_2+1-\frac{s}{3}]_r
[u_2-u_3+1-\frac{s}{3}]_r
[u_1-u_3+\frac{s}{3}]_r}.
\end{eqnarray}
Here $C$ is given by
$$|x^{\frac{2s}{3}}z_2|<|z_1|<|x^{-2+\frac{2s}{3}}z_2|,
|x^{\frac{2s}{3}}z_3|<|z_2|<|x^{-2+\frac{2s}{3}}z_3|,
|x^{2-\frac{2s}{3}}z_1|<|z_3|<|x^{-\frac{2s}{3}}z_1|.$$

~\\
The followings are some of {\bf Main Results}.

\begin{thm}~~\label{thm:Nonlocal-Com1}
The nonlocal integrals of motion ${\cal G}_n$
commute with each other.
\begin{eqnarray}
[{\cal G}_m,{\cal G}_n]=0,~~~(m,n=1,2,\cdots).
\end{eqnarray}
The nonlocal integrals of motion ${\cal G}^*_n$
commute with each other.
\begin{eqnarray}
[{\cal G}^*_m,{\cal G}^*_n]=0,~~~(m,n=1,2,\cdots).
\end{eqnarray}
\end{thm}

\begin{thm}~~\label{thm:Nonlocal-Com2}
The nonlocal integrals of motion ${\cal G}_n$
and ${\cal G}_n^*$
commute with each other
for regime $0<{\rm Re}(r)$ and ${\rm Re}(r^*)<0$.
\begin{eqnarray}
[{\cal G}_m,{\cal G}_n^*]=0,~~(m,n=1,2,\cdots).
\end{eqnarray}
\end{thm}

\begin{thm}\label{thm:Nonlocal-Com3}
~~~The local integrals of motion ${\cal I}_n$,
${\cal I}^*_n$
and nonlocal integrals of motion ${\cal G}_m$,
${\cal G}^*_m$
commute with each other.
\begin{eqnarray}
~[{\cal I}_n,{\cal G}_m]&=&0,~
~[{\cal I}_n,{\cal G}^*_m]=0,
~~(m,n=1,2,\cdots),\\
~[{\cal I}^*_n,{\cal G}_m]&=&0,~
~[{\cal I}^*_n,{\cal G}^*_m]=0,~~~(m,n=1,2,\cdots).
\end{eqnarray}
\end{thm}

\subsection{Proof of $[{\cal G}_m,{\cal G}_n]=0$}

In this section we study the commutation relations
$[{\cal G}_m,{\cal G}_n]=0$ for ${\rm Re}(r)>0$.
We omit details for other cases,
because they are similar.

\begin{prop}~~~\label{prop:thetaNonlocal}
For ${\rm Re}(r)>0$ we have
\begin{eqnarray}
&&\sum_{\sigma_1 \in S_{m+n}}
\sum_{\sigma_2 \in S_{m+n}}
\cdots \sum_{\sigma_N \in S_{m+n}}
\widehat{\vartheta}_\alpha
\left(\sum_{j=1}^m u_{\sigma_1(j)}^{(1)}\right|
\left.\sum_{j=1}^m u_{\sigma_2(j)}^{(2)}\right|
\cdots
\left|\sum_{j=1}^m u_{\sigma_N(j)}^{(N)}\right)\nonumber\\
&\times&
\widehat{\vartheta}_\beta
\left(\sum_{j=m+1}^{m+n} u_{\sigma_1(j)}^{(1)}\right|
\left.\sum_{j=m+1}^{m+n} u_{\sigma_2(j)}^{(2)}\right|
\cdots
\left|\sum_{j=m+1}^{m+n} u_{\sigma_N(j)}^{(N)}\right)
\nonumber\\
&\times&
\prod_{t=1}^N
\prod_{i=1}^m
\prod_{j=m+1}^{m+n}
\frac{\left[u_{\sigma_t(i)}^{(t)}-u_{\sigma_{t+1}(j)}^{(t+1)}-\frac{s}{N}\right]_r
\left[
u_{\sigma_{t+1}(i)}^{(t+1)}-
u_{\sigma_{t}(j)}^{(t)}+1-\frac{s}{N}\right]_r
}{
\left[u_{\sigma_t(i)}^{(t)}-u_{\sigma_{t}(j)}^{(t)}\right]_r
\left[u_{\sigma_{t}(j)}^{(t)}
-u_{\sigma_{t}(i)}^{(t)}-1\right]_r
}\nonumber
\\
&=&
\sum_{\sigma_1 \in S_{m+n}}
\sum_{\sigma_2 \in S_{m+n}}
\cdots \sum_{\sigma_N \in S_{m+n}}
\widehat{\vartheta}_\beta
\left(\sum_{j=1}^n u_{\sigma_1(j)}^{(1)}\right|
\left.\sum_{j=1}^n u_{\sigma_2(j)}^{(2)}\right|
\cdots
\left|\sum_{j=1}^n u_{\sigma_N(j)}^{(N)}\right)
\nonumber\\
&\times&
\widehat{\vartheta}_\alpha
\left(\sum_{j=n+1}^{m+n} u_{\sigma_1(j)}^{(1)}\right|
\left.\sum_{j=n+1}^{m+n} u_{\sigma_2(j)}^{(2)}\right|
\cdots
\left|\sum_{j=n+1}^{m+n} u_{\sigma_N(j)}^{(N)}\right)
\nonumber\\
&\times&
\prod_{t=1}^N
\prod_{i=1}^n
\prod_{j=n+1}^{m+n}
\frac{\left[u_{\sigma_t(i)}^{(t)}-u_{\sigma_{t+1}(j)}^{(t+1)}-\frac{s}{N}\right]_r
\left[u_{\sigma_{t+1}(i)}^{(t+1)}
-u_{\sigma_{t}(j)}^{(t)}+1-\frac{s}{N}\right]_r
}{
\left[u_{\sigma_t(i)}^{(t)}-u_{\sigma_{t}(j)}^{(t)}\right]_r
\left[u_{\sigma_t(j)}^{(t)}-u_{\sigma_{t}(i)}^{(t)}-1\right]_r
}.
\end{eqnarray}
Here $\widehat{\vartheta}_\alpha(u^{(1)}|u^{(2)}|\cdots|u^{(N)})$ and 
$\widehat{\vartheta}_\beta(u^{(1)}|u^{(2)}|\cdots|u^{(N)})$ 
are given by
\begin{eqnarray}
&&
\widehat{\vartheta}_{\alpha}(u^{(1)}|\cdots|u^{(t)}+r|\cdots|u^{(N)})
=\widehat{\vartheta}_{\alpha}(u^{(1)}|\cdots|u^{(t)}|\cdots|u^{(N)}),~~(1\leqq t \leqq N)\\
&&
\widehat{\vartheta}_\alpha(u^{(1)}|\cdots|u^{(t)}+r\tau|\cdots|u^{(N)})\\
&=&
e^{-2\pi i \tau+
\frac{2\pi i}{r}(u_{t-1}-2u_t+u_{t+1}+\sqrt{r(r-1)}P_{\alpha_t})
+\nu_{\alpha,t}}
\widehat{\vartheta}_\alpha(u^{(1)}|\cdots|u^{(t)}|\cdots|u^{(N)}),~~(1\leqq t \leqq N),\nonumber
\end{eqnarray}
\begin{eqnarray}
&&
\widehat{\vartheta}_{\beta}(u^{(1)}|\cdots|u^{(t)}+r|\cdots|u^{(N)})
=\widehat{\vartheta}_{\beta}(u^{(1)}|\cdots|u^{(t)}|\cdots|u^{(N)}),~~(1\leqq t \leqq N)\\
&&
\widehat{\vartheta}_\beta(u^{(1)}|\cdots|u^{(t)}+r\tau|\cdots|u^{(N)})\\
&=&
e^{-2\pi i \tau+
\frac{2\pi i}{r}(u_{t-1}-2u_t+u_{t+1}+\sqrt{r(r-1)}P_{\alpha_t})
+\nu_{\beta,t}}
\widehat{\vartheta}_\beta(u^{(1)}|\cdots|u^{(t)}|\cdots|u^{(N)}),~~(1\leqq t \leqq N).\nonumber
\end{eqnarray}
Here $\nu_{\alpha,t}, 
\nu_{\beta,t} \in {\mathbb C}$, 
$(1\leqq t \leqq N)$.
%%%%%%%%%%%%%%%%%%%%%%%%%%%%%%%%%%%%%%%%%%%%
\end{prop}
{\it Proof.}~~
In order to consider elliptic function,
we divide the above theta identity by
$\widehat{\vartheta}_{\gamma}(\sum_{j=1}^{m+n}u_j^{(1)}|\cdots|
\sum_{j=1}^{n+m}u_j^{(N)})$ with $\nu_{\gamma,t}\in {\mathbb C}$,
$(1\leqq t \leqq N)$:
\begin{eqnarray}
&&
\widehat{\vartheta}_{\gamma}
(u^{(1)}|\cdots|u^{(t)}+r|\cdots|u^{(N)})
=\widehat{\vartheta}_{\gamma}
(u^{(1)}|\cdots|u^{(t)}|\cdots|u^{(N)}),~~(1\leqq t \leqq N)\\
&&
\widehat{\vartheta}_\gamma
(u^{(1)}|\cdots|u^{(t)}+r\tau|\cdots|u^{(N)})\\
&=&
e^{-2\pi i \tau+
\frac{2\pi i}{r}(u_{t-1}-2u_t+u_{t+1}+\sqrt{r(r-1)}P_{\alpha_t})
+\nu_{\gamma,t}}
\widehat{\vartheta}_\gamma
(u^{(1)}|\cdots|u^{(t)}|\cdots|u^{(N)}),~~(1\leqq t \leqq N).\nonumber
\end{eqnarray}
Let us set
\begin{eqnarray}
{\rm LHS}(m,n)&=&
\sum_{K_1 \cup K_1^c=\{1,2,\cdots,n+m\}
\atop{|K_1|=m,
|K_1^c|=n}}
\cdots 
\sum_{K_N \cup K_N^c=\{1,2,\cdots,n+m\}
\atop{|K_N|=m,
|K_N^c|=n}}
\\
&\times&
\frac{\displaystyle
\widehat{\vartheta}_\alpha
\left(\sum_{j \in K_1}u_{j}^{(1)}\right|
\cdots
\left|\sum_{j \in K_N}u_{j}^{(N)}\right)
\widehat{\vartheta}_\beta
\left(\sum_{j \in K_1^c}u_{j}^{(1)}\right|
\cdots
\left|\sum_{j \in K_N^c}u_{j}^{(N)}\right)
}{\displaystyle
\widehat{\vartheta}_\gamma
\left(\sum_{j=1}^{m+n} u_{j}^{(1)}\right|
\cdots
\left|\sum_{j=1}^{m+n} u_{j}^{(N)}\right)
}\nonumber\\
&\times&
\prod_{t=1}^N
\frac{
\displaystyle
\prod_{i \in K_t}
\prod_{j \in K_{t+1}^c}
\left[u_i^{(t)}-u_j^{(t+1)}+1-\frac{s}{N}\right]_r
\prod_{i \in K_{t+1}}
\prod_{j \in K_t^c}
\left[u_i^{(t+1)}-u_j^{(t)}+\frac{s}{N}\right]_r}{
\displaystyle
\prod_{i \in K_t}
\prod_{j \in K_{t}^c}
\left[u_i^{(t)}-u_j^{(t)}\right]_r
\left[u_j^{(t)}-u_i^{(t)}-1\right]_r
},\nonumber
\end{eqnarray}
\begin{eqnarray}
{\rm RHS}(m,n)&=&
\sum_{K_1 \cup K_1^c=\{1,2,\cdots,n+m\}
\atop{|K_1|=n,
|K_1^c|=m}}
\cdots 
\sum_{K_N \cup K_N^c=\{1,2,\cdots,n+m\}
\atop{|K_N|=n,
|K_N^c|=m}}
\\
&\times&
\frac{\displaystyle
\widehat{\vartheta}_\beta
\left(\sum_{j \in K_1}u_{j}^{(1)}\right|
\cdots
\left|\sum_{j \in K_N}u_{j}^{(N)}\right)
\widehat{\vartheta}_\alpha
\left(\sum_{j \in K_1^c}u_{j}^{(1)}\right|
\cdots
\left|\sum_{j \in K_N^c}u_{j}^{(N)}\right)
}{\displaystyle
\widehat{\vartheta}_\gamma
\left(\sum_{j=1}^{m+n} u_{j}^{(1)}\right|
\cdots
\left|\sum_{j=1}^{m+n} u_{j}^{(N)}\right)
}\nonumber\\
&\times&
\prod_{t=1}^N
\frac{\displaystyle
\prod_{i \in K_t}
\prod_{j \in K_{t+1}^c}
\left[u_i^{(t)}-u_j^{(t+1)}+1-\frac{s}{N}\right]_r
\prod_{i \in K_{t+1}}
\prod_{j \in K_t^c}
\left[u_i^{(t+1)}-u_j^{(t)}+\frac{s}{N}\right]_r}{
\displaystyle
\prod_{i \in K_{t}}
\prod_{j \in K_{t}^c}
\left[u_i^{(t)}-u_j^{(t)}\right]_r
\left[u_j^{(t)}-u_i^{(t)}-1\right]_r
},\nonumber
\end{eqnarray}
Candidates of poles of both
${\rm LHS}(m,n)$ and ${\rm RHS}(m,n)$ are
$u_i^{(t)}=u_j^{(t)}$ and $u_i^{(t)}=u_j^{(t)}+1$
and $\vartheta_\gamma=0$.
Let us show that
the points $u_i^{(t)}=u_j^{(t)}$ are regular.
Take the residue of the ${\rm LHS}(m,n)$ at $u_1^{(1)}=u_2^{(1)}$.
We have
\begin{eqnarray}
&&{\rm Res}_{u_1^{(1)}=u_2^{(1)}}
\left(\frac{1}{[u_1^{(1)}-u_2^{(1)}]_r
[u_2^{(1)}-u_1^{(1)}-1]_r}+\frac{1}{[u_1^{(1)}-u_2^{(1)}]_r
[u_1^{(1)}-u_2^{(1)}-1]_r}\right)\nonumber\\
&\times&
\sum_{L_1\cup L_1^c=\{3,4,\cdots,n+m\}
\atop{|L_1|=m-1,|L_1^c|=n-1}}
\sum_{K_2\cup K_2^c=\{1,2,\cdots,n+m\}
\atop{|K_1|=m,|K_1^c|=n}}\cdots
\sum_{K_N\cup K_N^c=\{1,2,\cdots,n+m\}
\atop{|K_N|=m,|K_N^c|=n}}\nonumber\\
&\times&
\frac{\displaystyle
\widehat{\vartheta}_\alpha
(\sum_{j \in L_1\cup\{1\}}u_j^{(1)}|
\cdots|\sum_{j\in K_N}u_j^{(N)})
\widehat{\vartheta}_\beta
(\sum_{j \in L_1^c\cup\{1\}}u_j^{(1)}|
\cdots|\sum_{j\in K_N^c}u_j^{(N)})
}{\displaystyle
\widehat{\vartheta}_\gamma
(\sum_{j=1}^{m+n}
u_j^{(1)}|\cdots|\sum_{j=1}^{n+m}
u_j^{(N)})}\nonumber\\
&\times&
\frac{
\displaystyle
\prod_{j \in K_2^c}
\left[u_1^{(1)}-u_j^{(2)}+1-\frac{s}{N}\right]_r
\prod_{i \in K_2}
\left[u_i^{(2)}-u_2^{(1)}+\frac{s}{N}\right]_r
}{\displaystyle
\prod_{j \in L_1^c}\left[u_1^{(1)}-u_j^{(1)}\right]_r
\left[u_j^{(1)}-u_1^{(1)}-1\right]_r
}\nonumber\\
&\times&
\frac{\displaystyle
\prod_{j \in K_N}
\left[u_i^{(N)}-u_2^{(1)}+1-\frac{s}{N}\right]_r
\prod_{j \in K_N^c}
\left[u_1^{(1)}-u_j^{(N)}+\frac{s}{N}\right]_r
}{\displaystyle
\prod_{j \in L_1}
\left[u_i^{(1)}-u_2^{(1)}\right]_r
\left[u_2^{(1)}-u_i^{(1)}-1\right]_r
}\nonumber\\
&\times&
\frac{\displaystyle
\prod_{i \in L_1}\prod_{j \in K_2^c}
\left[u_i^{(1)}-u_j^{(2)}+1-\frac{s}{N}\right]_r
\prod_{i \in K_2}
\prod_{j \in L_1^c}
\left[u_i^{(2)}-u_j^{(1)}+\frac{s}{N}\right]_r
}{
\displaystyle
\prod_{i \in L_1}
\prod_{j \in L_1^c}
\left[u_i^{(1)}-u_j^{(1)}\right]_r
}\nonumber\\
&\times&
\frac{\displaystyle
\prod_{j \in K_N}
\prod_{j \in L_1^c}
\left[u_i^{(N)}-u_j^{(1)}+1-\frac{s}{N}\right]_r
\prod_{i \in L_1}
\prod_{j \in K_N^c}
\left[u_i^{(1)}-u_j^{(N)}+\frac{s}{N}\right]_r
}{\displaystyle
\prod_{i \in L_1}
\prod_{j \in L_1^c}
\left[u_j^{(1)}-u_i^{(1)}-1\right]_r
}\\
&\times&
\prod_{t=2}^N
\frac{\displaystyle
\prod_{i \in K_t}
\prod_{j \in K_{t+1}^c}
\left[u_i^{(t)}-u_j^{(t+1)}+1-\frac{s}{N}\right]_r
\prod_{i \in K_{t+1}}
\prod_{j \in K_t^c}
\left[u_i^{(t+1)}-u_j^{(t)}+\frac{s}{N}\right]_r
}{\displaystyle
\prod_{i \in K_t}
\prod_{j \in K_{t}^c}
\left[u_i^{(t)}-u_j^{(t)}\right]_r
\left[u_j^{(t)}-u_i^{(t)}-1\right]_r
}=0.\nonumber
\end{eqnarray}
Because the first term :
${\rm Res}_{u_1^{(1)}=u_2^{(1)}}
\left(\frac{1}{[u_1^{(1)}-u_2^{(1)}]_r
[u_2^{(1)}-u_1^{(1)}-1]_r}+\frac{1}{[u_1^{(1)}-u_2^{(1)}]_r
[u_1^{(1)}-u_2^{(1)}-1]_r}\right)=0$,
we have ${\rm Res}_{u_1^{(1)}=u_2^{(1)}}
{\rm LHS}(m,n)=0$.
Because ${\rm LHS}(m,n)$ is symmetric with 
respect to variables
$u_1^{(1)},
u_2^{(1)},\cdots,u_{m+n}^{(1)}$,
we have
${\rm Res}_{u_i^{(1)}=u_j^{(1)}}
{\rm LHS}(m,n)=0$ for $1\leqq i \neq j \leqq m+n$.
As the same manner as above,
we conclude
that points $u_i^{(t)}=u_j^{(t)}$
of ${\rm LHS}(m,n)$ and
${\rm RHS}(m,n)$
for $1\leqq t \leqq N$,
$1\leqq i \neq j \leqq m+n$ are regular.
Let us show ${\rm LHS}(m,n)={\rm RHS}(m,n)$ 
by induction for $m+n$.
Candidates of poles are only
$u_i^{(t)}=u_j^{(t)}+1$, $1\leqq t \leqq N$ and
$1\leqq i \neq j \leqq m+n$.
We assume $1\leqq m <n$ without loosing generality.
(The case $m=n$ is trivial.)
At first we show the starting point $1=m<n$ :
${\rm LHS}(1,n)={\rm RHS}(1,n)$ .
By straightfoeward calculations, we have
\begin{eqnarray}
&&
{\rm Res}_{u_2^{(1)}=u_1^{(1)}+1}\cdots
{\rm Res}_{u_2^{(N)}=u_1^{(N)}+1}{\rm LHS}(1,n)\nonumber\\
&=&
\prod_{t=1}^N
{\rm Res}_{u_2^{(t)}=u_1^{(t)}+1}
\frac{
\displaystyle
\left[u_1^{(t)}-u_2^{(t+1)}+1-\frac{s}{N}\right]_r
\left[u_1^{(t+1)}-u_2^{(t)}+\frac{s}{N}\right]_r}{
\displaystyle
\left[u_1^{(t)}-u_2^{(t)}\right]_r
\left[u_2^{(t)}-u_1^{(t)}-1\right]_r}
\nonumber\\
&\times&
\prod_{t=1}^N
\prod_{j=3}^{n+1}
\frac{\displaystyle
\left[u_1^{(t)}-u_j^{(t+1)}+1-\frac{s}{N}\right]_r
\left[u_1^{(t+1)}-u_j^{(t)}+\frac{s}{N}\right]_r}{
\displaystyle
\left[u_1^{(t)}-u_j^{(t)}\right]_r
\left[u_j^{(t)}-u_1^{(t)}-1\right]_r}
\nonumber\\
&\times&
\frac{\displaystyle
\widehat{\vartheta}_\alpha
(u_1^{(1)}|\cdots|u_1^{(N)})
\widehat{\vartheta}_\beta
\left(\left.
\sum_{j=2}^{n+1}u_j^{(1)}\right|
\cdots\left|\sum_{j=2}^{n+1}u_j^{(N)}\right.\right)
}{
\displaystyle
\widehat{\vartheta}_\gamma\left(\left.
\sum_{j=1}^{n+1}u_j^{(1)}\right|\cdots
\left|\sum_{j=1}^{n+1}u_j^{(N)}\right.\right)},
\end{eqnarray}
\begin{eqnarray}
&&
{\rm Res}_{u_2^{(1)}=u_1^{(1)}+1}\cdots
{\rm Res}_{u_2^{(N)}=u_1^{(N)}+1}{\rm RHS}(1,n)\nonumber\\
&=&
\prod_{t=1}^N
{\rm Res}_{u_2^{(t)}=u_1^{(t)}+1}
\frac{\displaystyle
\left[u_1^{(t)}-u_2^{(t+1)}+1-\frac{s}{N}\right]_r
\left[u_1^{(t+1)}-u_2^{(t)}+\frac{s}{N}\right]_r}{
\displaystyle
\left[u_1^{(t)}-u_2^{(t)}\right]_r
\left[u_2^{(t)}-u_1^{(t)}-1
\right]_r}
\nonumber\\
&\times&
\prod_{t=1}^N
\prod_{i=3}^{n+1}
\frac{
\displaystyle
\left[u_i^{(t)}-u_2^{(t+1)}+1-\frac{s}{N}\right]_r
\left[u_i^{(t+1)}-u_2^{(t)}+\frac{s}{N}\right]_r}{
\displaystyle
\left[u_i^{(t)}-u_2^{(t)}\right]_r
\left[u_2^{(t)}-u_i^{(t)}-1\right]_r}
\nonumber\\
&\times&
\frac{\displaystyle
\widehat{\vartheta}_\alpha
(u_2^{(1)}|\cdots|u_2^{(N)})
\widehat{\vartheta}_\beta\left(\left.
\sum_{j=1\atop{j \neq 2}}
^{n+1}u_j^{(1)}\right|
\cdots
\left|\sum_{j=1\atop{j\neq 2}}^{n+1}u_j^{(N)}\right.\right)
}{\displaystyle
\widehat{\vartheta}_\gamma
\left(\left.
\sum_{j=1}^{n+1}u_j^{(1)}\right|
\cdots
\left|\sum_{j=1}^{n+1}u_j^{(N)}\right.\right)}.
\end{eqnarray}
Upon specialization
$u_2^{(t)}=u_1^{(t)}+1$, $(1\leqq t \leqq N)$,
we have
$
\frac{[u_1^{(t)}-u_j^{(t+1)}+1-\frac{s}{N}]_r
[u_1^{(t+1)}-u_j^{(t)}+\frac{s}{N}]_r}{
[u_1^{(t)}-u_j^{(t)}]_r[u_j^{(t)}-u_1^{(t)}-1]_r}=
\frac{[u_i^{(t)}-u_2^{(t+1)}+1-\frac{s}{N}]_r
[u_i^{(t+1)}-u_2^{(t)}+\frac{s}{N}]_r}{
[u_i^{(t)}-u_2^{(t)}]_r[u_2^{(t)}-u_i^{(t)}-1]_r}$.
Hence we have
${\rm Res}_{u_2^{(1)}=u_1^{(1)}+1}\cdots
{\rm Res}_{u_2^{(N)}=u_1^{(N)}+1}{\rm LHS}(1,n)=
{\rm Res}_{u_2^{(1)}=u_1^{(1)}+1}\cdots
{\rm Res}_{u_2^{(N)}=u_1^{(N)}+1}{\rm RHS}(1,n)$,
using periodic condition
$
\widehat{\vartheta}_\alpha
(u_1^{(1)}+k|\cdots|u_1^{(N)}+k)=
\widehat{\vartheta}_\alpha
(u_1^{(1)}|\cdots|u_1^{(N)})$.
Both ${\rm LHS}(1,n)$ and
${\rm RHS}(1,n)$ are symmetric with respect to
$u_1^{(t)},u_2^{(t)},\cdots, u_{n+1}^{(t)}$,
we have
\begin{eqnarray}
{\rm Res}_{u_{i_1}^{(1)}=u_{j_1}^{(1)}+1}\cdots
{\rm Res}_{u_{i_N}^{(N)}=u_{j_N}^{(N)}+1}{\rm LHS}(1,n)=
{\rm Res}_{u_{i_1}^{(1)}=u_{j_1}^{(1)}+1}\cdots
{\rm Res}_{u_{i_N}^{(N)}=u_{j_N}^{(N)}+1}{\rm RHS}(1,n),
\nonumber\\
\label{NLeq1}
\end{eqnarray}
for $1\leqq i_t \neq j_t \leqq n+1$ and $1\leqq t \leqq N$.
After taking the residues finitely many times,
every residue relation which comes from 
${\rm LHS}(1,n)={\rm RHS}(1,n)$,
is redued to the above (\ref{NLeq1}).
Hence we have shown the starting relations
$n>m=1$.
For the second, we show the general $n>m \geqq 1$.
We assume the relation ${\rm LHS}(m-1,n-1)=
{\rm RHS}(m-1,n-1)$.
Let us take the residue at 
$u_{1}^{(t)}=u_{2}^{(t)}+1$, $(1\leqq t \leqq N)$.
We have
\begin{eqnarray}
&&{\rm Res}_{u_{2}^{(1)}=u_{1}^{(1)}+1}\cdots
{\rm Res}_{u_{2}^{(N)}=u_{1}^{(N)}+1}({\rm LHS}(m,n)
-{\rm RHS}(m,n))\nonumber\\
&=& 
\prod_{t=1}^N
{\rm Res}_{u_2^{(t)}=u_1^{(t)}+1}
\frac{\left[u_1^{(t)}-u_2^{(t+1)}+1-\frac{s}{N}\right]_r
\left[u_1^{(t+1)}-u_2^{(t)}+\frac{s}{N}\right]_r}
{\left[u_1^{(t)}-u_2^{(t)}\right]_r
\left[u_2^{(t)}-u_1^{(t)}-1\right]_r}
\nonumber\\
&\times&
\prod_{t=1}^N
\prod_{j=3}^{m+n}
\frac{\left[u_1^{(t)}-u_j^{(t+1)}+1-\frac{s}{N}\right]_r
\left[u_1^{(t+1)}-u_j^{(t)}+\frac{s}{N}\right]_r}
{\left[u_1^{(t)}-u_j^{(t)}\right]_r
\left[u_j^{(t)}-u_1^{(t)}-1\right]_r}
\nonumber\\
&\times&
\sum_{L_1 \cup L_1^c=\{3,4,\cdots,n+m\}
\atop{|L_1|=m-1,
|L_1^c|=n-1}}
\sum_{L_2 \cup L_2^c=\{3,4,\cdots,n+m\}
\atop{|L_2|=m-1,
|L_2^c|=n-1}}
\cdots
\sum_{L_N \cup L_N^c=\{3,4,\cdots,n+m\}
\atop{|L_N|=m-1,
|L_N^c|=n-1}}\nonumber\\
&\times&
\frac{\displaystyle
\widehat{\vartheta}_\alpha\left(
\left.\sum_{j \in L_1\cup\{1\}}
u_j^{(1)}
\right|\cdots\left|
\sum_{j \in L_N\cup\{1\}}
u_j^{(N)}
\right.\right)
\widehat{\vartheta}_\beta
\left(
\left.\sum_{j \in L_1^c \cup\{1\}}
u_j^{(1)}
\right|\cdots\left|
\sum_{j \in L_N^c \cup\{1\}}
u_j^{(N)}
\right.
\right)}{
\displaystyle
\widehat{\vartheta}_\gamma
\left(
\left.\sum_{j=1}^{m+n}
u_j^{(1)}
\right|\cdots\left|
\sum_{j=1}^{m+n}
u_j^{(N)}
\right.\right)
}\nonumber\\
&\times&
\left(
\prod_{t=1}^N
\frac{
\displaystyle
\prod_{i \in L_t}
\prod_{j\in L_{t+1}^c}
\left[u_i^{(t)}-u_j^{(t+1)}+1-\frac{s}{N}\right]_r
\prod_{i \in L_{t+1}}
\prod_{j\in L_{t}^c}
\left[u_i^{(t+1)}-u_j^{(t)}+\frac{s}{N}\right]_r
}{
\displaystyle
\prod_{i \in L_t}
\prod_{j \in L_{t+1}^c}
\left[u_i^{(t)}-u_j^{(t)}\right]_r
\left[u_j^{(t)}-u_i^{(t)}-1\right]_r
}
\right.\nonumber\\
&-&\left.
\prod_{t=1}^N
\frac{
\displaystyle
\prod_{i \in L_t^c}
\prod_{j\in L_{t+1}}
\left[u_i^{(t)}-u_j^{(t+1)}+1-\frac{s}{N}\right]_r
\prod_{i \in L_{t+1}^c}
\prod_{j\in L_{t}}
\left[u_i^{(t+1)}-u_j^{(t)}+\frac{s}{N}\right]_r
}{
\displaystyle
\prod_{i \in L_t^c}
\prod_{j \in L_{t+1}}
\left[u_i^{(t)}-u_j^{(t)}\right]_r
\left[u_j^{(t)}-u_i^{(t)}-1\right]_r
}
\right)
=0.\nonumber\\
\end{eqnarray}
We have already used the hypothesis for $(m-1,n-1)$.
Both ${\rm LHS}(m,n)$ and
${\rm RHS}(m,n)$ are symmetric with respect to
$u_1^{(t)},u_2^{(t)},\cdots, u_{m+n}^{(t)}$,
we have
\begin{eqnarray}
{\rm Res}_{u_{i_1}^{(1)}=u_{j_1}^{(1)}+1}\cdots
{\rm Res}_{u_{i_N}^{(N)}=u_{j_N}^{(N)}+1}{\rm LHS}(m,n)=
{\rm Res}_{u_{i_1}^{(1)}=u_{j_1}^{(1)}+1}\cdots
{\rm Res}_{u_{i_N}^{(N)}=u_{j_N}^{(N)}+1}{\rm RHS}(m,n),\nonumber\\\label{NLeq2}
\end{eqnarray}
for $1\leqq i_t \neq j_t \leqq m+n$ and $1\leqq t \leqq N$.
After taking the residues finitely many times,
every residue relation which comes from 
${\rm LHS}(m,n)={\rm RHS}(m,n)$,
is redued to the above (\ref{NLeq2}).
Hence we have shown ${\rm LHS}(m,n)={\rm RHS}(m,n)$
for $n>m\geqq 1$.
~~~Q.E.D.

~\\
Now let us show the commutation relation
$[{\cal G}_m,{\cal G}_n]=0$.

~\\
{\it Proof of Theorem \ref{thm:Nonlocal-Com1}}~~
We show $[{\cal G}_m, {\cal G}_n]=0$ for ${\rm Re}(r)>0$ and $0<{\rm Re}(s)<N$.
Others are shown by similar way.
We use the integral representation of the nonlocal integrals of motion.
In this regime, the integral contour exists in annulus.
Hence we can use the notation $[\cdots]_{1,z_1\cdots z_n}$.
The following operators in the integrand of the nonlocal
integrals of motion satisfies the $S_n$-invariance.
For $\sigma_1,\sigma_2,\cdots,\sigma_N \in S_{m+n}$, we have
\begin{eqnarray}
&&F_1(z_{\sigma_1(1)}^{(1)})\cdots F_1(z_{\sigma_1(m+n)}^{(1)})
F_2(z_{\sigma_2(1)}^{(2)})\cdots F_2(z_{\sigma_2(m+n)}^{(2)})\cdots
F_N(z_{\sigma_N(1)}^{(N)})\cdots F_N(z_{\sigma_N(m+n)}^{(N)})\nonumber\\
&\times&
\prod_{t=1}^N \prod_{1\leqq i<j \leqq m+n}
\left[u_{\sigma_t(i)}^{(t)}-u_{\sigma_t(j)}^{(t)}\right]_r
\left[u_{\sigma_t(j)}^{(t)}-u_{\sigma_t(i)}^{(t)}-1\right]_r\nonumber\\
&=&
F_1(z_1^{(1)})\cdots F_1(z_{m+n}^{(1)})
F_2(z_1^{(2)})\cdots F_2(z_{m+n}^{(2)})\cdots
F_N(z_1^{(N)})\cdots F_N(z_{m+n}^{(N)})\nonumber\\
&\times&
\prod_{t=1}^N \prod_{1\leqq i<j \leqq m+n}
\left[u_i^{(t)}-u_j^{(t)}\right]_r
\left[u_j^{(t)}-u_i^{(t)}-1\right]_r.
\end{eqnarray}
Hence we have
\begin{eqnarray}
{\cal G}_m \cdot {\cal G}_n&=&
\left[
F_1(z_1^{(1)})\cdots F_1(z_{m+n}^{(1)})
F_2(z_1^{(2)})\cdots F_2(z_{m+n}^{(2)})\cdots
F_N(z_1^{(N)})\cdots F_N(z_{m+n}^{(N)})
\right.
\nonumber\\
&\times&
\frac{\displaystyle
\prod_{t=1}^N \prod_{1\leqq i<j \leqq m+n}
\left[u_i^{(t)}-u_j^{(t)}\right]_r
\left[u_j^{(t)}-u_i^{(t)}-1\right]_r
}{\displaystyle
\prod_{t=1}^{N-1}\prod_{i,j=1}^{m+n} 
\left[u_i^{(t)}-u_j^{(t+1)}+1-\frac{s}{N}\right]_r}
\nonumber\\
&\times&
\frac{1}{\displaystyle
\prod_{i=1}^{m}\prod_{j=m+1}^{m+n} 
\left[u_i^{(N)}-u_j^{(1)}+\frac{s}{N}\right]_r
\prod_{i=m+1}^{m+n}
\prod_{j=1}^m
\left[u_j^{(N)}-u_i^{(1)}-\frac{s}{N}+1
\right]_r}\nonumber\\
&\times&\frac{1}{\displaystyle
\prod_{i,j=1}^m
\left[u_i^{(1)}-u_j^{(N)}+\frac{s}{N}\right]_r
\prod_{i,j=m+1}^{m+n}
\left[u_i^{(1)}-u_j^{(N)}+\frac{s}{N}\right]_r
}
\nonumber\\
&\times&\frac{1}{((m+n)!)^N}
\sum_{\sigma_1 \in S_{m+n}}
\cdots
\sum_{\sigma_N \in S_{m+n}}\nonumber\\
&\times&
\vartheta_\alpha(\sum_{j=1}^m
u_{\sigma_1(j)}^{(1)}|\cdots|\sum_{j=1}^m
u_{\sigma_N(j)}^{(N)})
\vartheta_\beta(\sum_{j=m+1}^{m+n}
u_{\sigma_1(j)}^{(1)}|\cdots|\sum_{j=m+1}^{m+n}
u_{\sigma_N(j)}^{(N)})\nonumber\\
&\times&\left.
\prod_{t=1}^N
\prod_{i=1}^m
\prod_{j=m+1}^{m+n}
\frac{\displaystyle
\left[
u_{\sigma_t(i)}^{(t)}-u_{\sigma_{t+1}(j)}^{(t+1)}
+1-\frac{s}{N}\right]_r
\left[
u_{\sigma_{t+1}(i)}^{(t+1)}-
u_{\sigma_{t}(j)}^{(t)}+\frac{s}{N}\right]_r
}{\displaystyle
\left[u_{\sigma_t(i)}^{(t)}-u_{\sigma_{t}(j)}^{(t)}\right]_r
\left[u_{\sigma_t(j)}^{(t)}-u_{\sigma_{t}(i)}^{(t)}-1\right]_r}
\right]
_{1,
z^{(1)}, \cdots, z^{(N)}}.
\end{eqnarray}
Therefore we have the 
following theta function identity as a sufficient
condition of the commutation relations
${\cal G}_m \cdot {\cal G}_n=
{\cal G}_n \cdot {\cal G}_m$.
\begin{eqnarray}
&&\sum_{\sigma_1 \in S_{m+n}}
\sum_{\sigma_2 \in S_{m+n}}
\cdots \sum_{\sigma_N \in S_{m+n}}
{\vartheta}_\alpha
\left(\sum_{j=1}^m u_{\sigma_1(j)}^{(1)}\right|
\left.\sum_{j=1}^m u_{\sigma_2(j)}^{(2)}\right|
\cdots
\left|\sum_{j=1}^m u_{\sigma_N(j)}^{(N)}\right)\nonumber\\
&\times&
{\vartheta}_\beta
\left(\sum_{j=m+1}^{m+n} u_{\sigma_1(j)}^{(1)}\right|
\left.\sum_{j=m+1}^{m+n} u_{\sigma_2(j)}^{(2)}\right|
\cdots
\left|\sum_{j=m+1}^{m+n} u_{\sigma_N(j)}^{(N)}\right)
\nonumber\\
&\times&
\prod_{t=1}^N
\prod_{i=1}^m
\prod_{j=m+1}^{m+n}
\frac{
\left[
u_{\sigma_t(i)}^{(t)}-u_{\sigma_{t+1}(j)}^{(t+1)}
-\frac{s}{N}\right]_r
\left[u_{\sigma_{t+1}(i)}^{(t+1)}-u_{\sigma_{t}(j)}^{(t)}
+\frac{s}{N}\right]_r
}{
\left[
u_{\sigma_{t}(i)}^{(t)}-u_{\sigma_{t}(j)}^{(t)}
\right]_r
\left[
u_{\sigma_t(j)}^{(t)}-u_{\sigma_{t}(i)}^{(t)}-1
\right]_r
}\nonumber
\\
&=&
\sum_{\sigma_1 \in S_{m+n}}
\sum_{\sigma_2 \in S_{m+n}}
\cdots \sum_{\sigma_N \in S_{m+n}}
{\vartheta}_\beta
\left(\sum_{j=1}^n u_{\sigma_1(j)}^{(1)}\right|
\left.\sum_{j=1}^n u_{\sigma_2(j)}^{(2)}\right|
\cdots
\left|\sum_{j=1}^n u_{\sigma_N(j)}^{(N)}\right)
\nonumber\\
&\times&
{\vartheta}_\alpha
\left(\sum_{j=n+1}^{m+n} u_{\sigma_1(j)}^{(1)}\right|
\left.\sum_{j=n+1}^{m+n} u_{\sigma_2(j)}^{(2)}\right|
\cdots
\left|\sum_{j=n+1}^{m+n} u_{\sigma_N(j)}^{(N)}\right)
\nonumber\\
&\times&
\prod_{t=1}^N
\prod_{i=1}^n
\prod_{j=n+1}^{m+n}
\frac{\left[u_{\sigma_t(i)}^{(t)}-u_{\sigma_{t+1}(j)}^{(t+1)}
-\frac{s}{N}\right]_r
\left[
u_{\sigma_{t+1}(j)}^{(t+1)}-u_{\sigma_{t}(i)}^{(t)}
+1-\frac{s}{N}\right]_r
}{
\left[u_{\sigma_t(i)}^{(t)}-u_{\sigma_{t}(j)}^{(t)}\right]_r
\left[u_{\sigma_t(j)}^{(t)}-u_{\sigma_{t}(i)}^{(t)}-1\right]_r
}.
\end{eqnarray}
This is a special case $\nu_{\alpha,t}=\nu_{\beta,t}=0$, $(1\leqq t \leqq N)$
of the theta identity in Proposition
\ref{prop:thetaNonlocal}.
Now we have shown Theorem.~~~Q.E.D.

%%%%%%%%%%%%%%%%%%%%%%%%%%%%%%%%%%%%%%%%%%%%%%%%%%%

\subsection{Proof of $[{\cal I}_m,{\cal G}_n]=0$}

In this section we give
proof of
the commutation relation
$[{\cal I}_m,{\cal G}_n]=0$.
The fundamental operators $\Lambda_j(z)$ and
$F_j(z)$ commute almost everywhere.
\begin{eqnarray}
~[\Lambda_j(z_1),F_j(z_2)]&=&(-x^{r^*}+x^{-r^*})\delta
\left(x^{\frac{2s}{N}j-r}\frac{z_2}{z_1}\right)
{\cal A}_j(x^{-r+\frac{2s}{N}j}z_1),
\nonumber
\\
~[\Lambda_{j+1}(z_1),F_j(z_2)]&=&(x^{r^*}-x^{-r^*})\delta
\left(x^{\frac{2s}{N}j+r}\frac{z_2}{z_1}\right)
{\cal A}_j(x^{r+\frac{2s}{N}j}z_1).\nonumber
\end{eqnarray}
Hence, in order to show the commutation relations,
remaining task for us is to check
whether delta-function factors cancell out or not.
The Dynkin-automorphism invariance $\eta({\cal I}_m)
={\cal I}_m$,
$\eta({\cal G}_m)={\cal G}_m$,
which we will show later,
plays an important role in proof of this commutation
relation $[{\cal I}_m,{\cal G}_n]=0$.

~\\
{\it Proof of Theorem \ref{thm:Nonlocal-Com3}}~~~
For a while we consider upon the regime: 
$0<{\rm Re}(s)<N$, $0<{\rm Re}(r)<1$.
At first we show the simple case,
$[{\cal I}_1, {\cal G}_n]=0$,
for reader's convenience.
Using  Leibnitz-rule of adjoint action 
$[A,BC]=[A,B]C+B[A,C]$ and the invariance 
$\eta({\cal I}_1)={\cal I}_1$,
we have
\begin{eqnarray}
[{\cal I}_1, {\cal G}_n]
&=&(x^{-r^*}-x^{r^*})\sum_{t=1}^{N-1}
\sum_{j=1}^n
\prod_{u=1}^N \prod_{k=1}^n
\int_{\widetilde{C}(t,j)}
\frac{dz_k^{(u)}}{2\pi \sqrt{-1} z_k^{(u)}}
F_1(z_1^{(1)})\cdots \nonumber\\
&\times&\cdots
F_t(z_1^{(t)})\cdots F_t(z_{j-1}^{(t)})
{\cal A}_t(x^{-r+\frac{2s}{N}t}z_j^{(t)})
F_t(z_{j+1}^{(t)})\cdots F_t(z_n^{(t)})
\cdots
F_N(z_n^{(N)})\nonumber\\
&\times&
\frac{\displaystyle
\prod_{t=1}^N \prod_{1\leqq i<j \leqq n}
\left[u_i^{(t)}-u_j^{(t)}\right]_r
\left[u_j^{(t)}-u_i^{(t)}-1\right]_r
}{
\displaystyle
\prod_{t=1}^{N-1}\prod_{i,j=1}^n 
\left[u_i^{(t)}-u_j^{(t+1)}+1-\frac{s}{N}\right]_r
\prod_{i,j=1}^n
\left[u_i^{(1)}-u_j^{(N)}+\frac{s}{N}\right]_r}\nonumber\\
&\times&
\vartheta\left(\sum_{j=1}^n u_j^{(1)}\right|
\left.\sum_{j=1}^n u_j^{(2)}\right|
\cdots
\left|\sum_{j=1}^n u_j^{(N)}\right)\nonumber\\
&+&
(x^{r^*}-x^{-r^*})\sum_{t=1}^{N-1}
\sum_{j=1}^n
\prod_{u=1}^N \prod_{k=1}^n
\int_{\widetilde{C}'(t,j)}
\frac{dz_k^{(u)}}{2\pi \sqrt{-1} z_k^{(u)}}
F_1(z_1^{(1)})\cdots \nonumber\\
&\times&\cdots
F_t(z_1^{(t)})\cdots F_t(z_{j-1}^{(t)})
{\cal A}_t(x^{r+\frac{2s}{N}t}z_j^{(t)})
F_t(z_{j+1}^{(t)})\cdots F_t(z_n^{(t)})
\cdots
F_N(z_n^{(N)})\nonumber\\
&\times&
\frac{\displaystyle
\prod_{t=1}^N \prod_{1\leqq i<j \leqq n}
\left[u_i^{(t)}-u_j^{(t)}\right]_r
\left[u_j^{(t)}-u_i^{(t)}-1\right]_r
}{
\displaystyle
\prod_{t=1}^{N-1}\prod_{i,j=1}^n
\left[u_i^{(t)}-u_j^{(t+1)}+1-\frac{s}{N}\right]_r
\prod_{i,j=1}^n 
\left[u_i^{(1)}-u_j^{(N)}+\frac{s}{N}\right]_r}\nonumber\\
&\times&
\vartheta\left(\sum_{j=1}^n u_j^{(1)}\right|
\left.\sum_{j=1}^n u_j^{(2)}\right|
\cdots
\left|\sum_{j=1}^n u_j^{(N)}\right)
\nonumber
\\
&+&\eta\left(
\sum_{j=1}^n
\prod_{u=1}^N \prod_{k=1}^n
\int_{C}
\frac{dz_k^{(u)}}{2\pi \sqrt{-1} z_k^{(u)}}
\right.
F_2(z_1^{(1)})\cdots F_2(z_n^{(1)})\cdots
F_N(z_1^{(N-1)})\cdots F_N(z_n^{(N-1)})\nonumber\\
&\times&
F_1(z_1^{(N)})\cdots F_1(z_{j-1}^{(N)})[{\cal I}_1, F_1(z_j^{(N)})]
F_1(z_{j+1}^{(N)})\cdots F_1(z_n^{(N)})\nonumber\\
&\times&
\frac{\displaystyle
\prod_{t=1}^N \prod_{1\leqq i<j \leqq n}
\left[u_i^{(t)}-u_j^{(t)}\right]_r
\left[u_j^{(t)}-u_i^{(t)}-1\right]_r
}{
\displaystyle
\prod_{t=1}^{N-1}\prod_{i,j=1}^n
\left[u_i^{(t)}-u_j^{(t+1)}+1-\frac{s}{N}\right]_r
\prod_{i,j=1}^n
\left[u_i^{(1)}-u_j^{(N)}+\frac{s}{N}\right]_r}\nonumber\\
&\times&\left.
\vartheta\left(\sum_{j=1}^n u_j^{(N)}\right|
\left.\sum_{j=1}^n u_j^{(1)}\right|\cdots
\left|\sum_{j=1}^n u_j^{(N-1)}\right)
\right).\label{NLeq3}
\end{eqnarray}
Here we have set
\begin{eqnarray}
\widetilde{C}(t,j)&:&
|x^{4r-2+\frac{2s}{N}t}z_{j+1}^{(t)}|,\cdots,
|x^{4r-2+\frac{2s}{N}t}z_m^{(t)}|<
|z_j^{(t)}|<|x^{-2r+2-\frac{2s}{N}t}z_1^{(t)}|,\cdots,
|x^{-2r+2-\frac{2s}{N}t}z_{j-1}^{(t)}|,
\nonumber\\
&&|x^{2r+2-\frac{2s}{N}}z_k^{(t-1)}|<|z_j^{(t)}|<|
x^{2r-\frac{2s}{N}}z_k^{(t-1)}|,~
(1\leqq k \leqq m),\nonumber\\
&&|x^{\frac{2s}{N}}z_k^{(t+1)}|<|z_j^{(t)}|<
|x^{-2+\frac{2s}{N}}z_k^{(t+1)}|,~
(1\leqq k \leqq m),
\\
\widetilde{C}'(t,j)&:&
|x^{2r-2+\frac{2s}{N}t}z_{j+1}^{(t)}|,\cdots,
|x^{2r-2+\frac{2s}{N}t}z_m^{(t)}|<
|z_j^{(t)}|<|x^{-4r+2-\frac{2s}{N}t}z_1^{(t)}|,\cdots,
|x^{-4r+2-\frac{2s}{N}t}z_{j-1}^{(t)}|,
\nonumber\\
&&|x^{2-\frac{2s}{N}}z_k^{(t-1)}|<|z_j^{(t)}|<|
x^{-\frac{2s}{N}}z_k^{(t-1)}|,~
(1\leqq k \leqq m),\nonumber\\
&&|x^{-2r+\frac{2s}{N}}z_k^{(t+1)}|<|z_j^{(t)}|<
|x^{-2r-2+\frac{2s}{N}}z_k^{(t+1)}|,~
(1\leqq k \leqq m).
\end{eqnarray}
Let us change the variable $z_j^{(t)}\to x^{2r}z_j^{(t)}$
in the first term of (\ref{NLeq3}).
Using the periodicity of the integrand, 
we deform the first term to the second term of (\ref{NLeq3}).
\begin{eqnarray}
&&\prod_{u=1}^N \prod_{k=1}^n
\int_{\widetilde{I}(t,j)}
\frac{dz_k^{(u)}}{2\pi \sqrt{-1} z_k^{(u)}}
F_1(z_1^{(1)})\cdots 
F_t(z_1^{(t)})\cdots F_t(z_{j-1}^{(t)})\nonumber\\
&\times&
{\cal A}_t(x^{-r}z_j^{(t)})
F_t(z_{j+1}^{(t)})\cdots F_t(z_n^{(t)})
\cdots
F_N(z_n^{(N)})\nonumber\\
&\times&
\frac{\displaystyle
\prod_{t=1}^N \prod_{1\leqq i<j \leqq n}
\left[u_i^{(t)}-u_j^{(t)}\right]_r
\left[u_j^{(t)}-u_i^{(t)}-1\right]_r
}{
\displaystyle
\prod_{t=1}^{N-1}\prod_{i,j=1}^n
\left[u_i^{(t)}-u_j^{(t+1)}+1-\frac{s}{N}\right]_r
\prod_{i,j=1}^n
\left[u_i^{(1)}-u_j^{(N)}+\frac{s}{N}\right]_r}
\vartheta\left(\sum_{j=1}^n u_j^{(1)}\right|
\left.\sum_{j=1}^n u_j^{(2)}\right|
\cdots
\left|\sum_{j=1}^n u_j^{(N)}\right)\nonumber\\
&=&
\prod_{u=1}^N \prod_{k=1}^n
\int_{\widetilde{I}'(t,j)}
\frac{dz_k^{(u)}}{2\pi \sqrt{-1} z_k^{(u)}}
F_1(z_1^{(1)})\cdots 
F_t(z_1^{(t)})\cdots F_t(z_{j-1}^{(t)})\nonumber\\
&\times&
{\cal A}_t(x^{r}z_j^{(t)})
F_t(z_{j+1}^{(t)})\cdots F_t(z_n^{(t)})
\cdots
F_N(z_n^{(N)})\nonumber\\
&\times&
\frac{\displaystyle
\prod_{t=1}^N \prod_{1\leqq i<j \leqq n}
\left[u_i^{(t)}-u_j^{(t)}\right]_r
\left[u_j^{(t)}-u_i^{(t)}-1\right]_r
}{
\displaystyle
\prod_{t=1}^{N-1}\prod_{i,j=1}^n
\left[u_i^{(t)}-u_j^{(t+1)}+1-\frac{s}{N}\right]_r
\prod_{i,j=1}^n 
\left[u_i^{(1)}-u_j^{(N)}+\frac{s}{N}\right]_r}
\vartheta\left(\sum_{j=1}^n u_j^{(1)}\right|
\left.\sum_{j=1}^n u_j^{(2)}\right|
\cdots
\left|\sum_{j=1}^n u_j^{(N)}\right).\nonumber\\
\end{eqnarray}
Hence we have
\begin{eqnarray}
&&\eta([{\cal I}_1,{\cal G}_n])\nonumber\\
&=&
(x^{-r^*}-x^{r^*})
\sum_{j=1}^n
\prod_{u=1}^N \prod_{k=1}^n
\int_{C(N,j)}
\frac{dz_k^{(u)}}{2\pi \sqrt{-1} z_k^{(u)}}
F_2(z_1^{(1)})\cdots F_2(z_n^{(1)})\cdots
F_N(z_1^{(N-1)})\cdots F_N(z_n^{(N-1)})\nonumber\\
&\times&
F_1(z_1^{(N)})\cdots F_1(z_{j-1}^{(N)})
{\cal A}_1(x^{-r+\frac{2s}{N}}z_j^{(N)})]
F_1(z_{j+1}^{(N)})\cdots F_1(z_n^{(N)})\nonumber\\
&\times&
\frac{\displaystyle
\prod_{t=1}^N \prod_{1\leqq i<j \leqq n}
\left[u_i^{(t)}-u_j^{(t)}\right]_r
\left[u_j^{(t)}-u_i^{(t)}-1\right]_r
}{
\displaystyle
\prod_{t=1}^{N-1}\prod_{i,j=1}^n
\left[u_i^{(t)}-u_j^{(t+1)}+1-\frac{s}{N}\right]_r
\prod_{i,j=1}^n 
\left[u_i^{(1)}-u_j^{(N)}+\frac{s}{N}\right]_r}\nonumber\\
&\times&
\vartheta\left(\sum_{j=1}^n u_j^{(1)}\right|
\left.\sum_{j=1}^n u_j^{(2)}\right|
\cdots
\left|\sum_{j=1}^n u_j^{(N)}\right)\nonumber\\
&-&(x^{-r^*}-x^{r^*})
\sum_{j=1}^n
\prod_{u=1}^N \prod_{k=1}^n
\int_{C'(N,j)}
\frac{dz_k^{(u)}}{2\pi \sqrt{-1} z_k^{(u)}}
F_2(z_1^{(1)})\cdots F_2(z_n^{(1)})\cdots
F_N(z_1^{(N-1)})\cdots F_N(z_n^{(N-1)})\nonumber\\
&\times&
F_1(z_1^{(N)})\cdots F_1(z_{j-1}^{(N)})
{\cal A}_1(x^{r+\frac{2s}{N}}z_j^{(N)})]
F_1(z_{j+1}^{(N)})\cdots F_1(z_n^{(N)})\nonumber\\
&\times&
\frac{\displaystyle
\prod_{t=1}^N \prod_{1\leqq i<j \leqq n}
\left[u_i^{(t)}-u_j^{(t)}\right]_r
\left[u_j^{(t)}-u_i^{(t)}-1\right]_r
}{
\displaystyle
\prod_{t=1}^{N-1}\prod_{i,j=1}^n
\left[u_i^{(t)}-u_j^{(t+1)}+1-\frac{s}{N}\right]_r
\prod_{i,j=1}^n 
\left[u_i^{(1)}-u_j^{(N)}+\frac{s}{N}\right]_r}\nonumber\\
&\times&
\vartheta\left(\sum_{j=1}^n u_j^{(N)}\right|
\left.\sum_{j=1}^n u_j^{(1)}\right|
\cdots
\left|\sum_{j=1}^n u_j^{(N-1)}\right).
\end{eqnarray}
Here we have set
\begin{eqnarray}
\widetilde{C}(N,j)&:&
|x^{4r-2+\frac{2s}{N}}z_{j+1}^{(N)}|,\cdots,
|x^{4r-2+\frac{2s}{N}}z_m^{(N)}|<
|z_j^{(N)}|<|x^{-2r+2-\frac{2s}{N}}z_1^{(N)}|,\cdots,
|x^{-2r+2-\frac{2s}{N}}z_{j-1}^{(N)}|,
\nonumber\\
&&|x^{2r+2-\frac{2s}{N}}z_k^{(N-1)}|<|z_j^{(N)}|<|
x^{2r-\frac{2s}{N}}z_k^{(N-1)}|,~
(1\leqq k \leqq m),\nonumber\\
&&|x^{\frac{2s}{N}}z_k^{(1)}|<|z_j^{(N)}|<
|x^{-2+\frac{2s}{N}}z_k^{(1)}|,~
(1\leqq k \leqq m),
\\
\widetilde{C}'(N,j)&:&
|x^{2r-2+\frac{2s}{N}}z_{j+1}^{(N-1)}|,\cdots,
|x^{2r-2+\frac{2s}{N}}z_m^{(N-1)}|\nonumber\\
&&<|z_j^{(N)}|<|x^{-4r+2-\frac{2s}{N}}z_1^{(N-1)}|,\cdots,
|x^{-4r+2-\frac{2s}{N}}z_{j-1}^{(N-1)}|,
\nonumber\\
&&|x^{2-\frac{2s}{N}}z_k^{(N-1)}|<|z_j^{(N)}|<|
x^{-\frac{2s}{N}}z_k^{(N-1)}|,~
(1\leqq k \leqq m),\nonumber\\
&&|x^{-2r+\frac{2s}{N}}z_k^{(1)}|<|z_j^{(N)}|<
|x^{-2r-2+\frac{2s}{N}}z_k^{(1)}|,~
(1\leqq k \leqq m).
\end{eqnarray}
Using the priodicity of the integtrand, we have
$[{\cal I}_1,{\cal G}_n]=0$.
For the second, we consider the commutation
relation $[{\cal I}_m,{\cal G}_n]=0$.
By using Leibnitz-rule of adjoint action
and the invariance $\eta({\cal I}_m)={\cal I}_m$,
we have
\begin{eqnarray}
&&[{\cal I}_m,{\cal G}_n]\nonumber\\
&=&(x^{-r^*}-x^{r^*})\sum_{t=1}^{N-1}
\sum_{i=1}^m \sum_{j=1}^n
\prod_{u=1}^N \prod_{k=1}^n
\int_{\widetilde{C}(t,j)}
\frac{dz_k^{(u)}}{2\pi \sqrt{-1}z_k^{(u)}}
\vartheta\left(\sum_{j=1}^n u_j^{(1)}\right|
\left.\sum_{j=1}^n u_j^{(2)}\right|
\cdots
\left|\sum_{j=1}^n u_j^{(N)}\right)
\nonumber\\
&\times&
\frac{\displaystyle
\prod_{t=1}^N \prod_{1\leqq i<j \leqq n}
\left[u_i^{(t)}-u_j^{(t)}\right]_r
\left[u_j^{(t)}-u_i^{(t)}-1\right]_r
}{
\displaystyle
\prod_{t=1}^{N-1}\prod_{i,j=1}^n
\left[u_i^{(t)}-u_j^{(t+1)}+1-\frac{s}{N}\right]_r
\prod_{i,j=1}^n
\left[u_i^{(1)}-u_j^{(N)}+\frac{s}{N}\right]_r}\nonumber\\
&\times&
\left(
\prod_{k=1}^m
\int_{I\cap 
\left\{\left|\frac{x^{-r+\frac{2s}{N}t}
z_j^{(t)}}{w_i}\right|<1\right\}}
\frac{dw_k}{2\pi \sqrt{-1}w_k}
\frac{1}{\left(1-\frac{x^{-r+\frac{2s}{N}t}
z_j^{(t)}}{w_i}\right)}\right.\nonumber\\
&&\left.+
\prod_{k=1}^m
\int_{I\cap \left\{\left|
\frac{x^{-r+\frac{2s}{N}t}z_j^{(t)}}{w_i}\right|>1\right\}}
\frac{dw_k}{2\pi \sqrt{-1}w_k}
\frac{\frac{x^{r-\frac{2s}{N}t}w_i}{z_j^{(t)}}}
{\left(1-\frac{x^{r-\frac{2s}{N}t}w_i}{z_j^{(t)}}\right)}
\right)\nonumber\\
&\times&
\prod_{1\leqq k<l \leqq m}h(v_k-v_l)
T_1(w_1)\cdots T_1(w_{i-1})
F_1(z_1^{(t)})\cdots
F_t(z_{j-1}^{(t)})\nonumber\\
&\times&
{\cal A}_t(x^{-r+\frac{2s}{N}t}
z_j^{(t)})F_t(z_{j+1}^{(t)})\cdots
F_N(z_n^{(t)})T_1(w_{i+1})\cdots
T_1(w_m)\nonumber\\
&-&
(x^{-r^*}-x^{r^*})\sum_{t=1}^{N-1}
\sum_{i=1}^m \sum_{j=1}^n
\prod_{u=1}^N \prod_{k=1}^n
\int_{\widetilde{C}'(t,j)}
\frac{dz_k^{(u)}}{2\pi \sqrt{-1}z_k^{(u)}}
\vartheta\left(\sum_{j=1}^n u_j^{(1)}\right|
\left.\sum_{j=1}^n u_j^{(2)}\right|
\cdots
\left|\sum_{j=1}^n u_j^{(N)}\right)
\nonumber\\
&\times&
\frac{\displaystyle
\prod_{t=1}^N \prod_{1\leqq i<j \leqq n}
\left[u_i^{(t)}-u_j^{(t)}\right]_r
\left[u_j^{(t)}-u_i^{(t)}-1\right]_r
}{
\displaystyle
\prod_{t=1}^{N-1}\prod_{i,j=1}^n
\left[u_i^{(t)}-u_j^{(t+1)}+1-\frac{s}{N}\right]_r
\prod_{i,j=1}^n
\left[u_i^{(1)}-u_j^{(N)}+\frac{s}{N}\right]_r}\nonumber\\
&\times&
\left(
\prod_{k=1}^m
\int_{I\cap 
\left\{\left|\frac{x^{r+\frac{2s}{N}t}
z_j^{(t)}}{w_i}\right|<1\right\}}
\frac{dw_k}{2\pi \sqrt{-1}w_k}
\frac{1}{\left(1-\frac{x^{r+\frac{2s}{N}t}z_j^{(t)}}{w_i}\right)}\right.\nonumber\\
&&\left.+
\prod_{k=1}^m
\int_{I\cap \left\{\left|\frac{x^{r+\frac{2s}{N}t}
z_j^{(t)}}{w_i}\right|>1\right\}}
\frac{dw_k}{2\pi \sqrt{-1}w_k}
\frac{\frac{x^{-r-\frac{2s}{N}t}w_i}{z_j^{(t)}}}
{\left(1-\frac{x^{-r-\frac{2s}{N}t}w_i}{z_j^{(t)}}\right)}
\right)\nonumber\\
&\times&
\prod_{1\leqq k<l \leqq m}h(v_k-v_l)
T_1(w_1)\cdots T_1(w_{i-1})
F_1(z_1^{(t)})\cdots
F_t(z_{j-1}^{(t)})\nonumber\\
&\times&
{\cal A}_t(x^{r+\frac{2s}{N}}z_j^{(t)})F_t(z_{j+1}^{(t)})\cdots
F_N(z_n^{(t)})T_1(w_{i+1})\cdots
T_1(w_m)\nonumber\\
&+&\eta
\left(
\sum_{i=1}^m \sum_{j=1}^n
\prod_{u=1}^N \prod_{k=1}^n
\int_{C}
\frac{dz_k^{(u)}}{2\pi \sqrt{-1}z_k^{(u)}}
\vartheta\left(\sum_{j=1}^n u_j^{(N)}\right|
\left.\sum_{j=1}^n u_j^{(1)}\right|
\cdots
\left|\sum_{j=1}^n u_j^{(N-1)}\right)\right.
\nonumber\\
&\times&
\frac{\displaystyle
\prod_{t=1}^N \prod_{1\leqq i<j \leqq m}
\left[u_i^{(t)}-u_j^{(t)}\right]_r
\left[u_j^{(t)}-u_i^{(t)}-1\right]_r
}{
\displaystyle
\prod_{t=1}^{N-1}\prod_{i,j=1}^n 
\left[u_i^{(t)}-u_j^{(t+1)}+1-\frac{s}{N}\right]_r
\prod_{i,j=1}^n
\left[u_i^{(1)}-u_j^{(N)}+\frac{s}{N}\right]_r}\nonumber\\
&\times&
\prod_{k=1}^m
\int_I \frac{dw_k}{2\pi \sqrt{-1}w_k}
\prod_{1\leqq k < l\leqq m}h(v_k-v_l)
T_1(w_1)\cdots T_1(w_{i-1})
\nonumber\\
&\times&
F_2(z_1^{(1)})\cdots F_2(z_n^{(1)})\cdots
F_N(z_1^{(N-1)})\cdots F_N(z_n^{(N-1)})
F_1(z_1^{(N)})\cdots F_1(z_{j-1}^{(N)})\nonumber\\
&\times&
[T_1(w_i),F_1(z_j^{(N)})]
F_1(z_{j+1}^{(N)})\cdots F_1(z_n^{(N)})
T_1(w_{i+1})\cdots T_1(w_m)).\label{NLeq4}
\end{eqnarray}
Here $\widetilde{C}(t,j)$, 
$\widetilde{C}'(t,j)$ are given as the same manner 
in proof of $[{\cal I}_1,{\cal G}_n]=0$.
Let us change the variable $z_j^{(t)}
\to x^{2r}z_j^{(t)}$ in the first term,
upon the conditions
$0<{\rm Re}(s)<N$, $0<{\rm Re}(r)<1$.
Using periodicity of the integrands, we have
the cancellation of the first and the second terms
of (\ref{NLeq4}).
\begin{eqnarray}
&&\prod_{u=1}^N \prod_{k=1}^n
\int_{\widetilde{C}(t,j)}
\frac{dz_k^{(u)}}{2\pi \sqrt{-1}z_k^{(u)}}
\vartheta\left(\sum_{j=1}^n u_j^{(1)}\right|
\left.\sum_{j=1}^n u_j^{(2)}\right|
\cdots
\left|\sum_{j=1}^n u_j^{(N)}\right)
\nonumber\\
&\times&
\frac{\displaystyle
\prod_{t=1}^N \prod_{1\leqq i<j \leqq n}
\left[u_i^{(t)}-u_j^{(t)}\right]_r
\left[u_j^{(t)}-u_i^{(t)}-1\right]_r
}{
\displaystyle
\prod_{t=1}^{N-1}\prod_{i,j=1}^n
\left[u_i^{(t)}-u_j^{(t+1)}+1-\frac{s}{N}\right]_r
\prod_{i,j=1}^n
\left[u_i^{(1)}-u_j^{(N)}+\frac{s}{N}\right]_r}\nonumber\\
&\times&
\left(
\prod_{k=1}^m
\int_{I\cap \left\{\left|\frac{x^{-r+\frac{2s}{N}t}
z_j^{(t)}}{w_i}\right|<1\right\}}
\frac{dw_k}{2\pi \sqrt{-1}w_k}
\frac{1}{\left(1-\frac{x^{-r+\frac{2s}{N}t}z_j^{(t)}}{w_i}\right)}\right.\nonumber\\
&&\left.
+
\prod_{k=1}^m
\int_{I\cap \left\{\left|\frac{x^{-r+\frac{2s}{N}t}
z_j^{(t)}}{w_i}\right|>1\right\}}
\frac{dw_k}{2\pi \sqrt{-1}w_k}
\frac{\frac{x^{r-\frac{2s}{N}t}w_i}{z_j^{(t)}}}
{\left(1-\frac{x^{r-\frac{2s}{N}t}w_i}{z_j^{(t)}}\right)}
\right)\nonumber\\
&\times&
\prod_{1\leqq k<l \leqq m}h(v_k-v_l)
T_1(w_1)\cdots T_1(w_{i-1})
F_1(z_1^{(t)})\cdots
F_t(z_{j-1}^{(t)})\nonumber\\
&\times&
{\cal A}_t(x^{-r+\frac{2s}{N}t}z_j^{(t)})F_t(z_{j+1}^{(t)})\cdots
F_N(z_n^{(t)})T_1(w_{i+1})\cdots
T_1(w_m)\nonumber\\
&=&
\prod_{u=1}^N \prod_{k=1}^n
\int_{\widetilde{C}(t,j)}
\frac{dz_k^{(u)}}{2\pi \sqrt{-1}z_k^{(u)}}
\vartheta\left(\sum_{j=1}^m u_j^{(1)}\right|
\left.\sum_{j=1}^n u_j^{(2)}\right|
\cdots
\left|\sum_{j=1}^n u_j^{(N)}\right)
\nonumber\\
&\times&
\frac{\displaystyle
\prod_{t=1}^N \prod_{1\leqq i<j \leqq m}
\left[u_i^{(t)}-u_j^{(t)}\right]_r
\left[u_j^{(t)}-u_i^{(t)}-1\right]_r
}{
\displaystyle
\prod_{t=1}^{N-1}\prod_{i,j=1}^n 
\left[u_i^{(t)}-u_j^{(t+1)}+1-\frac{s}{N}\right]_r
\prod_{i,j=1}^n
\left[u_i^{(1)}-u_j^{(N)}+\frac{s}{N}\right]_r}\nonumber\\
&\times&
\left(
\prod_{k=1}^m
\int_{I\cap \left\{\left|\frac{x^{r+\frac{2s}{N}t}
z_j^{(t)}}{w_i}\right|<1\right\}}
\frac{dw_k}{2\pi \sqrt{-1}w_k}
\frac{1}{\left(1-\frac{x^{r+\frac{2s}{N}t}z_j^{(t)}}{w_i}\right)}
\right.\nonumber\\
&&\left.+
\prod_{k=1}^m
\int_{I\cap \left\{\left|\frac{x^{r+\frac{2s}{N}t}
z_j^{(t)}}{w_i}\right|>1\right\}}
\frac{dw_k}{2\pi \sqrt{-1}w_k}
\frac{\frac{x^{-r-\frac{2s}{N}t}w_i}{z_j^{(t)}}}
{\left(1-\frac{x^{-r-\frac{2s}{N}t}w_i}{z_j^{(t)}}\right)}
\right)\nonumber\\
&\times&
\prod_{1\leqq k<l \leqq m}h(v_k-v_l)
T_1(w_1)\cdots T_1(w_{i-1})
F_1(z_1^{(t)})\cdots
F_t(z_{j-1}^{(t)})\nonumber\\
&\times&
{\cal A}_t(x^{r+\frac{2s}{N}t}z_j^{(t)})F_t(z_{j+1}^{(t)})\cdots
F_N(z_n^{(t)})T_1(w_{i+1})\cdots
T_1(w_m).
\end{eqnarray}
Hence we have
\begin{eqnarray}
&&\eta([{\cal I}_m,{\cal G}_n])\nonumber\\
&=&(x^{-r^*}-x^{r^*})
\sum_{i=1}^m \sum_{j=1}^n
\prod_{u=1}^N \prod_{k=1}^n
\int_{\widetilde{C}(N,j)}
\frac{dz_k^{(u)}}{2\pi \sqrt{-1}z_k^{(u)}}
\vartheta\left(\sum_{j=1}^n u_j^{(N)}\right|
\left.\sum_{j=1}^n u_j^{(1)}\right|
\cdots
\left|\sum_{j=1}^n u_j^{(N-1)}\right)
\nonumber\\
&\times&
\frac{\displaystyle
\prod_{t=1}^N \prod_{1\leqq i<j \leqq n}
\left[u_i^{(t)}-u_j^{(t)}\right]_r
\left[u_j^{(t)}-u_i^{(t)}-1\right]_r
}{
\displaystyle
\prod_{t=1}^{N-1}\prod_{i,j=1}^n
\left[u_i^{(t)}-u_j^{(t+1)}+1-\frac{s}{N}\right]_r
\prod_{i,j=1}^n
\left[u_i^{(1)}-u_j^{(N)}+\frac{s}{N}\right]_r}\nonumber\\
&\times&
\left(
\prod_{k=1}^m
\int_{I\cap 
\left\{\left|\frac{x^{-r+\frac{2s}{N}}
z_j^{(N)}}{w_i}\right|<1\right\}}
\frac{dw_k}{2\pi \sqrt{-1}w_k}
\frac{1}{\left(1-\frac{x^{-r+\frac{2s}{N}}z_j^{(N)}}{w_i}\right)}\right.\nonumber\\
&&\left.+
\prod_{k=1}^m
\int_{I\cap 
\left\{\left|\frac{x^{-r+\frac{2s}{N}}
z_j^{(N)}}{w_i}\right|>1\right\}}
\frac{dw_k}{2\pi \sqrt{-1}w_k}
\frac{\frac{x^{r-\frac{2s}{N}}w_i}{z_j^{(N)}}}
{\left(1-\frac{x^{r-\frac{2s}{N}}w_i}{z_j^{(N)}}\right)}
\right)\nonumber\\
&\times&
\prod_{1\leqq k < l\leqq m}h(v_k-v_l)
T_1(w_1)\cdots T_1(w_{i-1})
\nonumber\\
&\times&
F_2(z_1^{(1)})\cdots F_2(z_n^{(1)})\cdots
F_N(z_1^{(N-1)})\cdots F_N(z_n^{(N-1)})
F_1(z_1^{(N)})\cdots F_1(z_{j-1}^{(N)})\nonumber\\
&\times&
{\cal A}_1(x^{-r+\frac{2s}{N}}z_j^{(N)})
F_1(z_{j+1}^{(N)})\cdots F_1(z_n^{(N)})
T_1(w_{i+1})\cdots T_1(w_m)\nonumber\\
&-&(x^{-r^*}-x^{r^*})
\sum_{i=1}^m \sum_{j=1}^n
\prod_{u=1}^N \prod_{k=1}^n
\int_{\widetilde{C}'(N,j)}
\frac{dz_k^{(u)}}{2\pi \sqrt{-1}z_k^{(u)}}
\vartheta\left(\sum_{j=1}^n u_j^{(N)}\right|
\left.\sum_{j=1}^n u_j^{(1)}\right|
\cdots
\left|\sum_{j=1}^n u_j^{(N-1)}\right)
\nonumber\\
&\times&
\frac{\displaystyle
\prod_{t=1}^N \prod_{1\leqq i<j \leqq n}
\left[u_i^{(t)}-u_j^{(t)}\right]_r
\left[u_j^{(t)}-u_i^{(t)}-1\right]_r
}{
\displaystyle
\prod_{t=1}^{N-1}\prod_{i,j=1}^n
\left[u_i^{(t)}-u_j^{(t+1)}+1-\frac{s}{N}\right]_r
\prod_{i,j=1}^n
\left[u_i^{(1)}-u_j^{(N)}+\frac{s}{N}\right]_r}\nonumber\\
&\times&
\left(
\prod_{k=1}^m
\int_{I\cap 
\left\{\left|\frac{x^{r+\frac{2s}{N}}z_j^{(N)}}{w_i}\right|
<1\right\}}
\frac{dw_k}{2\pi \sqrt{-1}w_k}
\frac{1}{\left(1-\frac{x^{r+\frac{2s}{N}}z_j^{(N)}}{w_i}\right)}\right.\nonumber\\
&&\left.+
\prod_{k=1}^m
\int_{I\cap \left\{\left|\frac{x^{r+\frac{2s}{N}}
z_j^{(N)}}{w_i}\right|
>1\right\}}
\frac{dw_k}{2\pi \sqrt{-1}w_k}
\frac{\frac{x^{-r-\frac{2s}{N}}w_i}{z_j^{(N)}}}
{\left(1-\frac{x^{-r-\frac{2s}{N}}w_i}{z_j^{(N)}}\right)}
\right)\nonumber\\
&\times&
\prod_{1\leqq k < l\leqq m}h(v_k-v_l)
T_1(w_1)\cdots T_1(w_{i-1})
\nonumber\\
&\times&
F_2(z_1^{(1)})\cdots F_2(z_n^{(1)})\cdots
F_N(z_1^{(N-1)})\cdots F_N(z_n^{(N-1)})
F_1(z_1^{(N)})\cdots F_1(z_{j-1}^{(N)})\nonumber\\
&\times&
{\cal A}_1(x^{r+\frac{2s}{N}}z_j^{(N)})
F_1(z_{j+1}^{(N)})\cdots F_1(z_n^{(N)})
T_1(w_{i+1})\cdots T_1(w_m),
\end{eqnarray}
where $\widetilde{C}(N,j)$,
$\widetilde{C}'(N,j)$ are given as the same manner as the case
of $[{\cal I}_1,{\cal G}_n]=0$.
Changing variable $z_j^{(N)}\to x^{2r}z_j^{(N)}$, we have
the commutation relation
$[{\cal I}_m,{\cal G}_n]=0$.
Other commutation relations :
$[{\cal I}_m,{\cal G}_n^*]=
[{\cal I}_m^*,{\cal G}_n]=[{\cal I}_m^*,{\cal G}_n^*]=0$
are shown by similar way. 
We omit details.
~~~Q.E.D.

%%%%%%%%%%%%%%%%%%%%%%%%%%%%%%%%%%%%%%%%%%%%%%%%%

\subsection{Proof of $[{\cal G}_m,{\cal G}_n^*]=0$}

In this section we give
proof of
the commutation relation
$[{\cal I}_m,{\cal G}_n]=0$.
The fundamental operators $E_j(z)$ and
$F_j(z)$ commute almost everywhere.
\begin{eqnarray}
[E_j(z_1),F_j(z_2)]
&=&\frac{1}{x-x^{-1}}
(\delta(xz_2/z_1){H}_j(x^{r}z_2)
-\delta(xz_1/z_2){H}_j(x^{-r}z_2)).\nonumber
\end{eqnarray}
Hence, in order to show the commutation relations,
remaining task for us is to check
whether delta-function factors cancell out or not.

~\\
{\it Proof of Theorem \ref{thm:Nonlocal-Com2}}~~~
We consider the regime ${\rm Re}(r)>0$
and ${\rm Re}(r^*)<0$.
At first we consider the simple case :
$[{\cal G}_1,{\cal G}_1^*]=0$.
Using Leibnitz-rule of adjoint action and 
the commutation relations of screening currents
$E_j(z), F_j(z)$,
we have
\begin{eqnarray}
[{\cal G}_1^*,{\cal G}_1]
&=&\sum_{t=1}^N
\prod_{q=1}^N 
\prod_{p=1
\atop{p \neq t}}^N
\oint_{C^{(t)}} 
\frac{dz^{(q)}}{2\pi\sqrt{-1}z^{(q)}}
\frac{dw^{(p)}}{2\pi\sqrt{-1}w^{(p)}}
B^{(t)}(x^{r}z^{(t)};\{z^{(q)}\}_{1\leqq q \leqq N},
\{w^{(q)}\}_{1\leqq q\leqq N
\atop{q \neq t}})\nonumber\\
&-&
\sum_{t=1}^N
\prod_{q=1}^N \prod_{p=1
\atop{p \neq t}}^N
\oint_{\widetilde{C}^{(t)}} \frac{dz^{(q)}}{2\pi\sqrt{-1}
z^{(q)}}
\frac{dw^{(p)}}{2\pi\sqrt{-1}
w^{(p)}}
B^{(t)}(x^{-r}z^{(t)};\{z^{(q)}\}_{1\leqq q \leqq N},
\{w^{(q)}\}_{1\leqq q\leqq N
\atop{q \neq t}}).\nonumber\\
\end{eqnarray}            
Here we have set
\begin{eqnarray}
&&B^{(t)}(z;\{z^{(q)}\}_{1\leqq q \leqq N},
\{w^{(q)}\}_{1\leqq q\leqq N
\atop{q \neq t}})\nonumber\\
&=&\frac{1}{x-x^{-1}}F_1(z^{(1)})\cdots F_{t-1}(z^{(t-1)})
E_1(w^{(1)})\cdots E_{t-1}(w^{(t-1)})\nonumber\\
&\times&
H_t(z)
E_{t+1}(w^{(t+1)})\cdots E_{N}(w^{(N)})
F_{t+1}(z^{(t+1)})\cdots F_{N}(z^{(N)})\nonumber\\
&\times&
\frac{\vartheta(u^{(1)}|\cdots|u^{(N)})}
{\prod_{q=1}^{N-1}[u^{(q)}-u^{(q+1)}+1-\frac{s}{N}]_r
[u^{(1)}-u^{(N)}+\frac{s}{N}]_r }\nonumber\\
&\times&
\left.\frac{\vartheta^*(v^{(1)}|\cdots|v^{(N)})}{
\prod_{q=1}^{N-1}[v^{(q)}-v^{(q+1)}+1-\frac{s}{N}]_{-r^*}
[v^{(1)}-v^{(N)}+\frac{s}
{N}]_{-r^*}
}\right|_{v^{(t)}=u-\frac{r^*}{2}}
\end{eqnarray}
Here the contours $C^{(t)}$ and 
$\widetilde{C}^{(t)}$ are characterized by
\begin{eqnarray}
C^{(t)}&:&
|x^{2-\frac{2s}{N}}z^{(t-1)}|,
|x^{\frac{2s}{N}}z^{(t+1)}|<|z^{(t)}|<
|x^{-2r-\frac{2s}{N}}z^{(t-1)}|,
|x^{-2-2r+\frac{2s}{N}}z^{(t+1)}|,\nonumber\\
&&
|x^{-2r+3-\frac{2s}{N}}w^{(t-1)}|,
|x^{-2r+1+\frac{2s}{N}}w^{(t+1)}|<|z^{(t)}|<
|x^{-1-\frac{2s}{N}}w^{(t-1)}|,
|x^{-3+\frac{2s}{N}}w^{(t+1)}|,\nonumber\\
&&
|x^{\frac{2s}{N}}z^{(q+1)}|<|z^{(q)}|<|x^{-2+\frac{2s}{N}}z^{(q+1)}|,~
(1\leqq q (\neq t,t-1)\leqq N),\nonumber\\
&&
|x^{\frac{2s}{N}}w^{(q+1)}|<|w^{(q)}|<|x^{-2+\frac{2s}{N}}w^{(q+1)}|,~
(1\leqq q (\neq t,t-1)\leqq N),
\nonumber\\
\\
\widetilde{C}^{(t)}&:&
|x^{2r+2-\frac{2s}{N}}z^{(t-1)}|,
|x^{2r+\frac{2s}{N}}z^{(t+1)}|<|z^{(t)}|<
|x^{-\frac{2s}{N}}z^{(t-1)}|,
|x^{-2+\frac{2s}{N}}z^{(t+1)}|,\nonumber\\
&&
|x^{3-\frac{2s}{N}}w^{(t-1)}|,
|x^{1+\frac{2s}{N}}w^{(t+1)}|<|z^{(t)}|<
|x^{2r-1-\frac{2s}{N}}w^{(t-1)}|,
|x^{2r-3+\frac{2s}{N}}w^{(t+1)}|,\nonumber\\
&&|x^{\frac{2s}{N}}z^{(q+1)}|<|z^{(q)}|<|x^{-2+\frac{2s}{N}}z^{(q+1)}|,~
(1\leqq q (\neq t,t-1)\leqq N),\nonumber\\
&&
|x^{\frac{2s}{N}}w^{(q+1)}|<|w^{(q)}|<|x^{-2+\frac{2s}{N}}w^{(q+1)}|,~
(1\leqq q (\neq t,t-1)\leqq N).\nonumber\\
\end{eqnarray}
Let us change the variable $z^{(t)}\to x^{2r}z^{(t)}$ of
the integrand
$B^{(t)}(x^{-r}z^{(t)};\{z^{(q)}\}_{1\leqq q \leqq N},
\{w^{(q)}\}_{1\leqq q\leqq N
\atop{q \neq t}})$ and the contour $C^{(t)}$.
Using the periodic condition of theta function
$[u+r]_r=-[u]_r$,
we have
$B^{(t)}(x^{r}z^{(t)};\{z^{(q)}\}_{1\leqq q \leqq N},
\{w^{(q)}\}_{1\leqq q\leqq N
\atop{q \neq t}})$ and $\widetilde{C}^{(t)}$.
Hence we have
\begin{eqnarray}
&&
\prod_{q=1}^N 
\prod_{p=1
\atop{p \neq t}}^N
\oint_{C^{(t)}} 
\frac{dz^{(q)}}{2\pi\sqrt{-1}z^{(q)}}
\frac{dw^{(p)}}{2\pi\sqrt{-1}w^{(p)}}
B^{(t)}(x^{r}z^{(t)};\{z^{(q)}\}_{1\leqq q \leqq N},
\{w^{(q)}\}_{1\leqq q\leqq N
\atop{q \neq t}})\nonumber\\
&=&
\prod_{q=1}^N \prod_{p=1
\atop{p \neq t}}^N
\oint_{\widetilde{C}^{(t)}} \frac{dz^{(q)}}{2\pi\sqrt{-1}
z^{(q)}}
\frac{dw^{(p)}}{2\pi\sqrt{-1}
w^{(p)}}
B^{(t)}(x^{-r}z^{(t)};\{z^{(q)}\}_{1\leqq q \leqq N},
\{w^{(q)}\}_{1\leqq q\leqq N
\atop{q \neq t}}).
\end{eqnarray}            
Therefore we have $[{\cal G}_1^*,{\cal G}_1]=0$.
Generalization to generic parameter $0<{\rm Re}(r)<1$ and $s \in {\mathbb C}$
should be understood as analytic continuation.
For the second, we consider 
$[{\cal G}_m^*,{\cal G}_n]=0$.
Using Leibnitz-rule of adjoint action and 
the commutation relations of screening currents
$E_j(z), F_j(z)$,
we have
\begin{eqnarray}
[{\cal G}_m^*,{\cal G}_n]
&=&\sum_{t=1}^N
\sum_{i=1}^n
\sum_{j=1}^m
\prod_{q=1}^N 
\prod_{k=1}^n
\prod_{p=1}^N
\prod_{l=1
\atop{l\neq t}}^m
\oint_{C^{(t)}_{i,j}} 
\frac{dz^{(q)_k}}{2\pi\sqrt{-1}z^{(q)}_k}
\frac{dw^{(p)_l}}{2\pi\sqrt{-1}w^{(q)}_l}\nonumber\\
&\times&
B^{(t)}_{i,j}
\left(x^{-r}z^{(t)}_i;
\{z^{(q)}_j\}_{1\leqq q \leqq N
\atop{1\leqq k \leqq n}},
\{w^{(q)}\}_{1\leqq q\leqq N
\atop{1\leqq l \neq j \leqq m}}\right)\nonumber\\
&-&
\sum_{t=1}^N
\sum_{i=1}^n
\sum_{j=1}^m
\prod_{q=1}^N 
\prod_{k=1}^n
\prod_{p=1}^N
\prod_{l=1
\atop{l\neq t}}^m
\oint_{C^{(t)}_{i,j}} 
\frac{dz^{(q)_k}}{2\pi\sqrt{-1}z^{(q)}_k}
\frac{dw^{(p)_l}}{2\pi\sqrt{-1}w^{(q)}_l}\nonumber\\
&\times&B^{(t)}_{i,j}
\left(x^{r}z^{(t)}_i;
\{z^{(q)}_j\}_{1\leqq q \leqq N
\atop{1\leqq k \leqq n}},
\{w^{(q)}\}_{1\leqq q\leqq N
\atop{1\leqq l \neq j \leqq m}}\right).
\end{eqnarray}            
Here we have set
\begin{eqnarray}
&&B^{(t)}_{i,j}
\left(z;
\{z^{(q)}_j\}_{1\leqq q \leqq N
\atop{1\leqq k \leqq n}},
\{w^{(q)}\}_{1\leqq q\leqq N
\atop{1\leqq l \neq j \leqq m}}\right)\nonumber\\
&=&
\frac{1}{x-x^{-1}}
F_1(z_1^{(1)})\cdots F_1(z_n^{(1)})
E_1(w_1^{(1)})\cdots E_1(w_m^{(1)})\cdots\nonumber\\
&\times&
\cdots F_t(z_1^{(t)})\cdots F_t(z_{i-1}^{(t)})
E_t(w_1^{(t)})\cdots E_t(w_{j-1}^{(t)})\nonumber\\
&\times&H_t(z)
E_t(w_{j+1}^{(t)})\cdots E_t(w_{m}^{(t)})
F_t(z_{i+1}^{(t)})\cdots F_t(z_{n}^{(t)})\cdots\nonumber\\
&\times&
\cdots
E_1(w_1^{(N)})\cdots E_1(w_m^{(N)})
F_1(z_1^{(N)})\cdots F_1(z_n^{(N)})\nonumber\\
&\times&
\frac{\displaystyle
\prod_{q=1}^N \prod_{1\leqq k<l \leqq n}
\left[u_k^{(q)}-u_l^{(q)}\right]_r
\left[u_l^{(q)}-u_k^{(q)}-1\right]_r
}{
\displaystyle
\prod_{q=1}^{N-1}\prod_{k,l=1}^n
\left[u_k^{(q)}-u_l^{(q+1)}+1-\frac{s}{N}\right]_r
\prod_{k,l=1}^n 
\left[u_k^{(1)}-u_l^{(N)}+\frac{s}{N}\right]_r}
\nonumber\\
&\times&
\frac{\displaystyle
\prod_{q=1}^N \prod_{1\leqq k<l \leqq m}
\left[v_k^{(q)}-v_l^{(q)}\right]_{-r^*}
\left[v_l^{(q)}-v_k^{(q)}-1\right]_{-r^*}
}{
\displaystyle
\prod_{q=1}^{N-1}\prod_{k,l=1}^n
\left[v_k^{(q)}-v_l^{(q+1)}+1-\frac{s}{N}\right]_{-r^*}
\prod_{k,l=1}^n 
\left[v_k^{(1)}-v_l^{(N)}+\frac{s}{N}\right]_{-r^*}}
\nonumber\\
&\times&\left.
\vartheta\left(\sum_{k=1}^n u_k^{(1)}\right|
\left.\sum_{k=1}^n u_k^{(2)}\right|
\cdots
\left|\sum_{k=1}^n u_k^{(N)}\right)
\vartheta\left(\sum_{k=1}^m 
v_k^{(1)}\right|
\left.\sum_{k=1}^m v_k^{(2)}\right|
\cdots
\left|\sum_{k=1}^m v_k^{(N)}\right)\right|_{v_j=u-\frac{r^*}{2}}.
\nonumber\\
\end{eqnarray}
Here the contours $C^{(t)}_{i,j}$ and $C^{(t)}_{i,j}$ are characterized by
\begin{eqnarray}
C^{(t)}_{i,j}&:&
|x^{2-\frac{2s}{N}}z^{(t-1)}_k|,
|x^{\frac{2s}{N}}z^{(t+1)}_k|<|z^{(t)}_i|<
|x^{-2r-\frac{2s}{N}}z^{(t-1)}_k|,
|x^{-2-2r+\frac{2s}{N}}z^{(t+1)}_k|,(1\leqq k \leqq n)\nonumber\\
&&
|x^{-2r+3-\frac{2s}{N}}w^{(t-1)}_l|,
|x^{-2r+1+\frac{2s}{N}}w^{(t+1)}_l|<|z^{(t)}|<
|x^{-1-\frac{2s}{N}}w^{(t-1)}_i|,
|x^{-3+\frac{2s}{N}}w^{(t+1)}_i|,(1\leqq l \leqq m),\nonumber\\
&&|x^{\frac{2s}{N}}z^{(q+1)}_l|<|z^{(q)}_k|<|x^{-2+\frac{2s}{N}}z^{(q+1)}_l|,
(1\leqq q \leqq N; (q,k)\neq (t,i), (q,l)\neq (t-1,i)),\nonumber\\
&&
|x^{\frac{2s}{N}}w^{(q+1)}_l|<|w^{(q)}_k|<
|x^{-2+\frac{2s}{N}}w^{(q+1)}_l|,
(1\leqq q \leqq N; (q,k)\neq (t,j), (q,l)\neq (t-1,j)),
\nonumber\\
\\
\widetilde{C}^{(t)}&:&
|x^{2r+2-\frac{2s}{N}}z^{(t-1)}_k|,
|x^{2r+\frac{2s}{N}}z^{(t+1)}_k|<|z^{(t)}_i|<
|x^{-\frac{2s}{N}}z^{(t-1)}_k|,
|x^{-2+\frac{2s}{N}}z^{(t+1)}_k|,(1\leqq k \leqq n),
\nonumber\\
&&
|x^{3-\frac{2s}{N}}w^{(t-1)}_l|,
|x^{1+\frac{2s}{N}}w^{(t+1)}_l|<|z^{(t)}_i|<
|x^{2r-1-\frac{2s}{N}}w^{(t-1)}_l|,
|x^{2r-3+\frac{2s}{N}}w^{(t+1)}_l|,(1\leqq l \leqq m),\nonumber\\
&&|x^{\frac{2s}{N}}z^{(q+1)}_l|<|z^{(q)}_k|<|x^{-2+\frac{2s}{N}}z^{(q+1)}_l|,
(1\leqq q \leqq N; (q,k)\neq (t,i), (q,l)\neq (t-1,i)),\nonumber\\
&&
|x^{\frac{2s}{N}}w^{(q+1)}_l|<|w^{(q)}_k|<
|x^{-2+\frac{2s}{N}}w^{(q+1)}_l|,
(1\leqq q \leqq N; (q,k)\neq (t,j), (q,l)\neq (t-1,j)).\nonumber\\
\end{eqnarray}
Let us change the variable $z^{(t)}_i\to x^{2r}z^{(t)}_i$ of
the integrand
$B^{(t)}(x^{-r}z^{(t)}_i;
\{z^{(q)}_k\}_{1\leqq q \leqq N
\atop{1\leqq k \leqq n}},
\{w^{(q)}_k\}_{1\leqq q\leqq N
\atop{1\leqq k \neq j \leqq m}})$ and the contour $C^{(t)}_{i,j}$.
Using periodic condition of theta function
$[u+r]_r=-[u]_r$,
we have
$B^{(t)}(x^{r}z^{(t)}_i;
\{z^{(q)}_k\}_{1\leqq q \leqq N
\atop{1\leqq k \leqq n}},
\{w^{(q)}_k\}_{1\leqq q\leqq N
\atop{1\leqq k \neq j \leqq m}})$ and the contour 
$\widetilde{C}^{(t)}_{i,j}$.
Hence we have
\begin{eqnarray}
&&\prod_{q=1}^N 
\prod_{k=1}^n
\prod_{p=1}^N
\prod_{l=1
\atop{l\neq t}}^m
\oint_{C^{(t)}_{i,j}} 
\frac{dz^{(q)_k}}{2\pi\sqrt{-1}z^{(q)}_k}
\frac{dw^{(p)_l}}{2\pi\sqrt{-1}w^{(q)}_l}
B^{(t)}_{i,j}
\left(x^{-r}z^{(t)}_i;
\{z^{(q)}_j\}_{1\leqq q \leqq N
\atop{1\leqq k \leqq n}},
\{w^{(q)}\}_{1\leqq q\leqq N
\atop{1\leqq l \neq j \leqq m}}\right)\nonumber\\
&=&
\prod_{q=1}^N 
\prod_{k=1}^n
\prod_{p=1}^N
\prod_{l=1
\atop{l\neq t}}^m
\oint_{C^{(t)}_{i,j}} 
\frac{dz^{(q)_k}}{2\pi\sqrt{-1}z^{(q)}_k}
\frac{dw^{(p)_l}}{2\pi\sqrt{-1}w^{(q)}_l}B^{(t)}_{i,j}
\left(x^{r}z^{(t)}_i;
\{z^{(q)}_j\}_{1\leqq q \leqq N
\atop{1\leqq k \leqq n}},
\{w^{(q)}\}_{1\leqq q\leqq N
\atop{1\leqq l \neq j \leqq m}}\right).\nonumber\\
\end{eqnarray}            
Therefore we have shown the commutation relation
$[{\cal G}_m^*,{\cal G}_n]=0$.
Generalization to generic parameter $0<{\rm Re}(r)<1$ and $s \in {\mathbb C}$
should be understood as analytic continuation.~~~
Q.E.D.

%%%%%%%%%%%%%%%%%%%%%%%%%%%%%%%%%%%%
%%%%%%%%%%%%%%%%%%%%%%%%%%%%%%%%%%%%
\section{Dynkin-Automorphism Invariance}

In this section we consider
the Dynkin-automorphism invariance of
the integrals of motion.

\subsection{Dynkin-Automorphism Invariance}

The integrals of motion are invariant under
the action of the Dynkin-automorphism.

\begin{thm}~~~The local integrals of motion
${\cal I}_n$, ${\cal I}_n^*$ 
are invariant under the action of Dynkin-automorphism
$\eta$
\begin{eqnarray}
\eta({\cal I}_n)={\cal I}_n,~~
\eta({\cal I}_n^*)={\cal I}_n^*,~~(n \in {\mathbb N}).
\end{eqnarray}
\label{thm:DynkinLocal}
\end{thm}

\begin{thm}~~~The nonlocal integrals of motion
${\cal G}_n$, ${\cal G}_n^*$
are invariant ynder the action of Dynkin-automorphism
$\eta$
\begin{eqnarray}
\eta({\cal G}_n)={\cal G}_n,~~
\eta({\cal G}_n^*)={\cal G}_n^*,~~(n \in {\mathbb N}).
\end{eqnarray}
\label{thm:DynkinNonlocal}
\end{thm}
This theorem plays an important
role in proof of the commutation relation
$[{\cal I}_m,{\cal G}_n]=0$.

%%%%%%%%%%%%%%%%%%%%%%%%%%%%%%%%%%%%%%%%%%%%%%%%%%%%%%%%
%%%%%%%%%%%%%%%%%%%%%%%%%%%%%%%%%%%%%%%%%%%%%%%%%%%%%%%%

\subsection{Proof of Dynkin-Automorphism Invariance
$\eta({\cal I}_n)={\cal I}_n$}

In this section we show
Dynkin-Automorphism Invariance
$\eta({\cal I}_n)={\cal I}_n$,
by using Laurent series formulae
${\cal I}_n=[\prod_{j<k}s(z_k/z_j){\cal O}_n(z_1,\cdots,z_n)]$.
We have $\eta^N=id$.
Let us set the functions
$h_{J,K}^{\eta^p, \eta^q}(z)$ for $0\leqq J \leqq K \leqq N$,
$0\leqq p \leqq J-1$, $0\leqq q \leqq K-1$, 
\begin{eqnarray}
h_{J,K}^{\eta^p, \eta^q}(z)&=&
\prod_{j=1}^{J-p} \prod_{k=1}^{K-q}
h_{11}(x^{-K+J+2(k-j)+\frac{2s}{N}(q-p)}z)
\prod_{j=J-p+1}^J \prod_{k=K-q+1}^K 
h_{11}(x^{-K+J+2(k-j)+\frac{2s}{N}(q-p)}z)\nonumber\\
&\times&
\prod_{j=1}^{J-p} \prod_{k=K-q+1}^{K}
h_{11}(x^{-K+J+2(k-j)+\frac{2s}{N}(q-p)-2s}z)
\prod_{j=J-p+1}^J \prod_{k=1}^{K-q}
h_{11}(x^{-K+J+2(k-j)+\frac{2s}{N}(q-p)+2s}z).\nonumber\\
\end{eqnarray}
Here we have set $h_{1,1}(z)=h(u)$ for $z=x^{2u}$.
We use notataion  $h_{J,K}^{\eta^0,\eta^0}(z)=
h_{J,K}^{id,id}(z)=
h_{J,K}(z)$.

~\\
{\it Proof of Theorem \ref{thm:DynkinLocal}}~~
Let us study from
the invariance of
${\cal I}_2=[s(z_2/z_1){\cal O}_2(z_1,z_2)]_{1,z_1z_2}$.
The action of $\eta$ is given by
\begin{eqnarray}
&&\eta
\left([h_{1,1}(z_2/z_1)
T_1(z_1)T_1(z_2)]_{1,z_1,z_2}\right)\nonumber\\
&=&[h_{1,1}(z_2/z_1)T_1(z_1)T_1(z_2)]_{1,z_1,z_2}
+[h_{1,1}(z_2/z_1)(\Lambda_1(x^{-s}z_1)-
\Lambda_1(x^sz_1))\sum_{j=2}^N
\Lambda_j(x^sz_2)]_{1,z_1,z_2}
\nonumber\\
&+&[h_{1,1}(z_2/z_1)
\sum_{j=2}^N \Lambda_j(x^sz_1)
(\Lambda_1(x^{-s}z_2)-\Lambda_1(x^sz_2))]_{1,z_1,z_2}.
\end{eqnarray}
By using the relation
\begin{eqnarray}
&&h_{1,1}(z)-h_{1,1}(x^{2s}z)=
c_{11}(\delta(x^2z)-\delta(x^{2r-2+2s}z)),\\
&&c_{11}=-\frac{(x^2;x^{2s})_\infty
(x^{2r-2};x^{2s})_\infty
(x^{2s-2};x^{2s})_\infty
(x^{2s-2r+2};x^{2s})_\infty
}{
(x^{2r-4};x^{2s})_\infty
(x^{2s};x^{2s})_\infty
(x^{2s};x^{2s})_\infty
(x^{2s-2r+4};x^{2s})_\infty},\nonumber
\end{eqnarray}
we have
\begin{eqnarray}
&&[h_{1,1}(z_2/z_1)\Lambda_1(x^{-s}z_1)
\sum_{j=2}^N \Lambda_j(x^sz_2)]_{1,z_1 z_2}=
[h_{1,1}(z_2/z_1)\Lambda_1(z_1)
\sum_{j=2}^N \Lambda_j(z_2)]_{1,z_1 z_2}
\nonumber\\
&&+c s_{11}(x^{-2})
[\delta(x^2z_2/z_1)
:\Lambda_1(x^{-2s}z_1)
\sum_{j=2}^N \Lambda_j(x^{-2}z_1)
:
]_{1,z_1 z_2},\\
&&[h_{1,1}(z_2/z_1)
\sum_{j=2}^N \Lambda_j(x^sz_1)
\Lambda_1(x^{-s}z_2)]_{1,z_1 z_2}=
[h_{1,1}(z_2/z_1)
\sum_{j=2}^N \Lambda_j(z_1)
\Lambda_1(z_2)
]_{1,z_1 z_2}
\nonumber\\
&&-c s_{11}(x^{-2})
[\delta(x^{2-2s}z_2/z_1)
:\Lambda_1(x^{-2}z_1)
\sum_{j=2}^N \Lambda_j(z_1)
:
]_{1,z_1 z_2},
\end{eqnarray}
Here we have used
\begin{eqnarray}
&&\delta(x^{2r-2+2s}z_2/z_1)
\Lambda_1(x^{-s}z_1)\Lambda_j(x^sz_2)=
\delta(x^{2r-2}z_2/z_1)
\Lambda_j(x^{s}z_1)\Lambda_1(x^{-s}z_2)=0.
\end{eqnarray}
Summing up every terms, we have
\begin{eqnarray}
&&\eta
\left([h_{1,1}(z_2/z_1)T_1(z_1)T_1(z_2)]_{1,z_1,z_2}\right)
=[h_{1,1}(z_2/z_1)T_1(z_1)T_1(z_2)]_{1,z_1,z_2}\nonumber\\
&+&c s(x^{-2})[\delta(x^2z_2/z_1)
\eta(T_2(x^{-1}z_1))]_{1,z_1z_2}
-c s(x^{-2})
[\delta(x^2z_2/z_1)T_2(x^{-1}z_1)]_{1,z_1z_2}.
\end{eqnarray}
Summing up every termes of
$\eta([s(z_2/z_1){\cal O}_2(z_1,z_2)]_{1,z_1z_2})$,
we conclude $\eta({\cal I}_2)={\cal I}_2$.
Next we study $\eta({\cal I}_3)={\cal I}_3$.
We use weakly sense equation for
the basic operator $\Lambda_j(z)$
\begin{eqnarray}
&&g_{1,1}\left(\frac{z_2}{z_1}\right)\Lambda_j(z_1)\Lambda_i(z_2)
-g_{1,1}\left(\frac{z_1}{z_2}\right)
\Lambda_i(z_2)\Lambda_j(z_1)\nonumber\\
&&\sim c \delta\left(\frac{x^2z_2}{z_1}\right)
:\Lambda_i(z_2)\Lambda_j(z_1):, (i<j).
\end{eqnarray}
We have
\begin{eqnarray}
&&\eta\left
(\left[\prod_{1\leqq j<k \leqq 3}h_{1,1}(z_k/z_j)
T_1(z_1)T_1(z_2)T_1(z_3)\right]_{1,z_1z_2z_3}\right)\nonumber\\
&&=
\left[\prod_{1\leqq j<k \leqq 3}h_{1,1}(z_k/z_j)
T_1(z_1)T_1(z_2)T_1(z_3)\right.%\right]_{1,z_1z_2z_3}
\nonumber\\
&&-c s(x^{-2}) h_{1,2}(x^{-1}z_2/z_1)
T_1(z_1)T_2(x^{-1}z_2)\delta(x^2z_3/z_2)\nonumber\\
&&-c s(x^{-2}) h_{1,2}(x^{-1}z_1/z_2)
T_1(z_2)T_2(x^{-1}z_1)\delta(x^2z_3/z_1)\nonumber\\
&&-c s(x^{-2})h_{1,2}(x^{-1}z_1/z_3)
T_1(z_3)T_2(x^{-1}z_1)\delta(x^2z_2/z_1)\nonumber\\
%%%%%%%%%%%&&%%%%%%%%%%%%%%%%%%%%%%%%%%%%%%%%%%%%%%%%%%%%
&&+c s(x^{-2})h_{1,2}^{id,\eta}
(x^{-1}z_2/z_1)
T_1(z_1)\eta(T_2(x^{-1}z_2))
\delta(x^2z_3/z_2)\nonumber\\
&&+c s(x^{-2}) h_{1,2}^{id,\eta}(x^{-1}z_1/z_2)
T_1(z_2)\eta(T_2(x^{-1}z_1))
\delta(x^2z_3/z_1)\nonumber\\
&&+c s(x^{-2}) h_{1,2}^{id,\eta}(x^{-1}z_1/z_3)
T_1(z_3)\eta(T_2(x^{-1}z_1))
\delta(x^2z_2/z_1)\nonumber\\
&&+c^2 s(x^{-2})^2 s(x^{-4}) \Delta(x^3)
\delta(x^2z_1/z_2)\delta(x^2z_3/z_1)
\eta^2(T_3(z_1))\nonumber\\
&&+c^2 s(x^{-2})^2 s(x^{-4}) \Delta(x^3)
\delta(x^2z_1/z_3)
\delta(x^2z_2/z_1)
\eta^2(T_3(z_1))\nonumber\\
&&+c^2 s(x^{-2})^2 s(x^{-4})\Delta(x^3)
\delta(x^2z_2/z_1)
\delta(x^2z_3/z_2)
\eta^2(T_3(z_2))\\
&&+c^2 s(x^{-2})^2 s(x^{-4}) \Delta(x^3)
(\delta(x^2z_1/z_2)\delta(x^2z_3/z_1)+
\delta(x^2z_1/z_3)\delta(x^2z_2/z_1))T_3(z_1)
\nonumber\\
&&-c^2 s(x^{-2})^2 s(x^{-4}) \Delta(x^3)
(\delta(x^2z_1/z_2)\delta(x^2z_3/z_1)+
\delta(x^2z_1/z_3)\delta(x^2z_2/z_1))
\eta(T_3(z_1))]_{1,z_1z_2z_3}.\nonumber
\end{eqnarray}
%%%%%%%%%%%%%%%%%%%%%%%%%%%%%%%%%%%%%%
\begin{eqnarray}
&&\eta([c s(x^{-2}) h_{1,2}(x^{-1}z_2/z_1)
T_1(z_1)T_2(x^{-1}z_2)\delta(x^2z_3/z_2)
]_{1,z_1z_2z_3})\nonumber\\
&&=
[c s(x^{-2}) h_{1,2}^{id,\eta}(x^{-1}z_2/z_1)
T_1(z_1)\eta(T_2(x^{-1}z_2))\delta(x^2z_3/z_2)\nonumber\\
&&+c^2 s(x^{-2})^2 s(x^{-4})
\Delta(x^3)\delta(x^2z_3/z_2)\delta(x^2z_2/z_1)\eta^2(T_3(z_2))]_{1,z_1z_2z_3},
\\
%%%%%%%%%%%%%%%%%%%%%%%%%%%%%%%
&&\eta([\delta(x^2z_1/z_2)
\delta(x^2z_3/z_1)T_3(z_1)]_{1,z_1z_2z_3})=[\delta(x^2z_1/z_2)
\delta(x^2z_3/z_1)\eta(T_3(z_1))]_{1,z_1z_2z_3}.\nonumber\\
\end{eqnarray}
Summing up every term of
$\eta([s(z_2/z_1)s(z_3/z_1)s(z_3/z_2){\cal O}_3
(z_1,z_2,z_3)]_{1,z_1z_2z_3})$,
we have $\eta({\cal I}_3)={\cal I}_3$. 
%%%%%%%%%%%%%%%%%%%%%%%%%%%%%%%%%%%%%%%%%%%%%
%%%%%%%%%%%%%%%%%%%%%%%%%%%%%%%%%%%%%%%%%%%%%
We consider the case of general ${\cal I}_n$.
The action of $\eta$ is given  by
\begin{eqnarray}
&&\eta\left(\left[\prod_{1\leqq j<k \leqq n}h(z_k/z_j)
T_1(z_1)T_1(z_2)T_1(z_3)\cdots T_1(z_n)
\right]_{1,z_1\cdots z_n}\right)=
\sum_{
\alpha_1,\alpha_2,\cdots,\alpha_N \geqq 0
\atop{\alpha_1+2\alpha_2+\cdots+N\alpha_N=n}}\nonumber\\
&\times&
\sum_{\alpha_1^{(1)}=\alpha_1
\atop{
\alpha_2^{(1)},\alpha_2^{(2)}\geqq 0,~
\alpha_2^{(1)}+\alpha_2^{(2)}=\alpha_2
\atop{
\alpha_3^{(1)},\alpha_3^{(2)},\alpha_3^{(3)}\geqq 0,~
\alpha_3^{(1)}+\alpha_3^{(2)}+\alpha_3^{(3)}=\alpha_3
\atop{\cdots
\atop{
\alpha_N^{(1)},\alpha_N^{(2)},\cdots,
\alpha_N^{(N)}\geqq 0,~~
\alpha_N^{(1)}+\alpha_N^{(2)}+\cdots+
\alpha_N^{(N)}=\alpha_N
}}}}}
\sum_{\{A_j^{(t,q)}\}_{1\leqq q \leqq t \leqq N,~
1\leqq j \leqq \alpha_t^{(q)}}
\atop{
A_j^{(t,q)}\subset
\{1,2,\cdots,n\},~|A_j^{(t,q)}|=t,~
\cup_{t=1}^N \cup_{q=1}^t
\cup_{j=1}^{\alpha_t^{(q)}}
A_j^{(t,q)}
\atop{
Min(A_1^{(t,q)})<Min(A_2^{(t,q)})<\cdots
<Min(A_{\alpha_s^{(t)}}~^{(t,q)})}}}
\nonumber\\
&\times&
\sum_{\widetilde{A_j^{(3,3)}}\subset A_j^{(3,3)},~
|\widetilde{A_j^{(3,3)}}|=2
\atop{
\widetilde{A_j^{(4,3)}}\subset A_j^{(4,3)},~
\widetilde{A_j^{(4,4)}}\subset A_j^{(4,4)},~
|\widetilde{A_j^{(4,3)}}|=3,~
|\widetilde{A_j^{(4,4)}}|=2
\atop{\cdots
\atop{
\widetilde{A_j^{(N,3)}}\subset A_j^{(N,3)},~
\widetilde{A_j^{(N,4)}}\subset A_j^{(N,4)},
\cdots,\widetilde{A_j^{(N,N)}}\subset A_j^{(N,N)},~
|\widetilde{A_j^{(N,3)}}|=N-2,~
|\widetilde{A_j^{(N,4)}}|=N-3,~
\cdots,
|\widetilde{A_j^{(N,N)}}|=2
}}}}\nonumber\\
&\times&
(-1)^{\sum_{t=1}^{[\frac{N}{2}]}\sum_{u=1}^t \alpha_{2t}^{(2u-1)}
+\sum_{t=1}^{[\frac{N-1}{2}]}\sum_{u=1}^t
\alpha_{2t+1}^{(2u)}}
\prod_{t=2}^N\left(
c^{t-1}\prod_{u=1}^{t-2}
\Delta(x^{2u+1})^{t-u-1}
\prod_{u=1}^{t-1}s(x^{-2u})^{t-u}
\right)^{\alpha_t}\nonumber\\
&\times&
\left[
\prod_{t=1}^N
\prod_{q=1}^{t}
\prod_{j<k
\atop{j,k \in A_{Min}^{(t,q)}}}h_{t,t}^{\eta^{q-1},\eta^{q-1}}(z_k/z_j)
\right.\nonumber\\
&\times&
\prod_{1\leqq t<u \leqq N}
\prod_{q=1}^{t}\prod_{p=1}^{u}
\prod_{j \in A_{Min}^{(t,q)}}
\prod_{k \in A_{Min}^{(u,p)}}
h_{t,u}^{\eta^{q-1},\eta^{p-1}}(x^{u-t-2[\frac{u}{2}]+2[\frac{t}{2}]}
z_k/z_j)
\nonumber\\
&\times&
\left\{
\prod_{\longrightarrow
\atop{
j\in A_{Min}^{(1,1)}}}T_1(z_j)
\right\}
\left\{
\prod_{\longrightarrow
\atop{
j\in A_{Min}^{(2,1)}}}T_2(x^{-1}z_j)
\prod_{\longrightarrow
\atop{
j\in A_{Min}^{(2,2)}}}\eta(T_2(x^{-1}z_j))\right\}
\cdots
\nonumber\\
&\times&
\left\{\prod_{
\longrightarrow
\atop{
j\in A_{Min}^{(N,1)}}}
T_N(x^{-1+N-2[\frac{N}{2}]}z_j)
\cdots
\prod_{\longrightarrow
\atop{j\in A_{Min}^{(N,N)}}}
\eta^{N-1}(T_N(x^{-1+N-2[\frac{N}{2}]}z_j))
\right\}
\nonumber\\
&\times&
\prod_{q=1}^2
\prod_{j=1}^{\alpha_2^{(q)}}
\delta
\left(\frac{x^2z_{j_2}}{z_{j_1}}\right)
\prod_{t=3}^N
\prod_{q=1}^2
\prod_{j=1
\atop{
j_1=A_{j,1}^{(t,q)}
\atop{
\cdots
\atop{
j_t=A_{j,t}^{(t,q)}
}
}}}^{\alpha_t^{(q)}}
\sum_{\sigma \in S_{t}
\atop{\sigma(1)=1}}
\prod_{u=1\atop{u\neq [\frac{t}{2}]+1}}^t
\delta
\left(\frac{x^2z_{j_{\sigma(u+1)}}}{z_{j_{\sigma(u)}}}\right)
\nonumber\\
%%%%%%%%%%%%%%%%%%%%%%%%%%%%%%%%%%%%%%%%%
&\times&
\prod_{t=3}^N
\prod_{q=3}^{t}
\prod_{j=1
\atop{j_1=\widetilde{A_{j,1}^{(t,q)}}
\atop{\cdots
\atop{
j_{t-q+2}=\widetilde{A_{j,t-q+2}^{(t,q)}}}}}}^{\alpha_t^{(q)}}
\prod_{k_1,k_2,\cdots,k_{q-2}\in 
A_j^{(t,q)}-\widetilde{A_j^{(t,q)}}
\atop{k_1<k_2<\cdots <k_{q-2}}}
\sum_{\sigma \in S_{t-q+2}
\atop{\sigma(1)=1}}
\prod_{u=1
\atop{u\neq [\frac{t-q}{2}]+2}}^{t-q+2}
\delta
\left(\frac{x^2z_{j_{\sigma(u+1)}}}
{z_{j_{\sigma(u)}}}\right)
\sum_{\tau \in S_{q-2}}
\nonumber\\
%%%%%%%%%%%%%%%%%%%%%%%%%%%%%%%%%%%%%%%%%%%%%%%%%%%%%
&\times&
\prod_{
t-q\in 2{\mathbb Z}}
\delta\left(
\frac{x^2z_{k_{\tau([\frac{q+1}{2}])}}}
{z_{j_{\sigma([\frac{t-q}{2}]+2)}}}\right)
\delta
\left(
\frac{x^2z_{j_{\sigma([\frac{t-q}{2}]+3)}}}
{z_{k_{\tau(1)}}}\right)
\prod_{u=[\frac{q+3}{2}]}^{q-2}
\delta\left(\frac{x^2z_{k_{\tau(u)}}}{
z_{k_{\tau(u-1)}}}\right)
\prod_{u=2}^{[\frac{q-1}{2}]}
\delta\left(
\frac{x^2z_{k_{\tau(u-1)}}}{
z_{k_{\tau(u)}}}\right)\nonumber\\
&\times&
%%%%%%%%%%%%%%%%%%%%%%%%%%%%%%%%%
\prod_{t-q+1 \in 2{\mathbb Z}}
\delta\left(\frac{x^2z_{k_{\tau(1)}}}{z_{j_{\sigma([\frac{t-q}{2}]+2)}}}\right)
\delta\left(\frac{x^2z_{j_{\sigma([\frac{t-q}{2}]+3)}}}{z_{
k_{\tau([\frac{q+1}{2}])}}}\right)
\prod_{u=2}^{[\frac{q-1}{2}]}
\delta\left(\frac{x^2z_{k_{\tau(u)}}}{
z_{k_{\tau(u-1)}}}\right)
\prod_{u=[\frac{q+3}{2}]}^{q-2}
\delta\left(
\frac{x^2z_{k_{\tau(u-1)}}}{
z_{k_{\tau(u)}}}\right).\nonumber\\
\label{NLeq5}
\end{eqnarray}
%%%%%%%%%%%%%%%%%%%%%%%%%%%%%%%
Here we have set
$A_{Min}^{(t,q)}=\{Min(A_1^{(t,q)}),Min(A_2^{(t,q)}),
\cdots,Min(A_{\alpha_t^{(q,t)}}^{(t,q)})
\}$ for $q=0,1$,
and have set
$A_{Min}^{(t,q)}=\{Min(\widetilde{A_1^{(t,q)})},
\widetilde{Min(A_2^{(t,q)})},
\cdots,Min(\widetilde{A_{\alpha_t^{(q,t)}}^{(t,q)}})
\}$ for $q=2,\cdots,t$.
Here we have set
$A_{j,1}^{(u,t)},
A_{j,2}^{(u,t)},\cdots,
A_{j,u}^{(u,t)}$ such that
$A_j^{(u,t)}=\{
A_{j,1}^{(u,t)}<A_{j,2}^{(u,t)}<
\cdots<A_{j,u}^{(u,t)}\}$,
and have set
$\widetilde{A_{j,1}^{(u,t)}},
\widetilde{A_{j,2}^{(u,t)}},\cdots,
\widetilde{A_{j,N+2-t}^{(u,t)}}$ such that
$A_j^{(u,t)}=\{
\widetilde{A_{j,1}^{(u,t)}}<
\widetilde{A_{j,2}^{(u,t)}}<
\cdots<
\widetilde{A_{j,N+2-t}^{(u,t)}}\}$.
%%%%%%%%%%%%%%%%%%%%%%%%%%%%%%%%%%%%%%%%%%%%%%%%%%

We give the action of $\eta$ for more general case.
We prepare notations.
Let us set
$\beta_1,\beta_2,\cdots,\beta_N \geqq 0$ 
such that
$\beta_1+2\beta_2+3\beta_3+\cdots+N\beta_N=n$.
Let us set subset 
$B_j^{(t)}
\subset
\{1,2,\cdots,n \}$,
$(1\leqq t \leqq N, 1\leqq j \leqq \beta_t)$
such that
$|B_j^{(t)}|=t,~~
\cup_{t=1}^N \cup_{j=1}^{\alpha_t}
B_j^{(t)}=\{1,2,\cdots,n
\}$
and
$Min(B_1^{(t)})<Min(B_2^{(t)})<\cdots<Min(B_{\alpha_t}^{(t)})$.
Let us set
the index $B_{j,k}^{(t)}=j_k$
for
$B_j^{(t)}=
\{j_1,j_2,\cdots, j_t|j_1<j_2<\cdots<j_t\}$,
$(1\leqq t \leqq N, 1\leqq j \leqq \alpha_t)$,
and
$B_{Min}^{(t)}=\{B_{1,1}^{(t)},B_{2,1}^{(t)},
\cdots, B_{\alpha_t,1}^{(t)}\}$.
The action of $\eta$ is given as following.
\begin{eqnarray}
&&
\eta
\left(
\left[
\prod_{\longrightarrow
\atop{j \in B_{Min}^{(1)}}}T_1(z_j)
\prod_{\longrightarrow
\atop{j \in B_{Min}^{(2)}}}T_2(x^{-1}z_j)
\cdots
%\prod_{\longrightarrow
%\atop{j \in B_{Min}^{(t)}}}T_t(x^{-1+t-2[\frac{t}{2}]}z_j)
%\cdots
\prod_{\longrightarrow
\atop{j \in B_{Min}^{(N)}}}T_N(x^{-1+N-2[\frac{N}{2}]}z_j)\right.
\right.
\nonumber\\
&\times&
\prod_{t=1}^N
\left((-c)^{t-1}
\prod_{u=1}^{t-1}
\Delta(x^{2u+1})^{t-u-1}\right)^{\beta_t}
\prod_{t=2}^N
\prod_{j=1
\atop{j_1=B_{j,1}^{(t)}
\atop{\cdots
\atop{j_t=B_{j,t}^{(t)}}
}}}^{\beta_t}
\sum_{\sigma \in S_t
\atop{\sigma(1)=1}}
\prod_{u=1
\atop{u \neq [\frac{t}{2}]+1}}^t
\delta\left(\frac{x^2z_{j_{\sigma(u+1)}}}{z_{j_{\sigma(u)}}}\right)
\nonumber\\
&\times&
\left.
\left.\prod_{t=1}^N \prod_{j<k
\atop{j,k \in B_{Min}^{(t)}}}g_{t,t}\left(\frac{z_k}{z_j}\right)
\prod_{1\leqq t<u \leqq N}\prod_{j\in B_{Min}^{(t)}
\atop{k\in B_{Min}^{(u)}}}g_{t,u}\left(
x^{u-t-2[\frac{u}{2}]+2[\frac{t}{2}]}\frac{z_k}{z_j}
\right)\right]_{1,z_1\cdots z_n}\right)
\nonumber\\
&=&
\sum_{
\alpha_1,\alpha_2,\cdots,\alpha_N \geqq 0
\atop{\alpha_1+2\alpha_2+\cdots+N\alpha_N=n}}
\sum_{\alpha_1^{(1)}=\alpha_1
\atop{
\alpha_2^{(1)},\alpha_2^{(2)}\geqq 0,~
\alpha_2^{(1)}+\alpha_2^{(2)}=\alpha_2
\atop{
\alpha_3^{(1)},\alpha_3^{(2)},\alpha_3^{(3)}\geqq 0,~
\alpha_3^{(1)}+\alpha_3^{(2)}+\alpha_3^{(3)}=\alpha_3
\atop{\cdots
\atop{
\alpha_N^{(1)},\alpha_N^{(2)},\cdots,
\alpha_N^{(N)}\geqq 0,~~
\alpha_N^{(1)}+\alpha_N^{(2)}+\cdots+
\alpha_N^{(N)}=\alpha_N
}}}}}
\sum_{\{A_j^{(t,q)}\}_{1\leqq q \leqq t \leqq N,~
1\leqq j \leqq \alpha_t^{(q)}}
\atop{
A_j^{(t,q)}\subset
\{1,2,\cdots,n\},~|A_j^{(t,q)}|=t,~
\cup_{t=1}^N \cup_{q=1}^t
\cup_{j=1}^{\alpha_t^{(q)}}
A_j^{(t,q)}
\atop{
Min(A_1^{(t,q)})<Min(A_2^{(t,q)})<\cdots
<Min(A_{\alpha_s^{(t)}}~^{(t,q)})}}}
\nonumber\\
&\times&
\sum_{\widetilde{A_j^{(3,3)}}\subset A_j^{(3,3)},~
|\widetilde{A_j^{(3,3)}}|=2
\atop{
\widetilde{A_j^{(4,3)}}\subset A_j^{(4,3)},~
\widetilde{A_j^{(4,4)}}\subset A_j^{(4,4)},~
|\widetilde{A_j^{(4,3)}}|=3,~
|\widetilde{A_j^{(4,4)}}|=2
\atop{\cdots
\atop{
\widetilde{A_j^{(N,3)}}\subset A_j^{(N,3)},~
\widetilde{A_j^{(N,4)}}\subset A_j^{(N,4)},
\cdots,\widetilde{A_j^{(N,N)}}\subset A_j^{(N,N)},~
|\widetilde{A_j^{(N,3)}}|=N-2,~
|\widetilde{A_j^{(N,4)}}|=N-3,~
\cdots,
|\widetilde{A_j^{(N,N)}}|=2
}}}}\nonumber\\
&\times&
\sum_{
\{B_j^{(N-1)}\}_{1\leqq j \leqq \beta_{N-1}}
\subset
\{A_j^{(N,2)}\}_{1\leqq j \leqq \alpha_N^{(2)}}
\atop{
\{B_j^{(N)}\}_{1\leqq j \leqq \beta_{N}}
\subset
\{A_j^{(N,1)}\}_{1\leqq j \leqq \alpha_N^{(1)}}
}}
\sum_{
\{B_j^{(2)}\}_{1\leqq j \leqq \beta_2}
\subset
\{\widetilde{A_j^{(t,t)}}\}_
{3\leqq t\leqq N,1\leqq j \leqq \alpha_t^{(t)}}
\atop{
\{B_j^{(3)}\}_{1\leqq j \leqq \beta_3}
\subset
\{\widetilde{A_j^{(t,t-1)}}\}_
{4\leqq t\leqq N,1\leqq j \leqq \alpha_t^{(t-1)}}
\atop{\cdots
\atop{
\{B_j^{(N-2)}\}_{1\leqq j \leqq \beta_{N-2}}
\subset
\{\widetilde{A_j^{(N,3)}}\}_
{1\leqq j \leqq \alpha_N^{(3)}}
}
}}}\nonumber\\
&\times&
(-1)^{\sum_{t=1}^{[\frac{N}{2}]}\sum_{u=1}^t \alpha_{2t}^{(2u-1)}
+\sum_{t=1}^{[\frac{N-1}{2}]}\sum_{u=1}^t
\alpha_{2t+1}^{(2u)}
+\sum_{j=2}^N \beta_j}
\prod_{t=2}^N\left(
c^{t-1}\prod_{u=1}^{t-2}
\Delta(x^{2u+1})^{t-u-1}
\prod_{u=1}^{t-1}s(x^{-2u})^{t-u}
\right)^{\alpha_t}\nonumber\\
&\times&
\left[
\prod_{t=1}^N
\prod_{q=1}^{t}
\prod_{j<k
\atop{j,k \in A_{Min}^{(t,q)}}}h_{t,t}^{\eta^{q-1},\eta^{q-1}}(z_k/z_j)
\right.
\nonumber\\
&\times&\prod_{1\leqq t<u \leqq N}
\prod_{q=1}^{t}\prod_{p=1}^{u}
\prod_{j \in A_{Min}^{(t,q)}}
\prod_{k \in A_{Min}^{(u,p)}}
h_{t,u}^{\eta^{q-1},\eta^{p-1}}(x^{u-t-2[\frac{u}{2}]+2[\frac{t}{2}]}
z_k/z_j)
\nonumber\\
&\times&
\left\{
\prod_{\longrightarrow
\atop{
j\in A_{Min}^{(1,1)}}}T_1(z_j)
\right\}
\left\{
\prod_{\longrightarrow
\atop{
j\in A_{Min}^{(2,1)}}}T_2(x^{-1}z_j)
\prod_{\longrightarrow
\atop{
j\in A_{Min}^{(2,2)}}}\eta(T_2(x^{-1}z_j))\right\}
\cdots
\nonumber\\
&\times&
\left\{\prod_{
\longrightarrow
\atop{
j\in A_{Min}^{(N,1)}}}
T_N(x^{-1+N-2[\frac{N}{2}]}z_j)
\cdots
\prod_{\longrightarrow
\atop{j\in A_{Min}^{(N,N)}}}
\eta^{N-1}(T_N(x^{-1+N-2[\frac{N}{2}]}z_j))
\right\}
\nonumber\\
&\times&
\prod_{q=1}^2
\prod_{j=1}^{\alpha_2^{(q)}}
\delta
\left(\frac{x^2z_{j_2}}{z_{j_1}}\right)
\prod_{t=3}^N
\prod_{q=1}^2
\prod_{j=1
\atop{
j_1=A_{j,1}^{(t,q)}
\atop{
\cdots
\atop{
j_t=A_{j,t}^{(t,q)}
}
}}}^{\alpha_t^{(q)}}
\sum_{\sigma \in S_{t}
\atop{\sigma(1)=1}}
\prod_{u=1\atop{u\neq [\frac{t}{2}]+1}}^t
\delta
\left(\frac{x^2z_{j_{\sigma(u+1)}}}{z_{j_{\sigma(u)}}}\right)
\nonumber\\
%%%%%%%%%%%%%%%%%%%%%%%%%%%%%%%%%%%%%%%%%
&\times&
\prod_{t=3}^N
\prod_{q=3}^{t}
\prod_{j=1
\atop{j_1=\widetilde{A_{j,1}^{(t,q)}}
\atop{\cdots
\atop{
j_{t-q+2}=\widetilde{A_{j,t-q+2}^{(t,q)}}}}}}^{\alpha_t^{(q)}}
\prod_{k_1,k_2,\cdots,k_{q-2}\in 
A_j^{(t,q)}-\widetilde{A_j^{(t,q)}}
\atop{k_1<k_2<\cdots <k_{q-2}}}
\sum_{\sigma \in S_{t-q+2}
\atop{\sigma(1)=1}}
\prod_{u=1
\atop{u\neq [\frac{t-q}{2}]+2}}^{t-q+2}
\delta
\left(\frac{x^2z_{j_{\sigma(u+1)}}}
{z_{j_{\sigma(u)}}}\right)
\sum_{\tau \in S_{q-2}}\nonumber\\
%%%%%%%%%%%%%%%%%%%%%%%%%%%%%%%%%%%%%%%%%%%%%%%%%%%%%
&\times&
\prod_{
t-q\in 2{\mathbb Z}}
\delta\left(
\frac{x^2z_{k_{\tau([\frac{q+1}{2}])}}}
{z_{j_{\sigma([\frac{t-q}{2}]+2)}}}\right)
\delta
\left(
\frac{x^2z_{j_{\sigma([\frac{t-q}{2}]+3)}}}
{z_{k_{\tau(1)}}}\right)
\prod_{u=[\frac{q+3}{2}]}^{q-2}
\delta\left(\frac{x^2z_{k_{\tau(u)}}}{
z_{k_{\tau(u-1)}}}\right)
\prod_{u=2}^{[\frac{q-1}{2}]}
\delta\left(
\frac{x^2z_{k_{\tau(u-1)}}}{
z_{k_{\tau(u)}}}\right)\nonumber\\
&\times&
%%%%%%%%%%%%%%%%%%%%%%%%%%%%%%%%%
\prod_{t-q+1 \in 2{\mathbb Z}}
\delta\left(\frac{x^2z_{k_{\tau(1)}}}{z_{j_{\sigma([\frac{t-q}{2}]+2)}}}\right)
\delta\left(\frac{x^2z_{j_{\sigma([\frac{t-q}{2}]+3)}}}{z_{
k_{\tau([\frac{q+1}{2}])}}}\right)
\prod_{u=2}^{[\frac{q-1}{2}]}
\delta\left(\frac{x^2z_{k_{\tau(u)}}}{
z_{k_{\tau(u-1)}}}\right)
\prod_{u=[\frac{q+3}{2}]}^{q-2}
\delta\left(
\frac{x^2z_{k_{\tau(u-1)}}}{
z_{k_{\tau(u)}}}\right).\nonumber\\
\label{NLeq6}
\end{eqnarray}
We note
that differnces between the equations
(\ref{NLeq5}) and (\ref{NLeq6}) are
the signature and the restriction condition
$$
\sum_{
\{B_j^{(N-1)}\}_{1\leqq j \leqq \beta_{N-1}}
\subset
\{A_j^{(N,2)}\}_{1\leqq j \leqq \alpha_N^{(2)}}
\atop{
\{B_j^{(N)}\}_{1\leqq j \leqq \beta_{N}}
\subset
\{A_j^{(N,1)}\}_{1\leqq j \leqq \alpha_N^{(1)}}
}}
\sum_{
\{B_j^{(2)}\}_{1\leqq j \leqq \beta_2}
\subset
\{\widetilde{A_j^{(t,t)}}\}_
{3\leqq t\leqq N,1\leqq j \leqq \alpha_t^{(t)}}
\atop{
\{B_j^{(3)}\}_{1\leqq j \leqq \beta_3}
\subset
\{\widetilde{A_j^{(t,t-1)}}\}_
{4\leqq t\leqq N,1\leqq j \leqq \alpha_t^{(t-1)}}
\atop{\cdots
\atop{
\{B_j^{(N-2)}\}_{1\leqq j \leqq \beta_{N-2}}
\subset
\{\widetilde{A_j^{(N,3)}}\}_
{1\leqq j \leqq \alpha_N^{(3)}}
}
}}}.
$$
Hence, summing up
every terms of $\eta([\prod_{1\leqq j<k \leqq n}s(z_k/z_j)
{\cal O}_n(z_1,z_2,\cdots,z_n)]_{1,z_1\cdots z_n})$,
we have shown $\eta({\cal I}_n)={\cal I}_n$.~~~Q.E.D.

\subsection{Proof of Dynkin-Automorphism Invariance
$\eta({\cal G}_n)={\cal G}_n$}

In this section we show
Dynkin-automorphism invariance
$\eta({\cal G}_n)={\cal G}_n$.
\\\\
{\it Proof of Theorem \ref{thm:DynkinNonlocal}}~~For reader's convenience,
we explain $\eta({\cal G}_1)={\cal G}_1$ at first.
We have the action of $\eta$ as following.
\begin{eqnarray}
&&\eta\left(
\prod_{t=1}^N 
\oint_C \frac{dz^{(t)}}{2\pi \sqrt{-1}z^{(t)}}
\frac{
\displaystyle
F_1(z^{(1)})
F_2(z^{(2)})\cdots
F_N(z^{(N)})
\vartheta(u^{(1)}|u^{(2)}|\cdots|u^{(N)})}
{\displaystyle
\prod_{t=1}^{N-1} 
\left[u^{(t)}-u^{(t+1)}+1-\frac{s}{N}\right]_r
\left[u^{(1)}-u^{(N)}+\frac{s}{N}\right]_r}\right)\nonumber\\
&=&
\prod_{t=1}^N 
\oint_C \frac{dz^{(t)}}{2\pi \sqrt{-1}z^{(t)}}
\frac{\displaystyle
\eta(\vartheta(u^{(1)}|u^{(2)}|\cdots|u^{(N)}))
}
{\displaystyle
\prod_{t=2}^{N-1} 
\left[u^{(t)}-u^{(t+1)}+1-\frac{s}{N}\right]_r
\left[u^{(N)}-u^{(1)}+1-\frac{s}{N}\right]_r
\left[u^{(2)}-u^{(1)}+\frac{s}{N}\right]_r}
\nonumber\\
&\times&
F_2(z^{(1)})
F_3(z^{(2)})\cdots
F_N(z^{(N-1)})F_1(z^{(N)}).
\end{eqnarray}
Here we have used 
\begin{eqnarray}
\eta(F_1(z_1)F_2(z_2)\cdots F_N(z_N))=
F_2(z_1)\cdots F_N(z_{N-1})F_1(z_N).
\end{eqnarray}
Let us change variables 
$u^{(1)}\to u^{(N)},
u^{(2)}\to u^{(1)},u^{(2)}\to
u^{(1)},\cdots,u^{(N)}\to u^{(N-1)}$,
and move $F_1(z^{(1)})$ from the right to the left.
We have
\begin{eqnarray}
&&
\prod_{t=1}^N 
\oint_C \frac{dz^{(t)}}{2\pi \sqrt{-1}z^{(t)}}
\frac{\displaystyle
F_1(z^{(1)})
F_2(z^{(2)})\cdots
F_N(z^{(N)})
\eta(\vartheta(u^{(2)}|u^{(3)}|\cdots|u^{(N)}|u^{(1)}))}{
\displaystyle
\prod_{t=1}^{N-1} 
\left[u^{(t)}-u^{(t+1)}+1-\frac{s}{N}\right]_r
\left[u^{(1)}-u^{(N)}+\frac{s}{N}\right]_r}.
\end{eqnarray}
We conclude $\eta({\cal G}_1)={\cal G}_1$ from
theta property $
\eta(\vartheta(u^{(2)}|u^{(3)}|\cdots|u^{(N)}|u^{(1)}))=
\vartheta(u^{(1)}|u^{(2)}|\cdots|u^{(N)})$.
Let us show $\eta({\cal G}_m)={\cal G}_m$.
After changing
the variables
$u_j^{(1)} \to u_j^{(N)}$,
$u_j^{(2)} \to u_j^{(1)}$,
$u_j^{(3)} \to u_j^{(2)},\cdots$
$u_j^{(N)} \to u_j^{(N-1)}$,
we have $\eta({\cal G}_m)$ as following.
\begin{eqnarray}
&&
\prod_{t=1}^N \prod_{j=1}^m 
\oint_C \frac{dz_j^{(t)}}{2\pi \sqrt{-1}z_j^{(t)}}
F_2(z_1^{(1)})\cdots F_2(z_m^{(1)})
F_3(z_1^{(2)})\cdots F_3(z_m^{(2)})\cdots
F_N(z_1^{(N-1)})\cdots F_N(z_m^{(N-1)})\nonumber\\
&\times&
F_1(z_1^{(N)})\cdots F_1(z_m^{(N)})
\eta\left(\vartheta\left(\sum_{j=1}^m u_j^{(1)}\right|
\left.\sum_{j=1}^m u_j^{(2)}\right|
\cdots
\left|\sum_{j=1}^m u_j^{(N)}
\right)\right)
\nonumber\\
&\times&
\frac{\displaystyle
\prod_{t=1}^N \prod_{1\leqq i<j \leqq m}
\left[u_i^{(t)}-u_j^{(t)}\right]_r
\left[u_j^{(t)}-u_i^{(t)}-1\right]_r
}{\displaystyle
\prod_{t=2}^{N-1}\prod_{i,j=1}^m 
\left[u_i^{(t)}-u_j^{(t+1)}+1-\frac{s}{N}\right]_r
\prod_{i,j=1}^m 
\left[u_i^{(N)}-u_j^{(1)}+1-\frac{s}{N}\right]_r
\prod_{i,j=1}^m 
\left[u_i^{(1)}-u_j^{(2)}+\frac{s}{N}\right]_r}.\nonumber\\
\end{eqnarray}
Here we have used
\begin{eqnarray}
&&
\eta(
F_1(z_1^{(1)})\cdots F_1(z_m^{(1)})
\cdots
F_{N-1}(z_1^{(N-1)})\cdots F_{N-1}(z_m^{(N-1)})
F_N(z_1^{(N)})\cdots F_N(z_m^{(N)}))\\
&=&
F_2(z_1^{(1)})\cdots F_2(z_m^{(1)})
F_3(z_1^{(2)})\cdots F_3(z_m^{(2)})\cdots
F_N(z_1^{(N-1)})\cdots F_N(z_m^{(N-1)})
F_1(z_1^{(N)})\cdots F_1(z_m^{(N)}).
\nonumber
\end{eqnarray}
Let us change the variables
$u_j^{(1)} \to u_j^{(2)}$,
$u_j^{(2)} \to u_j^{(3)},\cdots$
$u_j^{(N-1)} \to u_j^{(N)}$,
$u_j^{(N)} \to u_j^{(1)}$,
and move
$F_1(z_1^{(N)})\cdots F_1(z_m^{(N)})$
from the right to the left.
We have
\begin{eqnarray}
&&\prod_{t=1}^N \prod_{j=1}^m 
\oint_C \frac{dz_j^{(t)}}{2\pi \sqrt{-1}z_j^{(t)}}
F_1(z_1^{(1)})\cdots F_1(z_m^{(1)})
F_2(z_1^{(2)})\cdots F_2(z_m^{(2)})\cdots
F_N(z_1^{(N)})\cdots F_N(z_m^{(N)})\nonumber\\
&\times&
\frac{\displaystyle
\prod_{t=1}^N \prod_{1\leqq i<j \leqq m}
\left[u_i^{(t)}-u_j^{(t)}\right]_r
\left[u_j^{(t)}-u_i^{(t)}-1\right]_r
}{\displaystyle
\prod_{t=1}^{N-1}\prod_{i,j=1}^m 
\left[u_i^{(t)}-u_j^{(t+1)}+1-\frac{s}{N}\right]_r
\prod_{i,j=1}^m 
\left[u_i^{(1)}-u_j^{(N)}+\frac{s}{N}\right]_r}\nonumber\\
&\times&
\eta\left(\vartheta\left(\sum_{j=1}^m u_j^{(2)}\right|
\left.\sum_{j=1}^m u_j^{(3)}\right|
\cdots\left|\sum_{j=1}^m u_j^{(N)}
\right.
\left|\sum_{j=1}^m u_j^{(1)}\right)\right).
\end{eqnarray}
We conclude $\eta({\cal G}_m)={\cal G}_m$
from
$\eta(\vartheta(u^{(1)}|\cdots|u^{(N)}))
=
\vartheta(u^{(N)}|u^{(1)})|\cdots|u^{(N-1)})$.
Proof of $\eta({\cal G}_m^*)={\cal G}_m^*$
is given as the same manner as above.
~~~
Q.E.D.

~\\
\\
{\bf Acknowledgements.}~~
We would like to thank 
Prof. B.Feigin,
Prof. M.Jimbo and
Mr. H.Watanabe
for usefull discussions.
We would like to thank to Prof.V.Bazhanov,
Prof. A.Belavin, 
Prof. S.Duzhin, 
Prof. E.Frenkel,
Prof. K.Hasegawa,
Prof. P.Kulish,
Prof. K.Takemura and 
Prof. M.Wadati for their
intersts in this work.
T.K. is partly supported by Grant-in Aid for 
Young Scientist {\bf B} (18740092) from JSPS.
J.S. is partly supported by Grant-in Aid for 
Scientific Research {\bf C} (16540183) from JSPS.
We would like to dedicate this paper
to Prof. Tetsuji Miwa 
on the occasion on the 60th birthday.

~\\

\begin{appendix}

\section{Normal Ordering}

We summarize the normal orderings of the
basic operators.
The normal orderings for $\Lambda_j(z)$ 
are given as followings.
\begin{eqnarray}
\Lambda_i(z_1)\Lambda_i(z_2)&=&::(1-z_2/z_1)
\frac{(x^2z_2/z_1;x^{2s})_\infty 
(x^{2r+2s-2}z_2/z_1;x^{2s})_\infty 
(x^{2s-2r}z_2/z_1;x^{2s})_\infty}{
(x^{2s-2}z_2/z_1;x^{2s})_\infty
(x^{2r}z_2/z_1;x^{2s})_\infty
(x^{2-2r}z_2/z_1;x^{2s})_\infty},
\nonumber\\
&&~~~~~~~~~~~~~~~~~~~~~~~~~~~~~~~~
~~~~~~~~~~~~~~~~~~~~~~~~~~~~~~~~~~(1\leqq i \leqq N),
\\
\Lambda_i(z_1)\Lambda_j(z_2)&=&::
\frac{(x^2z_2/z_1;x^{2s})_\infty 
(x^{-2r}z_2/z_1;x^{2s})_\infty 
(x^{2r-2}z_2/z_1;x^{2s})_\infty}{
(x^{-2}z_2/z_1;x^{2s})_\infty
(x^{2r}z_2/z_1;x^{2s})_\infty
(x^{2-2r}z_2/z_1;x^{2s})_\infty}~(1\leqq i<j \leqq N),\nonumber\\
\\
\Lambda_j(z_1)\Lambda_i(z_2)&=&::
\frac{(x^{2+2s}z_2/z_1;x^{2s})_\infty 
(x^{2s-2r}z_2/z_1;x^{2s})_\infty 
(x^{2s+2r-2}z_2/z_1;x^{2s})_\infty}{
(x^{2s-2}z_2/z_1;x^{2s})_\infty
(x^{2s+2r}z_2/z_1;x^{2s})_\infty
(x^{2s+2-2r}z_2/z_1;x^{2s})_\infty},~(1\leqq i<j\leqq N).\nonumber\\
\end{eqnarray}
%%%%%%%%%%%%%%%%%%%%%%%%%%%%%%%%%%%%%%
%%%%%%%%%%%%%%%%%%%%%%%%%%%%%%%%%%%%%%
For $N\geqq 3$, the normal orderings between 
$\Lambda_j(z)$ and $E_j(z), F_j(z)$
are given as followings.
The normal orderings for $N=2$ are summarized in appendix in \cite{FKSW1}.
\begin{eqnarray}
\Lambda_j(z_1)F_j(z_2)&=&
::x^{-2r^*}\frac{(1-x^{r-2+\frac{2s}{N}j}z_2/z_1)}{
(1-x^{-r+\frac{2s}{N}j}z_2/z_1)},~~~(1\leqq j \leqq N-1)\\
F_j(z_1)\Lambda_j(z_2)&=&
::\frac{(1-x^{2-r-\frac{2s}{N}j}z_2/z_1)}
{(1-x^{r-\frac{2s}{N}j}
z_2/z_1)},~~~
(1\leqq j \leqq N-1)\\
\Lambda_{j+1}(z_1)F_j(z_2)&=&::
x^{2r^*}\frac{(1-x^{2-r+\frac{2s}{N}j}z_2/z_1)}
{(1-x^{r+\frac{2s}{N}j}z_2/z_1)},~~~(1\leqq j \leqq N-1)
\\
F_j(z_1)\Lambda_{j+1}(z_2)&=&::
\frac{(1-x^{r-2-\frac{2s}{N}j}z_2/z_1)}
{(1-x^{-r-\frac{2s}{N}j}z_2/z_1)},~~~
(1\leqq j \leqq N-1)\\
%%%%%%%%%%%%%%%%%%%%%%%%%%%%%%%%%%%%%%%%%%%%%%%%%%%%%%%%%
\Lambda_1(z_1)F_N(z_2)&=&::
x^{2r^*}\frac{(1-x^{2-r}z_2/z_1)}{
(1-x^{r}z_2/z_1)},\\
F_N(z_1)\Lambda_1(z_2)&=&:
:\frac{(1-x^{r-2}z_2/z_1)}{(1-x^{-r}z_2/z_1)},\\
\Lambda_N(z_1)F_N(z_2)&=&::
x^{-2r^*}
\frac{(1-x^{r-2+2s}z_2/z_1)}{(1-x^{-r+2s}z_2/z_1)},
\\
F_N(z_1)\Lambda_N(z_2)&=&::
\frac{(1-x^{2-r-2s}z_2/z_1)}{(1-x^{r-2s}z_2/z_1)}.
\end{eqnarray}
%%%%%%%%%%%%%%%%%%%%%%%%%%%%%%%%%%%%%%%%%%%%%%%
\begin{eqnarray}
\Lambda_j(z_1)E_j(z_2)&=&
::
x^{2r}\frac{(1-x^{-r-1+\frac{2s}{N}j}z_2/z_1)}{
(1-x^{r-1+\frac{2s}{N}j}z_2/z_1)},~~~(1\leqq j \leqq N-1)\\
E_j(z_1)\Lambda_j(z_2)&=&
::
\frac{(1-x^{r+1-\frac{2s}{N}j}z_2/z_1)}
{(1-x^{-r+1-\frac{2s}{N}j}z_2/z_1)},~~~
(1\leqq j \leqq N-1)\\
\Lambda_{j+1}(z_1)E_j(z_2)&=&::
x^{-2r}\frac{(1-x^{r+1+\frac{2s}{N}j}z_2/z_1)}
{(1-x^{-r+1+\frac{2s}{N}j}z_2/z_1)},~~~
(\leqq j \leqq N-1)
\\
E_j(z_1)\Lambda_{j+1}(z_2)&=&::
\frac{(1-x^{-r-1-\frac{2s}{N}j}z_2/z_1)}
{(1-x^{r-1-\frac{2s}{N}j}z_2/z_1)},~~~
(1\leqq j \leqq N-1)\\
%%%%%%%%%%%%%%%%%%%%%%%%%%%%%%%%%%%%%%%%%%%%%%%%%%%%%%%%%%%%%%%%%
\Lambda_1(z_1)E_N(z_2)&=&
::
x^{-2r}\frac{(1-x^{r+1}z_2/z_1)}{
(1-x^{-r+1}z_2/z_1)},\\
E_N(z_1)\Lambda_1(z_2)&=&
::
\frac{(1-x^{-r-1}z_2/z_1)}{
(1-x^{r-1}z_2/z_1)},\\
\Lambda_N(z_1)E_N(z_2)&=&::
x^{2r}\frac{(1-x^{-r^*-2+2s}z_2/z_1)}{
(1-x^{r^*+2s}z_2/z_1)},
\\
E_N(z_1)\Lambda_N(z_2)
&=&::
\frac{(1-x^{r^*+2-2s}z_2/z_1)}{(1-x^{-r^*-2s}z_2/z_1)},
\end{eqnarray}
%%%%%%%%%%%%%%%%%%%%%%%%%%%%%%%%%%%%%%%%%%%%%%%%%%%%%%%
%%%%%%%%%%%%%%%%%%%%%%%%%%%%%%%%%%%%%%%%%%%%%%%%%%%%%%%
For $N\geqq 3$, the normal orderings of
$E_j(z), F_j(z)$
are given as followings.
The normal orderings for $N=2$ are summarized in appendix in \cite{FKSW1}.
For ${\rm Re}(r^*)>0$ we have
\begin{eqnarray}
E_j(z_1)E_j(z_2)&=&::
z_1^{\frac{2r}{r^*}}(1-z_2/z_1)\frac{(x^{-2}z_2/z_1;x^{2r^*})_\infty}{
(x^{2r^*+2}z_2/z_1;x^{2r^*})_\infty}
,~(1\leqq j \leqq N)\\
E_j(z_1)E_{j+1}(z_2)&=&::
(x^{\frac{2s}{N}-j}z_1)^{\frac{r}{r^*}}
\frac{(x^{2r-2+\frac{2s}{N}}\frac{z_2}{z_1};x^{2r^*})_\infty}
{(x^{\frac{2s}{N}-2}z_2/z_1;x^{2r^*})_\infty}
,~(1\leqq j \leqq N)\nonumber\\
E_{j+1}(z_1)E_{j}(z_2)&=&::
(x^{\frac{2s}{N}-j-1}z_1)^{\frac{r}{r^*}}
\frac{(x^{2r-\frac{2s}{N}}\frac{z_2}{z_1};x^{2r^*})_\infty}
{(x^{-\frac{2s}{N}}z_2/z_1;x^{2r^*})_\infty}
,~(1\leqq j \leqq N).
\end{eqnarray}
For ${\rm Re}(r^*)<0$ we have
\begin{eqnarray}
E_j(z_1)E_j(z_2)&=&::
z_1^{\frac{2r}{r^*}}(1-z_2/z_1)
\frac{(x^{2}z_2/z_1;x^{-2r^*})_\infty}{
(x^{-2r^*-2}z_2/z_1;x^{-2r^*})_\infty}
,~(1\leqq j \leqq N)\\
E_j(z_1)E_{j+1}(z_2)&=&::
(x^{\frac{2s}{N}-j}z_1)^{\frac{r}{r^*}}
\frac{(x^{-2r^*-2+\frac{2s}{N}}\frac{z_2}{z_1};x^{-2r^*})_\infty}
{(x^{\frac{2s}{N}}z_2/z_1;x^{-2r^*})_\infty}
,~(1\leqq j \leqq N)\nonumber\\
E_{j+1}(z_1)E_{j}(z_2)&=&::
(x^{\frac{2s}{N}-j-1}z_1)^{\frac{r}{r^*}}
\frac{(x^{-2r^*-\frac{2s}{N}}\frac{z_2}{z_1};x^{-2r^*})_\infty}
{(x^{2-\frac{2s}{N}}z_2/z_1;x^{-2r^*})_\infty}
,~(1\leqq j \leqq N).
\end{eqnarray}
%%%%%%%%%%%%%%%%%%%%%%%%%%%%%%%%%%%%%%%%%%%%%%%%%
For ${\rm Re}(r)>0$ we have
\begin{eqnarray}
F_j(z_1)F_j(z_2)&=&::x^{\frac{2r^*}{r}}
(1-z_2/z_1)\frac{(x^2z_2/z_1;x^{2r})_\infty}{
(x^{2r-2}z_2/z_1;x^{2r})_\infty},~~(1\leqq j \leqq N)\\
F_j(z_1)F_{j+1}(z_2)&=&::
(x^{\frac{2s}{N}-j}z_1)^{-\frac{r^*}{r}}
\frac{(x^{2r-2+\frac{2s}{N}}z_2/z_1;x^{2r})_\infty}
{(x^{\frac{2s}{N}}z_2/z_1;x^{2r})_\infty},~~
(1\leqq j \leqq N)\\
F_{j+1}(z_1)F_j(z_2)&=&::(x^{\frac{2s}{N}-j-1}z_1)^{-\frac{r^*}{r}}
\frac{(x^{2r-2+s}z_2/z_1;x^{2r})_\infty }{
(x^{2-\frac{2s}{N}}z_2/z_1;x^{2r})_\infty},~~(1\leqq j \leqq N).
\end{eqnarray}
%%%%%%%%%%%%%%%%%%%%%%%%%%%%%%%%%%%%%%%%%%%%
For ${\rm Re}(r)<0$ we have
\begin{eqnarray}
F_j(z_1)F_j(z_2)&=&::x^{\frac{2r^*}{r}}
(1-z_2/z_1)\frac{(x^{-2}z_2/z_1;x^{-2r})_\infty}{
(x^{2-2r}z_2/z_1;x^{-2r})_\infty},~~(1\leqq j \leqq N)\\
F_j(z_1)F_{j+1}(z_2)&=&::
(x^{\frac{2s}{N}-j}z_1)^{-\frac{r^*}{r}}
\frac{(x^{-2r+\frac{2s}{N}}z_2/z_1;x^{-2r})_\infty}
{(x^{-2+\frac{2s}{N}}z_2/z_1;x^{-2r})_\infty},~~
(1\leqq j \leqq N)\\
F_{j+1}(z_1)F_j(z_2)&=&::(x^{\frac{2s}{N}-j-1}z_1)^{-\frac{r^*}{r}}
\frac{(x^{-2r+2-\frac{2s}{N}}z_2/z_1;x^{-2r})_\infty }{
(x^{-\frac{2s}{N}}z_2/z_1;x^{-2r})_\infty},~~
(1\leqq j \leqq N).
\end{eqnarray}
%%%%%%%%%%%%%%%%%%%%%%%%%%%%%%%%%%%%%%%%%%%%%
For $N \geqq 3$ the normal orderings
of the screenings $E_j(z)$, $F_j(z)$ are given by
\begin{eqnarray}
E_j(z_1)F_j(z_2)&=&
::\frac{x^{(1-\frac{2s}{N})2j}}
{z_1^2(1-xz_2/z_1)(1-x^{-1}z_2/z_1)},~~
(1\leqq j \leqq N),
\\
F_j(z_1)E_j(z_2)&=&
::\frac{x^{(1-\frac{2s}{N})2j}}
{z_1^2(1-xz_2/z_1)(1-x^{-1}z_2/z_1)},~~
(1\leqq j \leqq N),
\\
E_j(z_1)F_{j+1}(z_2)
&=&::x^{(\frac{2s}{N}-1)j}z_1
(1-x^{-1+\frac{2s}{N}}z_2/z_1),~(1\leqq j \leqq N),\\
F_{j+1}(z_1)E_{j}(z_2)
&=&::x^{(\frac{2s}{N}-1)(j+1)}z_1
(1-x^{1-\frac{2s}{N}}z_2/z_1),~(1\leqq j \leqq N),\\
E_{j+1}(z_1)F_{j}(z_2)
&=&::x^{(\frac{2s}{N}-1)(j+1)}
z_1(1-x^{-1+\frac{2s}{N}}z_2/z_1),~(1\leqq j \leqq N),\\
E_j(z_1)F_{j+1}(z_2)
&=&::x^{(\frac{2s}{N}-1)j}z_1
(1-x^{1-\frac{2s}{N}}z_2/z_1),~(1\leqq j \leqq N).
\end{eqnarray}
%%%%%%%%%%%%%%%%%%%%%%%%%%%%%%%%%%%%%%%%%%%
For $N\geqq3$ we have
\begin{eqnarray}
{\cal A}_j(z_1)F_j(z_2)=::(x^{-j}z_1)^{\frac{2r^*}{r}}
\frac{(x^{-r+2+\frac{2s}{N}j}z_2/z_1;x^{2r})_\infty}{
(x^{3r-2+\frac{2s}{N}j}z_2/z_1;x^{2r})_\infty},\nonumber\\
(1\leqq j \leqq N-1),\\
F_j(z_1){\cal A}_j(z_2)=::
(x^{(\frac{2s}{N}-1)j}z_1)^{\frac{2r^*}{r}}
\frac{(x^{-r+2-\frac{2s}{N}j}z_2/z_1;x^{2r})_\infty}{
(x^{3r-2-\frac{2s}{N}j}z_2/z_1;x^{2r})_\infty},\nonumber\\
(1\leqq j \leqq N-1),
\end{eqnarray}
%%%%%%%%%%%%%%%%%%%%%%%%%%%%%%%%%%%%%%%%%%%%%%%%%
\begin{eqnarray}
{\cal A}_j(z_1)F_{j+1}(z_2)=::
x^{-r^*(1-\frac{1}{N})}(x^{-j}z_1)^{-\frac{r^*}{r}}
\frac{(x^{r-2+\frac{2s}{N}(j+1)}z_2/z_1;x^{2r})_\infty}{
(x^{-r+\frac{2s}{N}(j+1)}z_2/z_1;x^{2r})_\infty},\nonumber\\
(1\leqq j \leqq N-1),\\
F_{j+1}(z_1){\cal A}_j(z_2)=::
(x^{(\frac{2s}{N}-1)(j+1)}z_1)^{-\frac{r^*}{r}}
\frac{(x^{3r-\frac{2s}{N}(j+1)}z_2/z_1;x^{2r})_\infty}{
(x^{r+2-\frac{2s}{N}(j+1)}z_2/z_1;x^{2r})_\infty},\nonumber\\
(1\leqq j \leqq N-2),\\
F_N(z_1){\cal A}_{N-1}(z_2)=::
(x^{2s-N}z_1)^{-\frac{r^*}{r}(1-\frac{1}{N})}
\frac{(x^{3r-2s}z_2/z_1;x^{2r})_\infty}{
(x^{r+2-2s}z_2/z_1;x^{2r})_\infty},
\end{eqnarray}
%%%%%%%%%%%%%%%%%%%%%%%%%%%%%%%%%%%%%%%%%%%%%%%%
\begin{eqnarray}
{\cal A}_j(z_1)F_{j-1}(z_2)=::
x^{r^*(1-\frac{1}{N})}
(x^{-j}z_1)^{-\frac{r^*}{r}}
\frac{(x^{3r+\frac{2s}{N}(j-1)}z_2/z_1;x^{2r})_\infty}{
(x^{r+2+\frac{2s}{N}(j-1)}z_2/z_1;x^{2r})_\infty},\nonumber\\
(1\leqq j \leqq N-1),\\
F_{j-1}(z_1){\cal A}_j(z_2)=::
(x^{(\frac{2s}{N}-1)(j-1)}z_1)^{-\frac{r^*}{r}}
\frac{(x^{r-2-\frac{2s}{N}(j-1)}z_2/z_1;x^{2r})_\infty}{
(x^{-r-\frac{2s}{N}(j-1)}z_2/z_1;x^{2r})_\infty},\nonumber\\
(2\leqq j \leqq N-1),\\
F_N(z_1){\cal A}_{1}(z_2)=::
z_1^{-\frac{r^*}{r}(1-\frac{1}{N})}
\frac{(x^{r-2}z_2/z_1;x^{2r})_\infty}{
(x^{-r}z_2/z_1;x^{2r})_\infty},
\end{eqnarray}
%%%%%%%%%%%%%%%%%%%%%%%%%%%%%%%%%%%%%%%%%%%%%%%%
\begin{eqnarray}
{\cal B}_j(z_1)E_j(z_2)=::(x^{-j}z_1)^{\frac{2r}{r^*}}
\frac{(x^{-r^*-2+\frac{2s}{N}j}z_2/z_1;x^{2r^*})_\infty}{
(x^{3r^*+2+\frac{2s}{N}j}z_2/z_1;x^{2r^*})_\infty},\nonumber\\
(1\leqq j \leqq N-1),\\
E_j(z_1){\cal B}_j(z_2)=::
(x^{(\frac{2s}{N}-1)j})^{\frac{2r}{r^*}}
\frac{(x^{-r^*-2-\frac{2s}{N}j}z_2/z_1;x^{2r^*})_\infty}{
(x^{3r^*+2-\frac{2s}{N}j}z_2/z_1;x^{2r^*})_\infty},\nonumber\\
(1\leqq j \leqq N-1),
\end{eqnarray}
%%%%%%%%%%%%%%%%%%%%%%%%%%%%%%%%%%%%%%%%%%%%%%%%%
\begin{eqnarray}
{\cal B}_j(z_1)E_{j+1}(z_2)=::
x^{r(1-\frac{1}{N})}(x^{-j}z_1)^{-\frac{r}{r^*}}
\frac{(x^{3r^*+\frac{2s}{N}(j+1)}z_2/z_1;x^{2r^*})_\infty}{
(x^{r^*-2+\frac{2s}{N}(j+1)}z_2/z_1;x^{2r^*})_\infty},\nonumber\\
(1\leqq j \leqq N-1),\\
E_{j+1}(z_1){\cal B}_j(z_2)=::
(x^{(\frac{2s}{N}-1)(j+1)}z_1)^{-\frac{r}{r^*}}
\frac{(x^{r^*-\frac{2s}{N}(j+1)}z_2/z_1;x^{2r^*})_\infty}{
(x^{-r^*+2-\frac{2s}{N}(j+1)}z_2/z_1;x^{2r^*})_\infty},\nonumber\\
(1\leqq j \leqq N-2),\\
E_N(z_1){\cal B}_{N-1}(z_2)=::
(x^{2s-N}z_1)^{-\frac{r}{r^*}(1-\frac{1}{N})}
\frac{(x^{r^*-2s}z_2/z_1;x^{2r^*})_\infty}{
(x^{-r^*+2-2s}z_2/z_1;x^{2r^*})_\infty},
\end{eqnarray}
%%%%%%%%%%%%%%%%%%%%%%%%%%%%%%%%%%%%%%%%%%%%%%%%
\begin{eqnarray}
{\cal B}_j(z_1)E_{j-1}(z_2)=::
x^{-r(1-\frac{1}{N})}(x^{-j}z_1)^{-\frac{r}{r^*}}
\frac{(x^{r^*+2+\frac{2s}{N}(j-1)}z_2/z_1;x^{2r^*})_\infty}{
(x^{-r^*+\frac{2s}{N}(j-1)}z_2/z_1;x^{2r^*})_\infty},\nonumber\\
(1\leqq j \leqq N-1),\\
E_{j-1}(z_1){\cal B}_j(z_2)=::
(x^{(\frac{2s}{N}-1)(j-1)})^{-\frac{r}{r^*}}
\frac{(x^{3r^*-\frac{2s}{N}(j-1)}z_2/z_1;x^{2r^*})_\infty}{
(x^{r^*-2-\frac{2s}{N}(j-1)}z_2/z_1;x^{2r^*})_\infty},\nonumber\\
(2\leqq j \leqq N-1),\\
F_N(z_1){\cal A}_{1}(z_2)=::
z_1^{-\frac{r}{r^*}(1-\frac{1}{N})}
\frac{(x^{3r^*}z_2/z_1;x^{2r^*})_\infty}{
(x^{r^*-2}z_2/z_1;x^{2r^*})_\infty}.
\end{eqnarray}
%%%%%%%%%%%%%%%%%%%%%%%%%%%%%%%%%%%%%%%%%%%%%%%%
For ${\rm Re}(r)>0$ we have
\begin{eqnarray}
F_{j-1}(z_1)H_j(z_2)&=&::
(x^{(\frac{2s}{N}-1)(j-1)}z_1)^{\frac{1}{r}}
\frac{(x^{r-2+\frac{2s}{N}}z_2/z_1;x^{2r})_\infty}
{(x^{r+\frac{2s}{N}}z_2/z_1;x^{2r})_\infty},\\
F_j(z_1)H_j(z_2)&=&::
(x^{(\frac{2s}{N}-1)j}z_1)^{-\frac{2}{r}}
\frac{(x^{r+2}z_2/z_1;x^{2r})_\infty}
{(x^{r-2}z_2/z_1;x^{2r})_\infty}
,\\
H_j(z_1)F_{j}(z_2)&=&::
(x^{(\frac{2s}{N}-1)j}z_1)^{-\frac{2}{r}}
\frac{(x^{r+2}z_2/z_1;x^{2r})_\infty}
{(x^{r-2}z_2/z_1;x^{2r})_\infty},\\
H_j(z_1)F_{j+1}(z_2)&=&::
(x^{(\frac{2s}{N}-1)j}z_1)^{\frac{1}{r}}
\frac{(x^{r-2+\frac{2s}{N}}z_2/z_1;x^{2r})_\infty}
{(x^{r+\frac{2s}{N}}z_2/z_1;x^{2r})_\infty},
\end{eqnarray}
For ${\rm Re}(r^*)<0$ we have
\begin{eqnarray}
E_{j-1}(z_1)H_j(z_2)&=&::
(x^{(\frac{2s}{N}-1)(j-1)}z_1)^{-\frac{1}{r^*}}
\frac{(x^{-r^*-2+\frac{2s}{N}}z_2/z_1;x^{-2r^*})_\infty}
{(x^{-r^*+\frac{2s}{N}}z_2/z_1;x^{-2r^*})_\infty},\\
E_j(z_1)H_j(z_2)&=&::(x^{(\frac{2s}{N}-1)j}z_1)^{\frac{2}{r^*}}
\frac{(x^{-r^*+2}z_2/z_1;x^{-2r^*})_\infty}
{(x^{-r^*-2}z_2/z_1;x^{-2r^*})_\infty}
,\\
H_j(z_1)E_{j}(z_2)&=&::
(x^{(\frac{2s}{N}-1)j}z_1)^{\frac{2}{r^*}}
\frac{(x^{-r^*+2}z_2/z_1;x^{-2r^*})_\infty}
{(x^{-r^*-2}z_2/z_1;x^{-2r^*})_\infty},\\
H_j(z_1)E_{j+1}(z_2)&=&::
(x^{(\frac{2s}{N}-1)j}z_1)^{-\frac{1}{r^*}}
\frac{(x^{-r^*-2+\frac{2s}{N}}z_2/z_1;x^{-2r^*})_\infty}
{(x^{-r^*+\frac{2s}{N}}z_2/z_1;x^{-2r^*})_\infty}.
\end{eqnarray}

\end{appendix}

\end{document}